\documentclass[aos,preprint]{imsart}
\pdfoutput=1                    

\usepackage[OT1]{fontenc}
\usepackage{amsthm,amsmath,graphicx,enumerate}
\usepackage{natbib}
\usepackage{amssymb,tabularx,multicol,multirow,booktabs}
\usepackage{mhequ}
\usepackage{relsize}
\usepackage[title,toc,page]{appendix}
\usepackage{xr-hyper} 
\usepackage[colorlinks,citecolor=blue,urlcolor=blue,filecolor=blue]{hyperref}  
\usepackage{color,soul,subfigure}
\usepackage{rotating}

\usepackage{float}
\restylefloat{table}
\restylefloat{figure}

\arxiv{arXiv:1701.04889}

\startlocaldefs
\numberwithin{equation}{section}
\theoremstyle{plain}

\endlocaldefs

\newtheorem{theorem}[]{Theorem}[section]
\newtheorem{lemma}[]{Lemma}[section]
\newtheorem{assumption}[]{Assumption}[section]
\newtheorem{definition}[]{Definition}[section]
\newtheorem{remark}[]{Remark}[section]

\def\bx{\mathbf{x}}
\def\bX{\mathbf{X}}

\def\bz{\mathbf{z}}
\def\bZ{\mathbf{Z}}
\def\bu{\mathbf{u}}
\def\bU{\mathbf{U}}
\def\bv{\mathbf{v}}
\def\bV{\mathbf{V}}
\def\bp{\mathbf{p}}
\def\bP{\mathbf{P}}

\def\bM{\mathbf{M}}

\def\K{\mathbb{K}}
\def\E{\mathbb{E}}
\def\G{\mathbb{G}}
\def\M{\mathbb{M}}

\def\P{\mathbb{P}}
\def\T{\mathbb{T}}
\def\A{\mathbb{A}}

\def\bxv{\overrightarrow{\bx}}
\def\bXv{\overrightarrow{\bX}}

\def\btheta{\boldsymbol{\theta}}

\def\bnabla{\boldsymbol{\nabla}}
\def\bpsi{\boldsymbol{\psi}}
\def\bSigma{\boldsymbol{\Sigma}}
\def\bxi{\boldsymbol{\xi}}
\def\blambda{\boldsymbol{\lambda}}

\def\bthetahat{\widehat{\btheta}}

\def\etapr{\boldsymbol{\eta}_{\bP_r}}
\def\etahatprk{\widehat{\boldsymbol{\eta}}_{(\bP_{r},\K)}}
\def\thetahatprk{\bthetahat_{(\bP_{r},\K)}}

\def\Phat{\widehat{\mathbf{P}}}
\def\Deltahat{\widehat{\Delta}}
\def\deltahat{\widehat{\delta}}

\def\deltabar{\bar{\delta}}

\def\Mhat{\widehat{\M}}

\def\bXPr{\bX_{\bP_r}}

\def\chiPr{\mathcal{X}_{\bP_r}}

\def\mhat{\widehat{m}}
\def\lhat{\widehat{l}}
\def\fhat{\widehat{f}}

\def\mtil{\widetilde{m}}
\def\ltil{\widetilde{l}}
\def\ftil{\widetilde{f}}

\def\mPr{m_{{\bP_r}}}
\def\lPr{l_{{\bP_r}}}
\def\fPr{f_{\bP_r}}

\def\mPrtil{\widetilde{m}_{\bP_r}}
\def\lPrtil{\widetilde{l}_{\bP_r}}
\def\fPrtil{\widetilde{f}_{\bP_r}}

\def\vphiPr{\varphi_{\bP_r}}
\def\vphiPrtil{\vphitil_{\bP_r}}
\def\vphitil{\widetilde{\varphi}}
\def\vphihat{\widehat{\varphi}}

\def\bMhat{\widehat{\bM}}

\def\bpsil{\bpsi_{[l]}}
\def\bpsizerol{\bpsi_{0[l]}}

\newcommand\ind{\protect\mathpalette{\protect\independenT}{\perp}}
\def\independenT#1#2{\mathrel{\rlap{$#1#2$}\mkern4mu{#1#2}}}

\newcommand\given[1][]{\:#1\vert\:}
\newcommand\medgiven{\hspace{0.25mm}\vert\hspace{0.25mm}}
\newcommand\smallgiven{\hspace{0.1mm}\vert\hspace{0.1mm}}

\def\tcr{\textcolor{red}}

\def\Ssc{\mathcal{S}}
\def\Lsc{\mathcal{L}}
\def\Usc{\mathcal{U}}
\def\Isc{\mathcal{I}}
\def\Xsc{\mathcal{X}}
\def\Psc{\mathcal{P}}
\def\Pschat{\widehat{\Psc}}

\def\nk{n_\K}
\def\nkminus{n_\K^-}

\def\cnkminus{c_{n_\K^-}}

\def\Dhat{\widehat{D}}
\def\deltatil{\widetilde{\delta}}

\def\bTtil{\widetilde{\bT}}
\def\bThat{\widehat{\bT}}

\def\bzeta{\boldsymbol{\zeta}}
\def\bzetahat{\widehat{\bzeta}}

\def\ahat{\widehat{a}}
\def\atil{\widetilde{a}}
\def\bhat{\widehat{b}}
\def\btil{\widetilde{b}}

\def\Zhat{\widehat{\mathbb{Z}}}
\def\Ztil{\widetilde{\mathbb{Z}}}
\def\Util{\widetilde{\mathbb{U}}}

\def\ihat{\widehat{i}}

\def\Var{\mbox{Var}}

\def\half{\frac{1}{2}}

\def\nhalf{n^{\half}}
\def\nnhalf{n^{-\half}}
\def\ninv{n^{-1}}

\def\Nnhalf{N^{-\half}}
\def\Ninv{N^{-1}}

\def\Khalf{\K^{\half}}
\def\Knhalf{\K^{-\half}}
\def\Kinv{\K^{-1}}

\def\bone{\mathbf{1}}
\def\bzero{\mathbf{0}}
\def\bb{\mathbf{b}}

\def\Tsc{\mathcal{T}}

\def\Msc{\mathcal{M}}
\def\bGamma{\boldsymbol{\Gamma}}
\def\bphi{\boldsymbol{\phi}}

\def\bT{\mathbf{T}}
\def\bThat{\widehat{\bT}}
\def\bR{\mathbf{R}}
\def\bRhat{\widehat{\bR}}
\def\bS{\mathbf{S}}
\def\bShat{\widehat{\bS}}

\def\bg{\mathbf{g}}
\def\bG{\mathbf{G}}
\def\bghat{\widehat{\bg}}
\def\bGhat{\widehat{\bG}}

\def\Gschat{\widehat{\mathcal{G}}}

\def\bGbar{\overline{\bG}}
\def\Gscbar{\overline{\mathcal{G}}} %
\def\S{\mathbb{S}}
\def\That{\widehat{T}}

\def\eff{\mbox{\small eff}}
\def\bDelta{\boldsymbol{\Delta}}
\def\bDeltahat{\widehat{\bDelta}}
\def\bDeltabar{\overline{\bDelta}}
\def\Vtil{\widetilde{\mathbb{V}}}
\def\Hbb{\mathbb{H}}
\def\bH{\mathbf{H}}
\def\bw{\mathbf{w}}
\def\bwix{\bw_{i,\bx}}
\def\bMhat{\widehat{\bM}}

\def\deltabar{\overline{\delta}}

\def\bmu{\boldsymbol{\mu}}

\def\R{\mathbb{R}}

\def\Msc{\mathcal{M}}
\def\Nsc{\mathcal{N}}
\def\T{\mathbb{T}}

\def\S{\mathbb{S}}
\def\G{\mathbb{G}}

\def\Nsc{\mathcal{N}}

\def\Z{\mathbb{Z}}

\def\Ztil{\widetilde{\Z}}
\def\bxi{\boldsymbol{\xi}}

\def\Lsc{\mathcal{L}}

\def\bSigma{\boldsymbol{\Sigma}}

\def\bGammainv{\bGamma^{-1}}
\def\bGammahat{\widehat{\bGamma}}
\def\X{\mathbb{X}}
\def\bPbar{\overline{\bP}}
\def\bSigmahat{\widehat{\bSigma}}
\def\bmuhat{\widehat{\bmu}}

\def \hs2{\hspace{2mm}}

\numberwithin{table}{section}
\numberwithin{equation}{section}

\definecolor{jcolor}{RGB}{041,122,000}
\definecolor{darkred}{RGB}{100,000,000}
\definecolor{purple}{RGB}{200,000,200}

\def\boxit#1{\vbox{\hrule\hbox{\vrule\kern6pt  \vbox{\kern6pt#1\kern6pt}\kern6pt\vrule}\hrule}}

\def\PZ{\P_{\bZ}}

\def\muhat{\widehat{\mu}}
\def\bthetahatmar{\bthetahat_{\mbox{\tiny{MAR}}}}

\def\bPhat{\widehat{\bP}}
\def\muhat{\widehat{\mu}}

\def\RN{R}
\def\piN{\pi_\Nbar}
\def\PNbar{\P_\Nbar}

\def\sumiNbar{\sum_{i=1}^\Nbar}
\def\Sbb{\mathbb{S}}
\def\muhat{\widehat{\mu}}
\def\bthetahatmar{\bthetahat_{\mbox{\tiny{MAR}}}}
\def\behat{\widehat{\be}}
\def\be{\mathbf{e}}

\def\Nbar{{\bar{N}}}
\def\Nbarinv{{\bar{N}}^{-1}}

\def\Nbarnhalf{\Nbar^{-\half}}
\def\bT{\mathbf{T}}
\def\bS{\mathbf{S}}
\def\bG{\mathbf{G}}

\begin{document}

\begin{frontmatter}

\title{Efficient and Adaptive Linear Regression in Semi-Supervised Settings}
\runtitle{Semi-Supervised Linear Regression}

\begin{aug}
 \author{\fnms{Abhishek} \snm{Chakrabortty}\corref{}\ead[label=e1]{abhich@mail.med.upenn.edu}}\thanksref{t1,t2}
 \and
 \author{\fnms{Tianxi} \snm{Cai}\corref{}\ead[label=e2]{tcai@hsph.harvard.edu}\thanksref{t2}}
\thankstext{t1}{Corresponding author; previously at Harvard University during the time of this work.} \thankstext{t2}{This research was partially supported by National Institutes of Health grants R01 HL089778, U54 HG007963, and U01 HL121518.}
 \affiliation{University of Pennsylvania and Harvard University}
\runauthor{A. Chakrabortty \and T. Cai}

\address{Abhishek Chakrabortty\\
Department of Statistics\\
University of Pennsylvania\\
3730 Walnut Street\\
Jon M. Huntsman Hall, 4th Floor\\
Philadelphia, PA 19104, USA.\\
\printead{e1}\\
\phantom{E-mail:\ }}

\address{Tianxi Cai\\
Department of Biostatistics\\
Harvard University\\
655 Huntington Avenue\\
Building 2, 4th Floor\\
Boston, MA 02115, USA.\\
\printead{e2}\\
\phantom{E-mail:\ }}
\end{aug}



\begin{abstract}
We consider the linear regression problem under semi-supervised settings wherein the available data typically consists of: (i) a small or moderate sized `labeled' data, and (ii) a \emph{much larger} sized `unlabeled' data. Such data arises naturally from settings where the outcome, unlike the covariates, is expensive to obtain, a frequent scenario in modern studies involving large databases like electronic medical records (EMR). Supervised estimators like the ordinary least squares (OLS) estimator utilize only the labeled data. It is often of interest to investigate if and when the unlabeled data can be exploited to improve estimation of the regression parameter in the adopted linear model.

In this paper, we propose a class of `Efficient and Adaptive Semi-Supervised Estimators' (EASE) to improve estimation efficiency. The EASE are two-step estimators adaptive to model mis-specification, leading to improved (optimal in some cases) efficiency under model mis-specification, and equal (optimal) efficiency under a linear model. This adaptive property, often unaddressed in the existing literature, is 
crucial for advocating `safe' use of the unlabeled data. The construction of EASE primarily involves a flexible `semi-non-parametric' imputation, including a smoothing step that works well even when the number of covariates is not small; and a follow up `refitting' step along with a cross-validation (CV) strategy both of which have useful practical as well as theoretical implications towards addressing two important issues: under-smoothing and over-fitting. 
We establish asymptotic results including consistency, asymptotic normality and the adaptive properties of EASE. We also provide influence function expansions and a `double' CV strategy for inference. The results are further validated through extensive simulations, followed by application to an EMR study on auto-immunity.
\end{abstract}

\end{frontmatter}

\section{Introduction}\label{intro}

In recent years, semi-supervised learning (SSL) has emerged as an exciting new area of research in statistics and machine learning. A detailed discussion on SSL including its practical relevance, the primary question of interest in SSL, and the existing relevant literature can be found in \citet{Chapelle_2006} and \citet{Zhu_2008}. A typical semi-supervised (SS) setting is characterized by two types of available data: (i) a small or moderate sized `labeled' data, $\Lsc$, containing observations for both an outcome $Y$ and a set of covariates $\bX$ of interest, and (ii) an `unlabeled' data, $\Usc$, of \emph{much larger} size but having observations \emph{only} for the covariates $\bX$.  By virtue of its large size, $\Usc$ essentially gives us the distribution of $\bX$, denoted henceforth by $\P_{\bX}$. Such a setting arises naturally whenever the covariates are easily available so that unlabeled data is plentiful, but the outcome is costly or difficult to obtain, thereby limiting the size of $\Lsc$. This scenario is directly relevant to a variety of practical problems, especially in the modern `big data' era, with massive unlabeled datasets (often electronically recorded) becoming increasingly available and tractable. A few familiar examples include machine learning problems like text mining, web page classification, speech recognition, natural language processing etc.

Among biomedical applications, a particularly interesting problem where SSL can be of great use is the statistical
analysis of electronic medical records (EMR) data. Endowed with a wealth of de-identified clinical and phenotype data for large patient cohorts, EMR linked with bio-repositories are increasingly gaining popularity as rich resources of data for discovery research \citep{Kohane_2011}. Such large scale datasets obtained in a cost-effective and timely manner are of great importance in modern medical research for addressing important questions such as the biological role of genetic variants in disease susceptibility and progression \citep{Kohane_2011}. However, one major bottleneck impeding EMR driven research is the difficulty in obtaining validated phenotype information \citep{Liao_2010} since they are labor intensive or expensive to obtain. Thus, gold standard labels and genomic measurements are typically available only for a small subset nested within a large cohort. In contrast, digitally recorded data on the clinical variables are often available on all subjects, highlighting the necessity and utility of developing robust SSL methods that can leverage such rich source of auxiliary information to improve phenotype definition and estimation precision.

SSL primarily distinguishes from standard supervised methods by making use of $\Usc$, an information that is ignored by the latter. The ultimate question of interest in SSL is to investigate if and when the information on $\P_{\bX}$ in $\Usc$ 
can be exploited to improve the efficiency over a given supervised approach. In recent years, several graph based non-parametric SSL approaches have been proposed \citep{Zhu_2005a, Belkin_2006} for regression or classification. These approaches essentially target non-parametric SS estimation of $\E(Y \medgiven \bX)$ and therefore, for provable improvement guarantees, must rely implicitly or explicitly on assumptions relating $\P_{\bX}$ to $\P_{Y \smallgiven \bX}$ (the conditional distribution of $Y$ given $\bX$), as duly noted and characterized more formally in 
\citet{Lafferty_2007}. For non-parametric classification problems, the theoretical underpinnings of SSL including its scope and the consequences of using $\Usc$ have been also studied earlier by 
\citet{Castelli_1995,Castelli_1996}. More parametric SS approaches, still aimed mostly at prediction, have also been studied for classification, including the `generative model' approach \citep{Nigam_2000,Nigam_2001} which is based on modeling the joint distribution of $(Y,\bX)$ as an identifiable mixture of parametric models, thereby implicitly relating $\P_{Y \smallgiven \bX}$ and $\P_{\bX}$. However, these approaches depend strongly on the validity of the assumed mixture model, violation of which can actually \emph{degrade} their performance compared to the supervised approach \citep{Cozman_2001,Cozman_2003}.

%

However SS \emph{estimation problems}, especially from a semi-parametric point of view, has been somewhat less studied in SSL. Such problems are generally aimed at estimating some (finite-dimensional) parameter $\theta_0 \equiv \theta_0(\P)$, where $\P = (\P_{Y \smallgiven \bX},\P_{\bX})$, and the key to the potential usefulness of $\Usc$ in improving estimation of $\theta_0$ lies in understanding when $\theta_0(\P)$ relates to $\P_{\bX}$. For simple parameters like $\theta_0(\P) = \E(Y)$, unless $\E(Y \medgiven \bX)$ is a constant, $\theta_0$ clearly depends on $\P_{\bX}$ and hence, improved SS estimation is possible compared to the supervised estimator $\overline{Y}_{\Lsc}$, the sample mean of $Y$ based on $\Lsc$.
The situation is however more subtle for other choices of $\theta_0$, especially those where $\theta_0$ is the target parameter corresponding to an underlying parametric \emph{working} model for $\P_{Y \smallgiven \bX}$. This includes the least squares parameter, as studied in this paper, targeted by a working linear model for $\E(Y \medgiven \bX)$. Such models are often adopted due to their appealing simplicity and interpretability.

In general, for such cases, if the adopted working model for $\P_{Y \smallgiven \bX}$ is correct and $\theta_0$ is not related to $\P_{\bX}$, then one \emph{cannot} possibly gain through SSL by using the knowledge of $\P_{\bX}$ \citep{Zhang_2000, Seeger_2002}. On the other hand, under model mis-specification, $\theta_0$ 
may inherently \emph{depend} on $\P_{\bX}$, and thus imply the potential utility of $\Usc$ in improving the estimation. However, inappropriate use of $\Usc$ may lead to degradation of the estimation precision. This therefore signifies the need for \emph{robust} and efficient SS estimators that are \emph{adaptive} to model mis-specification, so that they are as efficient as the supervised estimator under the correct model and more efficient under model mis-specification. To the best of our knowledge, work done along these lines is relatively scarce in the SSL literature, one notable exception being the recent work of \citet{Kawakita_2013}, where they use a very different approach based on density ratio estimation, building on the more restrictive approach of \citet{Sokolovska_2008}. 
However, as we observe in our simulation studies, the extent of the efficiency gain actually achieved by these approaches can be quite incremental, 
at least in finite samples. Further, the seemingly unclear choice of the ideal (nuisance) model to be used for density ratio estimation can also have a significant impact on the performance, both finite sample and asymptotic, of these estimators. 

We propose here a class of Efficient and Adaptive Semi-Supervised Estimators (EASE) in the context of linear regression problems. We essentially adopt a semi-parametric perspective wherein the adopted linear `working' model can be potentially mis-specified, and the goal is to obtain efficient and adaptive SS estimators of the regression parameter through robust usage of $\Usc$. The EASE are two-step estimators with a simple and scalable construction based on a first step of `semi-non-parametric' (SNP) imputation which includes a smoothing step and a follow-up `refitting' step. In the second step, we regress the imputed outcomes against the covariates using the unlabeled data to obtain our SNP imputation based SS estimator, and then further combine it optimally with the supervised estimator to obtain the final EASE. Dimension reduction methods are also employed in the smoothing step to accommodate higher dimensional $\bX$, if necessary. Further, we extensively adopt cross-validation (CV) techniques in the imputation, leading to some useful theoretical properties (apart from practical benefits) typically not observed for smoothing based two-step estimators. We demonstrate that EASE is guaranteed to be efficient and adaptive in the sense discussed above, and also achieves semi-parametric optimality whenever the SNP imputation is `sufficient' or the linear model holds. We also provide data adaptive methods to optimally select the directions for smoothing when dimension reduction is desired, and tools for inference with EASE.

The rest of this paper is organized as follows. In Section \ref{psetup}, we 
formulate the SS linear regression problem. In Section \ref{imp_ss_est}, we construct a family of SS estimators based on SNP imputation and establish all their properties, and further propose the EASE as a refinement of these estimators. For all our proposed estimators, we also address their associated inference procedures based on `double' CV methods. In Section \ref{kern_smth_impl}, we discuss a kernel smoothing based implementation of the SNP imputation and establish all its properties. In Section \ref{dim_red}, we discuss SS dimension reduction techniques, useful for implementing the SNP imputation. Simulation results and an application to an EMR study are shown in Section \ref{num_analysis}, followed by 
concluding discussions in Section \ref{discus}. Proofs of all theoretical results and associated technical materials, and further numerical results and discussions are distributed 
in the \hyperref[app]{Appendix} and the \hyperref[supp_mat]{Supplementary Material} [\citet{Own_Supp_2017}]. 

\section{Problem Set-up}\label{psetup}

%
\paragraph*{Data Representation}\label{data_rep} Let $Y \in \mathbb{R}$ denote the outcome random
variable and $\bX \in \mathbb{R}^p$ denote the covariate vector, where $p$ is fixed, and let $\bZ = (Y,\bX')'$. Then the entire data available for analysis can be represented as $\mathbb{S} = (\Lsc \cup \Usc)$, where $\Lsc =\{\bZ_i \equiv (Y_i, \bX_i')': i=1,\hdots,n\}$ consists of 
$n$ independent and identically distributed (i.i.d.) observations from the joint distribution $\P_\bZ$ of $\bZ$, $\Usc = \{\bX_i:$ $i=n+1,\hdots, n+N\}$ consists of 
$N$ i.i.d. observations from $\P_{\bX}$, 
and $\Lsc \ind \Usc$. Throughout, for notational convenience, we use the subscript `$j$' to denote the unlabeled observations, and re-index without loss of generality (w.l.o.g.) the $N$ observations in $\Usc$ as: $\Usc = \{\bX_j : j = n+1, \hdots, n+N\}$. 

\begin{assumption}[Basic Assumptions]\label{basic_assmpn}
\emph{
(a) We assume that $\bZ$ has finite $2^{nd}$ moments and $\bSigma \equiv \Var (\bX)$ is positive definite, denoted as $\bSigma \succ 0$. We also assume, for simplicity, that $\bX$ has a compact support $\mathcal{X} \subseteq \mathbb{R}^{p}$.
}
\emph{
\paragraph*{}(b) We assume $N \gg n$ i.e. $n/N \rightarrow 0$ as $n,N \rightarrow \infty$, and $\Lsc$ and $\Usc$ arise from the same underlying distribution, i.e. $\bZ \sim \P_{\bZ}$ for all subjects in $\mathcal{S}$.
}
\emph{
\paragraph*{Notations} Let $\bGamma = \E(\bXv\bXv') \succ 0$, where $\forall\; \bv \in \R^p$, $\overrightarrow{\bv} = (1,\bv')' \in \R^{(p+1)}$. Let $\Lsc_2(\P_\bX)$ denote the space of all $\mathbb{R}$-valued measurable functions of $\bX$ having finite $L_2$ norm with respect to (w.r.t.) $\P_{\bX}$, and for any $g(.) \in \Lsc_2(\P_\bX)$, let $\bSigma(g) \succ 0$ denote the $(p+1) \times (p+1)$ matrix $\bGammainv \E[\bXv\bXv'\{Y - g(\bX)\}^2] \bGammainv$. Lastly, let $\| \cdot \|$ denote the $L_2$ vector norm, and for any integer $a \geq 1$, let $I_a$ denote the identity matrix of order $a$, and $\Nsc_a[\bmu,\boldsymbol{\Omega}]$ denote the $a$-variate Gaussian distribution with mean $\bmu \in \R^a$ and covariance matrix $\boldsymbol{\Omega}_{a \times a} \succ 0$. 
}
\end{assumption}
\begin{remark}\label{rem_missingdata}
\emph{
Assumption \ref{basic_assmpn} (b) enlists some fundamental characteristics of SS settings. 
Indeed, the condition of $\Lsc$ and $\Usc$ being equally distributed 
has usually been an integral part of the \emph{definition} of SS settings \citep{Chapelle_2006,Kawakita_2013}. Interpreted in missing data terminology, it entails that $Y$ in $\Usc$ are `missing completely at random' (MCAR), with the missingness/labeling being typically by design. 
Interestingly, the crucial assumption of MCAR, although commonly required, has often stayed implicit in the SSL literature \citep{Lafferty_2007}. It is important to note that while the SS set-up can be viewed as a missing data problem, it is quite \emph{different} from standard ones, since with $n/N \rightarrow 0$ i.e. $|\Usc| \gg |\Lsc|$, the proportion of $Y$ observed in $\Ssc$ tends to $0$ in SSL. Hence, the `positivity assumption' typical in missing data theory, requiring this proportion to be bounded away from $0$,
is violated here. It is also worth noting that owing to such violations, the analysis of SS settings under more general missingness mechanisms such as `missing at random' (MAR) is considerably more complicated and to our knowledge, the literature for SS \emph{estimation} problems under such settings is virtually non-existent. Furthermore, for such problems, the traditional goal in SSL, that of improving upon a `supervised' estimator, can become unclear without MCAR, unless an appropriately weighted version of the supervised estimator is considered. 
Given these subtleties and the traditional assumptions (often implicit) in SSL, the MCAR condition is assumed for most of this paper, although a brief discussion on possible extensions of our proposed SS estimators to MAR settings is provided in the \hyperref[supp_mat]{Supplementary Material}. 
}
\end{remark}

\subsection{The Target Parameter and Its Supervised Estimator}\label{wm_tp_ols}

We consider the linear regression \emph{working model} given by:
\begin{equation}\label{model}
Y = \bXv'\btheta + \epsilon, \quad \mbox{with} \quad \E(\epsilon \given \bX) = 0,
\end{equation}
where, 
$\btheta \in \mathbb{R}^{(p+1)}$ is an unknown regression parameter. Accounting for the potential mis-specification of the working model (\ref{model}), we define the target parameter of interest as a model free parameter, as follows:
\begin{definition}\label{tp_defn}\emph{
The target parameter $\btheta_0$ for linear regression may be defined as the solution to the normal equations: $\E\{\bXv(Y-\bXv'\btheta)\} = \mathbf{0}$ in $\btheta \in \mathbb{R}^{(p+1)}$, or equivalently, $\btheta_0 = \underset{\btheta \in \mathbb{R}^{(p+1)}}{\mbox{argmin}}\; \E(Y - \bXv'\btheta)^2$.
}
\end{definition}
\noindent Existence and uniqueness of $\btheta_0$ in \ref{tp_defn} is clear. Further, $\bXv'\btheta_0$ is the $L_2$ projection of $\E(Y \medgiven \bX) \in \Lsc_2(\P_\bX)$ onto the subspace of all linear functions of $\bX$ and hence, is the best linear predictor of $Y$ given $\bX$. The linear model (\ref{model}) is \emph{correct} (else, \emph{mis-specified}) if and only if $\E(Y \medgiven \bX)$ lies in this space (in which case, $\E(Y \medgiven \bX) = \bXv'\btheta_0$).
When the model is correct, $\btheta_0$ depends only on $\P_{Y \smallgiven \bX}$, not on $\P_\bX$. Hence, improved estimation of $\btheta_0$ through SSL is impossible in this case unless further assumptions relating $\btheta_0$ to $\P_\bX$ are made. On the other hand, under model mis-specification, the normal equations defining $\btheta_0$ inherently depend on $\P_\bX$, thereby implying the potential utility of SSL in improving the estimation of $\btheta_0$ in this case.

The usual supervised estimator of $\btheta_0$ is the OLS estimator $\bthetahat$, the solution in $\btheta$ to the equation: $n^{-1}\sum_{i=1}^{n}\bXv_i(Y_i - \bXv_i'\btheta) = \mathbf{0}$, the normal equations based on $\Lsc$. Under Assumption \ref{basic_assmpn} (a), 
it is well known that as $n \rightarrow \infty$,
\begin{equation}\label{ols_expnsn}
\nhalf(\bthetahat - \btheta_0) = \nnhalf\sum_{i=1}^{n} \bpsi_{0}(\bZ_i)+ O_p\left(\nnhalf\right) \;\; \stackrel{d}{\rightarrow} \mathcal{N}_{(p+1)}[\mathbf{0},\bSigma(g_{\btheta_0})],
\end{equation}
where $\bpsi_{0}(\bZ) = \bGammainv \{\bXv(Y - \bXv'\btheta_0)\}$ and $g_{\btheta}(\bX) = \bXv'\btheta \;\; \forall \; \btheta \in \mathbb{R}^{(p+1)}$.

Our primary goal is to obtain efficient SS estimators of $\btheta_0$ using the \emph{entire} data $\mathcal{S}$ and compare their efficiencies to that of $\bthetahat$. It is worth noting that the estimation efficiency of $\btheta_0$ also relates to the predictive performance of the fitted linear model since its out-of-sample prediction error is directly related to the mean squared error (w.r.t. the $\bSigma$ metric) of the parameter estimate.

\section{A Family of Imputation Based Semi-Supervised Estimators}\label{imp_ss_est}

If $Y$ in $\Usc$ were actually observed, then one would simply fit the working model to the entire data in $\mathcal{S}$ for estimating $\btheta_0$. Our general approach is precisely motivated by this intuition. We first attempt to impute the missing $Y$ in $\Usc$ based on suitable training of $\Lsc$ in step (I). Then in step (II), we fit the linear model (\ref{model}) to $\Usc$ with the imputed outcomes. Clearly, the imputation is critical. Inaccurate imputation would lead to biased estimate of $\btheta_0$, while inadequate imputation would result in loss of efficiency. We next consider SS estimators constructed under two imputation strategies for step (I) including a fully non-parametric imputation based on kernel smoothing (KS), and a semi-non-parametric (SNP) imputation that involves a smoothing step and a follow up `refitting' step. Although the construction of the final EASE is based on the SNP imputation strategy, it is helpful to begin with a discussion of the first strategy in order to appropriately motivate and elucidate the discussion on EASE and the SNP imputation strategy.

\subsection{A Simple SS Estimator via Fully Non-Parametric Imputation}\label{fnp_est}

We present here an estimator based on a fully non-parametric imputation involving KS when $p$ is small. For simplicity, we shall assume here that $\bX$ is continuous with a density $f(\cdot)$. Let $m(\bx) = \E(Y \given \bX =\bx)$ and $l(\bx) = m(\bx) f(\bx)$. Consider the local constant KS estimator of $m(\bx)$,
\begin{equation}\label{mhat_defn}
\mhat(\bx) \; = \; \frac{\frac{1}{nh^p}\sum_{i=1}^{n} \{K_{h}(\bX_i,\bx)\}Y_i}
{\frac{1}{nh^p}\sum_{i=1}^{n} K_{h}(\bX_i,\bx)} \; = \; \frac{\widehat{l}(\bx)}{\widehat{f}(\bx)},
\end{equation}where $K_{h}(\bu,\bv) = K\{(\bu -\bv)/h\}$ with $K: \mathbb{R}^p \rightarrow \mathbb{R}$ being some suitable kernel function and $h$ $= h(n)$ $> 0$ being the bandwidth. With $\mhat(\cdot)$ as defined in (\ref{mhat_defn}), we now fit (\ref{model}) to the imputed unlabeled data: $[\{\mhat(\bX_j), \bX_j'\}':j= n+1,...,n+N]$ and obtain a SS estimator $\bthetahat_{np}$ of $\btheta_0$ as the solution in $\btheta$ to:
\begin{equation}\label{thetahat_np_defn}
 \frac{1}{N}\sum_{j=n+1}^{n+N}\bXv_j\{\mhat(\bX_j) - \bXv_j'\btheta\} = \mathbf{0}.
\end{equation}
Here and throughout in our constructions of SS estimators, $\Lsc$ with either the true or the imputed $Y$ is \emph{not} included in the final fitting step mostly due to technical convenience in the asymptotic analysis of our estimators, and also due to the fact that the contribution of $\Lsc$, included in any form, in the final fitting step is asymptotically negligible since $n/N \rightarrow 0$.

In order to study the properties of $\bthetahat_{np}$, we require uniform (in $L_\infty$ norm) convergence of $\mhat(\cdot)$ to $m(\cdot)$, a problem that has been extensively studied in the non-parametric statistics literature \citep{Newey_1994, Andrews_1995, Masry_1996, Hansen_2008} under fairly general settings and assumptions. In particular, we would assume the following regularity conditions to hold:
\begin{assumption}\label{assmpn_np}
\emph{(i) $K(\cdot)$ is a symmetric $q^{th}$ order kernel for some integer $q \geq 2$. (ii) $K(\cdot)$ is bounded, Lipschitz continuous and has a bounded support $\mathcal{K} \subseteq \mathbb{R}^p$. 
(iii) $\E(|Y|^s) < \infty$ for some $s > 2$. $\E(|Y|^s \given \bX = \bx)f(\bx)$ and $f(\bx)$ are bounded on $\mathcal{X}$. (iv) $f(\bx)$ is bounded away from $0$ on $\mathcal{X}$. (v) $m(\cdot)$ and $f(\cdot)$ are $q$ times continuously differentiable with bounded $q^{th}$ derivatives on some open set $\mathcal{X}_0 \supseteq \mathcal{X}$. (vi) For any $\delta > 0$, let $A_\delta \subseteq \mathbb{R}^p$ denote the set $\left\{(\bx-\bX)/\delta : \bx \in \mathcal{X}\right\}$. Then, for small enough $\delta$, $A_\delta \supseteq \mathcal{K}$ almost surely (a.s.).} 
\end{assumption}
\noindent Conditions (i)-(v) are fairly standard in the literature. In (v), the set $\mathcal{X}_0$ is needed mostly to make the notion of differentiability well-defined, with both $m(\cdot)$ and $f(\cdot)$ understood to have been analytically extended over $(\mathcal{X}_0\backslash\mathcal{X})$. Condition (vi) implicitly controls the tail behaviour of $\bX$, requiring that perturbations of $\bX$ in the form of $(\bX + \delta\boldsymbol{\phi})$ with $\boldsymbol{\phi} \in \mathcal{K}$ (bounded) and $\delta$ small enough, belong to $\mathcal{X}$ a.s. $[\P_{\bX}]$. We now present our result on $\bthetahat_{np}$. 
\begin{theorem}\label{thm1}
Suppose $\nhalf h^q \rightarrow 0$ and $(\log n)/(\nhalf h^p) \rightarrow 0$ as $n \rightarrow \infty$, and let $r_n = \nhalf h^q + (\log n)/(\nhalf h^p) + (n/N)^{\half}$. Then, under Assumption \ref{assmpn_np},
\begin{equation}\label{np_exp}
  \nhalf \left(\bthetahat_{np} - \btheta_{0}\right) = \nnhalf \sum_{i=1}^{n} \bpsi_{\eff}(\bZ_i)+ O_p(r_n) \;\; \stackrel{d}{\rightarrow} \mathcal{N}_{(p+1)}[\mathbf{0},\bSigma(m)],
\end{equation} where $\bpsi_{\eff}(\bZ) = \bGammainv [\bXv\{ Y - m(\bX)\}]$. 
\end{theorem}
\begin{remark}\label{rem_thm1_good}
\emph{Theorem \ref{thm1} establishes the efficient and adaptive nature of $\bthetahat_{np}$. The asymptotic variance $\bSigma(m)$ of $\bthetahat_{np}$ satisfies $\bSigma(g) - \bSigma(m) \succeq 0 \; \forall \; g(\cdot) \in \Lsc^2(\bX)$ and the inequality is strict unless $g(\cdot) = m(\cdot)$ a.s. $[\P_{\bX}]$. Hence, $\bthetahat_{np}$ is asymptotically \emph{optimal} among the class of all regular and asymptotically linear (RAL) estimators of $\btheta_0$ with influence function (IF) of the form: $\bGammainv [\bXv\{Y-g(\bX)\}]$ with $g(\cdot) \in \Lsc_2(\P_\bX)$. In particular, $\bthetahat_{np}$ is more efficient than $\bthetahat$ whenever (\ref{model}) is mis-specified, and equally efficient when (\ref{model}) is correct i.e. $m(\cdot) = g_{\btheta_0}(\cdot)$.}
\emph{Further, it can also be shown that $\bpsi_{\eff}(\bZ)$ is the `efficient' IF for estimating $\btheta_0$ under the semi-parametric model $\mathcal{M}_{\bX} \equiv \{(\P_{Y \smallgiven \bX},\P_\bX): \P_\bX \; \mbox{is known}, \; \P_{Y \smallgiven \bX} \; \mbox{is unrestricted upto Assumption \ref{basic_assmpn} (a)}\}$.
Thus, $\bthetahat_{np}$ also globally \emph{achieves} the semi-parametric efficiency bound under $\mathcal{M}_{\bX}$. Lastly, note that at any parametric sub-model in $\mathcal{M}_{\bX}$ that corresponds to (\ref{model}) being correct, $\bthetahat$ also achieves optimality, thus showing that under $\mathcal{M}_{\bX}$, it is not possible to improve upon $\bthetahat$ if the linear model is correct.}
\end{remark}
\begin{remark}\label{rem_thm1_bad}
\emph{The asymptotic results in Theorem \ref{thm1} require a kernel of order $q > p$ and $h$ smaller in order than the `optimal' bandwidth order $h_{opt} = O(n^{-1/(2q + p)})$. This \emph{under-smoothing} requirement, often encountered in two-step estimators involving a first-step smoothing \citep{Newey_1998}, generally results in sub-optimal performance of $\mhat(.)$. The optimal under-smoothed bandwidth order for Theorem \ref{thm1} is given by: $O(n^{-1/(q + p)})$.}
\end{remark}

\subsection{SS Estimators Based on Semi-Non-Parametric (SNP) Imputation}\label{snp_est}

The simple and intuitive imputation strategy in Section \ref{fnp_est} based on a fully non-parametric $p$-dimensional KS is however often undesirable in practice owing to the curse of dimensionality. In order to accommodate larger $p$, we now propose a more flexible SNP imputation method involving a dimension reduction, if needed, followed by a non-parametric calibration. An additional `refitting' step is proposed to reduce the impact of bias from non-parametric estimation and possibly inadequate imputation due to dimension reduction. We also introduce some flexibility in terms of the smoothing methods, apart from KS, that can be used for the non-parametric calibration.

Let $r \leq p$ be a fixed positive integer and let $\bP_r =$ $[\bp_1,..,\bp_r]_{p\times r}$ be any rank $r$ transformation matrix. Let $\bXPr = \bP_r'\bX$. Given $(r,\bP_r)$, we may now consider approximating the regression function $\E(Y \medgiven \bX)$ by smoothing $Y$ over the $r$ dimensional $\bXPr$ instead of the original $\bX \in \mathbb{R}^p$. In general, $\bP_r$ can be user-defined and data dependent. A few reasonable choices of $\bP_r$ are discussed in Section \ref{dim_red}. If $\bP_r$ depends only on the distribution of $\bX$, it may be assumed to be known given the SS setting considered. If $\bP_r$ also depends on the distribution of $Y$, then it needs to be estimated from $\Lsc$ and the smoothing needs to be performed using the estimated $\bP_r$.

For approximating $\E(Y \medgiven \bX)$, we may consider \emph{any} reasonable smoothing technique $\Tsc$. Some examples of $\Tsc$ include KS, kernel machine regression and smoothing splines. Let $m(\bx;\bP_r)$ denote the `target function' for smoothing $Y$ over $\bX_{\bP_r}$ using $\Tsc$.
For notational simplicity, the dependence of $m(\bx;\bP_r)$ and other quantities on $\Tsc$ is suppressed throughout.
%
For $\Tsc :=$ KS, the appropriate target function is given by: $m(\bx;\bP_r) = m_{\bP_r} (\bP_r'\bx)$, where  $m_{\bP_r}(\bz) \equiv \E(Y \given \bXPr = \bz)$. For basis function expansion based methods, $m(\bx;\bP_r)$ will typically correspond to the $L_2$ projection of $m(\bx) \equiv \E(Y \given \bX=\bx)\in \Lsc_2(\P_\bX)$ onto the functional space spanned by the basis functions associated with $\Tsc$. The results in this section apply to any choice of $\Tsc$ that satisfies the required conditions. In Section \ref{kern_smth_impl}, we provide more specific results for the implementation of our methods using $\Tsc :=$ KS.

Note that we do \emph{not} assume $m(\bx;\bP_r) = m(\bx)$ anywhere, and hence the name `semi-non-parametric'. 
Clearly, with $\bP_r = I_p$ and $\Tsc :=$ KS, it reduces to a fully non-parametric approach.
%
We next describe the two sub-steps involved in step (I) of the SNP imputation: (Ia) smoothing, and (Ib) refitting.
\paragraph*{(Ia) Smoothing Step}\label{smooth_step_snp} With $\bP_r$ and $m(\bx;\bP_r)$ as defined above, let $\Phat_r$ and $\mhat(\bx;\Phat_r)$ respectively denote their estimators based on $\Lsc$. In order to address potential overfitting issues in the subsequent steps, we further consider generalized versions of these estimators based on $\K$-fold CV for a given fixed  integer $\K\geq 1$. For any $\K \geq 2$, let $\{\Lsc_k\}_{k=1}^{\K}$ denote a random partition of $\Lsc$ into $\K$ disjoint subsets of equal sizes, $\nk = n/\K$, with index sets $\{\Isc_k\}_{k=1}^{\K}$. Let $\Lsc_k^-$ denote the set excluding $\Lsc_k$ with size $\nkminus$ $= n-\nk$ and respective index set $\Isc_k^-$. Let $\Phat_{r,k}$ and $\mhat_k(\bx;\Phat_{r,k})$ denote the corresponding estimators based on $\Lsc_k^-$. Further, for notational consistency, we define for $\K =1$, $\Lsc_k$ $= \Lsc_k^-$ $= \Lsc$; $\Isc_k$  $= \Isc_k^-$ $= \{1,...,n\}$; $\nk =$ $\nkminus = $ $n$; $\Phat_{r,k} = \Phat_r$ and $\mhat_k(\bx;\Phat_{r,k}) = \mhat(\bx;\Phat_r)$.
\paragraph*{(Ib) Refitting Step}\label{refit_step_snp} In this step, we fit the linear model to $\Lsc$ using $\bX$ as predictors and
the estimated $m(\bX; \bP_r)$ as an \emph{offset}. To motivate this, we recall that the fully non-parametric imputation given in Section \ref{fnp_est} consistently estimates $\E(Y|\bX)$, the $L_2$ projection onto a space that always contains the working model space, i.e. the linear span of $\bXv$. This need not be true for the SNP imputation, since we do not assume $m(\bX; \bP_r) = m(\bX)$ necessarily. The refitting step essentially `adjusts' for this so that the final imputation, combining the predictions from these two steps, targets a space that contains the working model space. In particular, for $\Tsc :=$ KS with $r < p$, this step is critical to remove potential bias due to inadequate imputation.

Interestingly, it turns out that the refitting step should \emph{always} be performed, \emph{even when} $m(\bX; \bP_r) = m(\bX)$. It plays a crucial role in reducing the bias of the resulting SS estimator due to the inherent bias from non-parametric curve estimation. In particular, for $\Tsc :=$ KS with \emph{any} $r \leq p$, it ensures that a bandwidth of the optimal order can be used, thereby \emph{eliminating the under-smoothing issue} as encountered in Section \ref{fnp_est}.
\hypertarget{eta_def}{}The target parameter for the refitting step is simply the regression coefficient obtained from regressing the residual $Y- m(\bX;\bP_r)$ on $\bX$ and may be defined as: $\etapr$, the solution in $\boldsymbol{\eta} \in \mathbb{R}^{(p+1)}$ to the equation: $\E[\bXv\{Y- m(\bX;\bP_r) - \bXv'\boldsymbol{\eta}\}] = \mathbf{0}$. For any $\K \geq 1$, we estimate $\etapr$ as $\etahatprk$, the solution in $\boldsymbol{\eta}$ to the equation:
\begin{equation}\label{eq-etahat}
\ninv\sum_{k=1}^{\K}\sum_{i\in\Isc_k}\bXv_i\{Y_i - \mhat_k(\bX_i;\Phat_{r,k}) - \bXv_i'\boldsymbol{\eta}\} = \mathbf{0}.
\end{equation} For $\bX_i \in \Lsc_k$, the estimate of $m(\bX_i;\bP_r)$ to be used as an offset is obtained from $\mhat_k(\cdot\;;\Phat_{r,k})$ that is based on data in $\Lsc_k^-$. For $\K \geq 2$, with $\Lsc_k^- \ind \Lsc_k$, the residuals are thus estimated in a cross-validated manner. For $\K = 1$ however, $\mhat(\cdot\;;\Phat_r)$ is estimated using the entire $\Lsc$ which can lead to considerable underestimation of the true residuals owing to over-fitting and consequently, substantial finite sample bias in the resulting SS estimator of $\btheta_0$. This bias can be effectively reduced by using the CV approach with $\K \geq 2$. We next estimate the \emph{target function} for the SNP imputation given by: 
\begin{eqnarray}
\mu(\bx;\bP_r) &=& m(\bx;\bP_r) + \bxv'\etapr \;\; \mbox{as:}\label{eq-mu}\\
\widehat{\mu}(\bx; \Pschat_{r,\K}) &=& \Kinv \sum_{k=1}^{\K} \mhat_{k}(\bx;\Phat_{r,k}) + \bxv'\etahatprk,\label{eq-muhat}
\end{eqnarray} where $\Pschat_{r,\K} = \{\Phat_{r,k}\}_{k=1}^{\K}$. For notational simplicity, we suppress throughout the inherent dependence of $\widehat{\mu}(\cdot\;;\;\cdot)$ itself on $\K$ and $\{\Lsc_k^-\}_{k=1}^{\K}$. 
Note that similar to $m(\bX;\bP_r)$, we also do \emph{not} assume $\mu(\bX;\bP_r) = m(\bX)$. Apart from the geometric motivation for the refitting step and its technical role in bias reduction, it also generally ensures the condition: $\E[\bXv\{Y - \mu(\bX;\bP_r\}] = \bzero$, \emph{regardless} of the true underlying $m(\bX)$. This condition is a key requirement for the asymptotic expansions, in Theorem \ref{thm2}, of our resulting SS estimators.
Using $\widehat{\mu}(\cdot \;; \Pschat_{r,\K})$, we now construct our final SS estimator as follows.

\paragraph*{SS Estimator from SNP Imputation}\label{ss_est_snp} In step (II), we fit the linear model to the SNP imputed unlabeled data: $[\{\widehat{\mu}(\bX_j; \Pschat_{r,\K}), \bX_j'\}': j=n+1,...,n+N]$ and obtain a SS estimator $\thetahatprk$ of $\btheta_0$ given by:
\begin{equation}\label{thetahat_snp_defn}
\thetahatprk \; \text{is the solution in} \; \btheta \; \text{to} \;\; \frac{1}{N}\sum_{j=n+1}^{n+N}\bXv_j\{\widehat{\mu}(\bX_j; \Pschat_{r,\K}) - \bXv_j'\btheta\} = \mathbf{0}.
\end{equation}
For convenience of further discussion, let us define: $\forall \; k \in \{1,\hdots,\K\}$,
\begin{eqnarray}
&& \Deltahat_{k}(\bx;\bP_r,\Phat_{r,k}) = \mhat_{k}(\bx;\Phat_{r,k})-m(\bx;\bP_r) \;\; \forall \; \bx \in \Xsc, \quad \mbox{and} \label{deltakhat_def}\\
&& \bGhat_{k}(\bx) = \bxv\Deltahat_{k}(\bx;\bP_r,\Phat_{r,k}) - \E_{\bX} \{\bXv\Deltahat_{k}(\bX;\bP_r,\Phat_{r,k})\} \;\; \forall \; \bx \in \Xsc,\label{Gkhat_def}
\end{eqnarray} where $\E_\bX(\cdot)$ denotes expectation w.r.t. $\bX \in \Usc$. The dependence of $\bGhat_{k}(\cdot)$ on $(\bP_r,\Phat_{r,k})$ and $\P_{\bX}$ is suppressed here for notational simplicity. We now present our main result summarizing the properties of $\thetahatprk$.
\begin{theorem}\label{thm2}
Suppose that $\Tsc$ satisfies: (i) 
$\text{sup}_{\bx\in \mathcal{X}}|m(\bx;\bP_r)| < \infty$ and (ii)
$\text{sup}_{\bx\in \mathcal{X}}|\mhat(\bx;\Phat_r)-m(\bx;\bP_r)| = O_p(c_n)$ for some $c_n=o(1)$. With $\bGhat_{k}(.)$ as in (\ref{Gkhat_def}), define $\G_{n,\K} = \nnhalf \sum_{k=1}^{\K}\sum_{i\in\Isc_k}\bGhat_{k}(\bX_i)$. Then, for any $\K \geq 1$,
\begin{equation}\label{snp_fund_exp}
  \nhalf \left(\thetahatprk - \btheta_{0}\right) = \nnhalf \sum_{i=1}^{n} \bpsi(\bZ_i; \bP_r) - \bGammainv\G_{n,\K} + O_p(c_{n,\K}^{*}),
\end{equation} where $\bpsi(\bZ; \bP_r) = \bGammainv [\bXv\{ Y - \mu(\bX;\bP_r)\}]$ and
$c_{n,\K}^{*}$ $= \cnkminus + \nnhalf + (n/N)^{\half}$ $= o(1)$. Further, for any fixed $\K \geq 2$, $\G_{n,\K} =$ $O_p(\cnkminus)$, so that
\begin{equation}\label{snp_k2_exp}
  \nhalf \left(\thetahatprk - \btheta_{0}\right) = \nnhalf \sum_{i=1}^{n}\bpsi(\bZ_i; \bP_r)+ O_p(\cnkminus + c_{n,\K}^{*}),
\end{equation} which converges in distribution to $\mathcal{N}_{(p+1)}[\mathbf{0},\bSigma\{\mu(\cdot \;;\bP_r)\}]$.
\end{theorem}
\begin{remark}\label{rem_thm2_1}
\emph{If the imputation is `sufficient' so that $\mu(\bx;\bP_r) = m(\bx)$, then $\thetahatprk$, for any $\K \geq 2$, enjoys the same set of optimality properties as those noted in Remark \ref{rem_thm1_good} for $\bthetahat_{np}$ (while requiring less stringent assumptions about $K(\cdot)$ and $h$, if KS is used). If $\mu(\bx;\bP_r) \neq m(\bx)$, then it is however unclear whether $\thetahatprk$ is always more efficient than $\bthetahat$. This will be addressed in Section \ref{EASE} where we develop the final EASE.}
\end{remark}
\begin{remark}\label{rem_thm2_2}
\emph{Apart from the fairly mild condition \hyperref[thm2]{(i)}, Theorem \ref{thm2} \emph{only} requires uniform consistency of $\mhat(\cdot \;;\Phat_{r})$ w.r.t. $m(\cdot \;;\bP_r)$ for establishing the $\nhalf$-consistency and asymptotic normality (CAN) of $\thetahatprk$ for any $\K \geq 2$. The uniform consistency typically holds for a wide range of smoothing methods $\Tsc$ under fairly general conditions. For $\Tsc :=$ KS in particular, we provide explicit results in Section \ref{kern_smth_impl} under mild regularity conditions that allow the use of any kernel order and the associated optimal bandwidth order. This is a notable relaxation from the stringent requirements for Theorem \ref{thm1} that necessitate under-smoothing and the use of higher order kernels.}
\end{remark}
%
%
\begin{remark}\label{rem_thm2_3} 
\emph{
The CAN property of $\bthetahat_{(\bP_r,1)}$  has \emph{not} yet been established. The term $\G_{n,\K}$ in (\ref{snp_fund_exp}) behaves quite \emph{differently} when $\K = 1$, compared to $\K \geq 2$ when it has a nice structure due to the inherent `cross-fitting' involved, and can be controlled easily, and quite generally, under mild conditions as noted in Remark \ref{rem_thm2_2}. For $\K =1$ however, $\G_{n,\K}$ is simply a centered empirical process devoid of any such structure and in general, controlling it requires stronger conditions and the use of empirical process theory (see for instance \citet{VdVaart_2000} for relevant results). We derive the properties of $\bthetahat_{(\bP_r,1)}$ 
for the case of $\Tsc :=$ KS in Theorem \ref{thm4} using a different approach however, specialized for KS estimators, in order to control $\G_{n,1}$. 
}
\end{remark}

\subsection{Efficient and Adaptive Semi-Supervised Estimators (EASE)}\label{EASE}

To ensure adaptivity even when $\mu(\bx;\bP_r) \neq m(\bx)$, we now define the final EASE as an optimal linear combination of $\bthetahat$ and $\thetahatprk$. Specifically, for any fixed $(p+1)\times(p+1)$ matrix $\bDelta$, 
$\thetahatprk(\bDelta)=\bthetahat + \bDelta(\thetahatprk - \bthetahat)$ is a CAN estimator of $\btheta_0$ whenever $\bthetahat$ and $\thetahatprk$ are, and an optimal $\bDelta$ can be selected easily to minimize the asymptotic variance of the combined estimator. For simplicity, we focus here on $\bDelta$ being a diagonal matrix with $\bDelta = \mbox{diag}(\delta_1, ..., \delta_{p+1})$. Then the EASE is defined as $\thetahatprk^E \equiv \thetahatprk(\bDeltahat)$ with $\bDeltahat$ being any consistent estimator (see Section \ref{inference} for details) of the minimizer $\bDeltabar = \mbox{diag}(\deltabar_1,...,\deltabar_{p+1})$, where $\forall \; 1 \leq l \leq (p+1)$,
\begin{equation}
\deltabar_l = - \; \underset{\epsilon \downarrow 0}{\mbox{lim}} \; \frac{\text{Cov}\left\{\bpsizerol(\bZ),\; \bpsil(\bZ;\bP_r)-\bpsizerol(\bZ)\right\}}
{\Var\left\{\bpsil(\bZ;\bP_r)-\bpsizerol(\bZ)\right\} \; + \; \epsilon},\label{ease_defn}
\end{equation}
and for any vector $\mathbf{a}$, $\mathbf{a}_{[l]}$ denotes its $l^{th}$ component. Note that in (\ref{ease_defn}), the $\epsilon$ and the limit outside are included to formally account for the case: $\bpsizerol(\bZ) = \bpsi_{[l]}(\bZ,\bP_r)$ a.s. $[\P_{\bZ}]$, when we define $\deltabar_l = 0$ for identifiability.

It is straightforward to show that $\thetahatprk^E$ and $\thetahatprk(\bDeltabar)$ are asymptotically equivalent, so that $\thetahatprk^E$ is a RAL estimator of $\btheta_0$ satisfying:
\begin{equation*}
\nhalf \left(\thetahatprk^E - \btheta_{0}\right) = \nnhalf \sum_{i=1}^{n}\bpsi(\bZ_i;\bP_r, \bDeltabar) + o_p(1) \;\; \stackrel{d}{\rightarrow} \mathcal{N}_{(p+1)}[\mathbf{0},\bSigma_{\bP_r}(\bDeltabar)] ,
\end{equation*}
as $n \to \infty$, where $\bpsi(\bZ;\bP_r, \bDeltabar) = \bpsi_0(\bZ) + \bDeltabar\{\bpsi(\bZ;\bP_r) - \bpsi_0(\bZ)\}$ and
$\bSigma_{\bP_r}(\bDeltabar) = \Var\{\bpsi(\bZ;\bP_r, \bDeltabar)\}.$ Note that when either the linear model holds or the SNP imputation is sufficient, then $\bpsi(\bZ;\bP_r, \bDeltabar) = \bpsi_{\eff}(\bZ)$, so that $\thetahatprk^E$ is asymptotically optimal in the sense of Remark \ref{rem_thm1_good}. Further, when neither cases hold, $\thetahatprk^E$ is no longer optimal, but is \emph{still} efficient and adaptive compared to $\bthetahat$. Lastly, if the imputation is certain to be sufficient (for example, if $r = p$ and $\Tsc:=$ KS), we may  simply define $\thetahatprk^E = \thetahatprk$.

\begin{remark}\label{rem_IF_class}
\emph{
It can be shown that under $\Msc_{\bX}$, defined in Remark \ref{rem_thm1_good}, the class of all possible IFs achievable by RAL estimators of $\btheta_0$
is given by: 
%
\def\bg{\mathbf{g}}
$\mathcal{IF}_{\btheta_0,\Msc_{\bX}} = \{ \bpsi_{\bg}(\bZ) \equiv \bpsi_{\eff} (\bZ) + \bg(\bX) : \E\{\bg(\bX)\} = \bzero , \hspace{0.02in} \bg_{[j]}(\cdot) \in \Lsc_2(\P_{\bX}) \; \forall \; j \}$. The IFs achieved by $\bthetahat$, $\thetahatprk$ and $\thetahatprk^{E}$ are clearly members of this class. 
The SNP imputation, for various choices of the imputation function $\mu(\cdot\;;\bP_r)$, therefore equips us with a \emph{family} of RAL estimator pairs $\{\thetahatprk,\thetahatprk^E\}$ for estimating $\btheta_0$. The IF of $\thetahatprk^E$ is further guaranteed to dominate that of $\bthetahat$, and when $\mu(\cdot\;;\bP_r) = m(\cdot)$, it also dominates all other IFs $\in$ $\mathcal{I}\mathcal{F}_{\btheta_0,M_{\bX}}$.
}
\end{remark}

\subsection{Inference for EASE and the SNP Imputation Based SS Estimators}\label{inference}

We now provide procedures for making inference about $\btheta_0$ based on $\thetahatprk$ and $\thetahatprk^E$ obtained using $\K \geq 2$. We also employ a `double' CV to overcome bias in variance estimation due to over-fitting. A key step involved in the variance estimation is to obtain reasonable estimates of $\{\mu(\bX_i;\bP_r)\}_{i=1}^n$. Although $\etahatprk$ in (\ref{eq-etahat}) was constructed via CV, the corresponding estimate, $\widehat{\mu}(\bx;\Pschat_{r,\K})$ in (\ref{eq-muhat}), of $\mu(\bx;\bP_r)$ is likely to be over-fitted for $\bX_i \in \Lsc$. To construct bias corrected estimates of  $\mu(\bX_i;\bP_r)$, we first obtain $\K$  separate {\em doubly cross-validated} estimates of $\etapr$, $\{\etahatprk^k: k = 1, ..., \K\}$, with $\etahatprk^k $, for each $k$, being the solution in $\boldsymbol{\eta}$ to $\sum_{k' \ne k}\Ssc_{k'}(\boldsymbol{\eta}) = \mathbf{0}$, where
$$
\Ssc_{k'}(\boldsymbol{\eta})  = {\textstyle \sum_{i\in\Isc_{k'}}} \bXv_i\{Y_i - \mhat_{k'}(\bX_i;\Phat_{r,k'}) - \bXv_i'\boldsymbol{\eta}\} \quad \forall \; k' \in \{1,\hdots,\K\}.
$$
For each $k$ and $k' \neq k$, $\Ssc_{k'}(\boldsymbol{\eta})$ is constructed such that $\{\bZ_i: i \in \Isc_{k'}\}$ used for obtaining $\etahatprk^k $ is independent of $\mhat_{k'}(\cdot \;;\Phat_{r,k'})$ that is based on $\Lsc_{k'}^- \ind \Lsc_{k'}$. Then, for each $\bX_i \in \Lsc_k$ and $k \in \{1,\hdots,\K\}$, we may estimate $\mu(\bX_i;\bP_r)$ as:
$$
\widehat{\mu}_{k}(\bX_i;\Pschat_{r,\K}) = \mhat_{k}(\bX_i;\Phat_{r,k}) + \bXv_i'\etahatprk^k.
$$
We exclude $\Ssc_{k} (\boldsymbol{\eta})$ in the construction of $\etahatprk^k$ to reduce over-fitting bias in the residuals $\{Y_i - \widehat{\mu}_{k}(\bX_i;\Pschat_{r,\K})\}$ which we now use for estimating the IFs.

For each $\bZ_i \in \Lsc_{k}$ and $k \in \{1,..,\K\}$, we estimate $\bpsi_0(\bZ_i)$ and $\bpsi(\bZ_i;\bP_r)$, the corresponding IFs of $\bthetahat$ and $\thetahatprk$, respectively as:
$$\widehat{\bpsi}_0(\bZ_i) = \bGammahat^{-1}\{\bXv_i(Y_i - \bXv_i'\bthetahat)\} \ \mbox{and} \ \widehat{\bpsi}_{k}(\bZ_i;\bP_r) = \bGammahat^{-1} [\bXv_i\{Y_i-\widehat{\mu}_{k}(\bX_i;\Pschat_{r,\K})\}],$$
where $\bGammahat$ denotes any consistent estimator of $\bGamma$ from $\Lsc$ and/or $\Usc$ (for example, $\bGammahat = \bGamma_n \equiv \ninv\sum_{i=1}^n \bXv_i\bXv_i'$ based on $\Lsc$, or  $\bGammahat = \bGamma_N \equiv N^{-1} \sum_{j=n+1}^{n+N} \bXv_j\bXv_j'$ based on $\Usc$). Then, $\bSigma\{\mu(\cdot \;;\bP_r)\}$ in (\ref{snp_k2_exp}) may be consistently estimated as:
$$
\widehat{\bSigma}\{\mu(\cdot\;;\bP_r)\} = \ninv \sum_{k=1}^{\K}\sum_{i \in \Isc_k} \widehat{\bpsi}_{k}(\bZ_i;\bP_r)\widehat{\bpsi}_{k}'(\bZ_i;\bP_r).
$$
To estimate the combination matrix $\bDeltabar$ in (\ref{ease_defn}) and the asymptotic variance, $\bSigma_{\bP_r}(\bDeltabar)$, of EASE consistently, let us define, $\forall \; 1 \leq l \leq (p+1)$,
\begin{align*}
\widehat{\sigma}_{l,12} &= -\ \ninv {\textstyle \sum_{k=1}^\K \sum_{i\in\Isc_k}}  \widehat{\bpsi}_{0[l]}(\bZ_i) \{\widehat{\bpsi}_{k[l]}(\bZ_i;\bP_r) -
\widehat{\bpsi}_{0[l]}(\bZ_i)\},\\
\widehat{\sigma}_{l,22}  & = \ninv {\textstyle \sum_{k=1}^\K\sum_{i\in\Isc_k}} \{\widehat{\bpsi}_{k[l]}(\bZ_i;\bP_r) - \widehat{\bpsi}_{0[l]}(\bZ_i)\}^2,
\end{align*}
and $\deltahat_l = \widehat{\sigma}_{l,12} / (\widehat{\sigma}_{l,22} + \epsilon_n)$ for some sequence $\epsilon_n \to 0$ with $n^{\half}\epsilon_n \to \infty$.
Then, we estimate $\bDeltabar$ and $\bSigma_{\bP_r}(\bDeltabar)$ respectively as: $\bDeltahat = \mbox{diag}(\deltahat_1,...,\deltahat_{p+1})$ and
$$
\widehat{\bSigma}_{\bP_r}(\bDeltahat) =  \ninv\sum_{k=1}^{\K}\sum_{i\in\Isc_k} \widehat{\bpsi}_{k}(\bZ_i;\bP_r,\bDeltahat)\widehat{\bpsi}_{k}'(\bZ_i;\bP_r,\bDeltahat),
$$
where $\widehat{\bpsi}_{k}(\bZ;\bP_r,\bDeltahat) = \widehat{\bpsi}_0(\bZ)  + \bDeltahat\{\widehat{\bpsi}_{k}(\bZ;\bP_r) - \widehat{\bpsi}_0(\bZ)\}$ $\forall \; k \in \{1, \hdots, \K\}$. Normal confidence intervals (CIs) for the parameters of interest can also be constructed accordingly based on these variance estimates.

\section{Implementation Based on KS}\label{kern_smth_impl}

We next detail the specific implementation of the SNP imputation based on KS estimators. With $\Tsc :=$ KS, the target function for the smoothing is given by: $m(\bx;\bP_r) = m_{\bP_r} (\bP_r'\bx) \equiv \E(Y \given \bXPr = \bP_r'\bx)$.
For simplicity, we assume that $\bX_{\bP_r}$ is continuous with a density $\fPr(\cdot)$ and support $\chiPr \equiv \{\bP_r'\bx: \bx \in \mathcal{X}\} \subseteq$ $\mathbb{R}^r$. Let us now consider the following class of local constant KS estimators for $m(\bx;\bP_r)$:
\begin{equation}\label{mhatprk_kern_defn}
\mhat_{k}(\bx;\Phat_{r,k}) \; = \; \frac{\frac{1}{\nkminus h^r}\sum_{i \in \Isc_{k}^{-}} \{K_{h}(\Phat_{r,k}'\bX_i,\Phat_{r,k}'\bx)\}Y_i}
{\frac{1}{\nkminus h^r}\sum_{i \in \Isc_{k}^-}K_{h}(\Phat_{r,k}'\bX_i,\Phat_{r,k}'\bx)} \hspace{4mm} \forall \;\; 1 \leq k \leq \K ,
\end{equation}
where $K_{h}(\cdot)$ and $h$ are as in Section \ref{fnp_est} with $K(\cdot)$ now being a suitable kernel on $\mathbb{R}^r$. In the light of Theorem \ref{thm2}, we focus primarily on establishing the uniform consistency of $\mhat(\bx;\Phat_r)$ $\equiv \mhat_1(\bx; \Phat_{r,1})$ in (\ref{mhatprk_kern_defn}) with $\K=1$, \emph{accounting} for the additional estimation error from $\Phat_r$.
For establishing the desired result, we shall assume the following regularity conditions to hold:
\begin{assumption}\label{assmpn_unifconv_snp}
\emph{(i) $K(\cdot)$ is a symmetric kernel of order $q \ge 2$ with finite $q^{th}$ moments. (ii) $K(\cdot)$ is bounded, integrable and is either Lipschitz continuous with a compact support or, has a bounded derivative $\bnabla K(\cdot)$ which satisfies: $\|\bnabla K(\bz)\| \leq \Lambda\|\bz\|^{-\rho} \; \forall \; \bz \in \mathbb{R}^r$ with $\|\bz\| > L$, where $\Lambda >0$, $L > 0$ and $\rho >1$ are some fixed constants, and $\|.\|$ denotes the standard $L_2$ vector norm. (iii) $\chiPr \subseteq \mathbb{R}^r$ is compact. $\E(|Y|^s) < \infty$ for some $s > 2$. $\E(|Y|^s \given \bXPr = \bz)\fPr(\bz)$ and $\fPr(\bz)$ are bounded on $\chiPr$. (iv) $\fPr(\bz)$ is bounded away from $0$ on $\chiPr$. (v) $\mPr(\bz)$ and $\fPr(\bz)$ are both $q$ times continuously differentiable with bounded $q^{th}$ derivatives on some open set $\mathcal{X}_{0,\bP_r} \supseteq \chiPr$. \emph{Additional Conditions (required only when $\bP_r$ needs to be estimated):} (vi) $K(\cdot)$ has a bounded and integrable derivative $\bnabla {K} (\cdot)$. (vii) $\bnabla K(\cdot)$ satisfies: $\|\bnabla K(\bz_1) - \bnabla K(\bz_2)\| \leq \|\bz_1 - \bz_2\| \; \phi(\bz_1) \; \forall \; \bz_1, \bz_2 \in \mathbb{R}^r$ such that $\|\bz_1-\bz_2\| \leq L^*$, for some fixed constant $L^* > 0$, and some bounded and integrable function $\phi: \mathbb{R}^r \rightarrow \mathbb{R}^{+}$. (viii) $\bnabla K(\cdot)$ is Lipschitz continuous on $\mathbb{R}^r$. (ix) $\E(\bX \given \bXPr = \bz)$ and $\E(\bX Y \given \bXPr = \bz)$ are both continuously differentiable with bounded first derivatives on $\mathcal{X}_{0,\bP_r} \supseteq \chiPr$.}
\end{assumption}
Assumption \ref{assmpn_unifconv_snp}, mostly adopted from \citet{Hansen_2008}, imposes some mild smoothness and moment conditions most of which are fairly standard, except perhaps the conditions on $K(\cdot)$ in (vi)-(viii) all of which are however satisfied by the Gaussian kernel among others. We now propose the following result.
\begin{theorem}\label{thm3}
Suppose $(\Phat_r - \bP_r) = O_p(\alpha_n)$ for some $\alpha_n =$ $o(1)$ with $\alpha_n$ $= 0$ identically if $\bP_r$ is known. Let $q$ be the order of the kernel $K(.)$ in (\ref{mhatprk_kern_defn}) for some integer $q \geq 2$. Define: $$a_{n,1} = \alpha_n \left(\frac{\log n}{nh^{r+2}}\right)^{\frac{1}{2}}+ \alpha_n^2 h^{-(r+2)} + \alpha_n, \quad a_{n,2} = \left(\frac{\log n}{nh^r}\right)^{\frac{1}{2}} + h^q $$ and assume that each of the terms involved in $a_{n,1}=o(1)$ and $a_{n,2} = o(1)$.  Then, under Assumption \ref{assmpn_unifconv_snp}, 
$\mhat(\bx;\Phat_r)$, based on (\ref{mhatprk_kern_defn}), satisfies:
\begin{equation}\label{rate_kern_unif_conv}
{\text{sup}}_{\bx\in \mathcal{X}}\;|\mhat(\bx;\Phat_r)-m(\bx;\bP_r)| = O_p(a_{n,1} + a_{n,2}).
\end{equation}
\end{theorem}
\begin{remark}\label{rem_thm3}
\emph{Theorem \ref{thm3} establishes the $L_{\infty}$ error rate of $\mhat(\bx;\Phat_r)$ under mild regularity conditions and restrictions on $h$. Among its various implications, the rate also ensures uniform consistency of $\mhat(\bx;\Phat_r)$ at the optimal bandwidth order: $h_{opt} = O(n^{-1/(2q+r)})$ for any kernel order $q \geq 2$ and any $r \leq p$, as long as $\alpha_n=o(n^{-(r+2)/(4q+2r)})$ which always includes: $\alpha_n = O(\nnhalf)$ and $\alpha_n = 0$. These two cases are particularly relevant in practice as $\bP_r$ being finite dimensional, $\nhalf$-consistent estimators of $\bP_r$ should typically exist. For both cases, using $h_{opt}$ results in $a_{n,1}$ to be of lower order (for $q > 2$) or the same order (for $q = 2$) compared to that of the main term $a_{n,2}$, so that the usual optimal rate prevails as the overall error rate.}
\end{remark}

\subsubsection*{Properties of $\thetahatprk$ for $\K = 1$}\label{prop_k1} We now address the CAN property of $\thetahatprk$ for $\K = 1$ under the KS framework. Based on (\ref{snp_fund_exp}) and Remark \ref{rem_thm2_3}, the only step required for this is to effectively control the term $\G_{n,\K}$ in (\ref{snp_fund_exp}). 
The following result is in this regard. It involves Lemmas \ref{lemma1}-\ref{lemma2} as the main technical tools which may themselves be of independent interest.
\begin{theorem}\label{thm4}
Let $\K = 1$, $\Tsc :=$ KS, $\G_{n,\K}$ be as in (\ref{snp_fund_exp}), and $\mhat(\bx;\Phat_r)$ be the KS estimator based on (\ref{mhatprk_kern_defn}). Let $\alpha_n$, $a_{n,1}$ and $a_{n,2}$ be as in Theorem \ref{thm3} with $(\Phat_r - \bP_r) = O_p(\alpha_n)$. Assume that $a_{n,1}^*$ and $a_{n,2}^*$ are $o(1)$, where $$a_{n,1}^* = \alpha_n + \frac{\alpha_n}{ \nhalf h^{(r+1)}} + \nhalf\alpha_n^2 h^{-2} + \nhalf a_{n,1}^2 + \nhalf a_{n,1}a_{n,2}  \quad \mbox{and} \quad a_{n,2}^* = \nhalf a_{n,2}^2. $$
Then, under Assumption \ref{assmpn_unifconv_snp},
$\G_{n,\K} = O_p(a_{n,1}^*+a_{n,2}^*)$ $= o_p(1)$. Further, let $c_{n,\K}^*$ be as in Theorem \ref{thm2} with $c_n$ $= (a_{n,1}+a_{n,2})$. Then, using (\ref{snp_fund_exp}),
\begin{equation}\label{snp_k1_exp}
  \nhalf \left(\thetahatprk - \btheta_{0}\right) = \nnhalf \sum_{i=1}^{n} \bpsi(\bZ_i,\bP_r) 
  + O_p(c_{n,\K}^{*}+d_n),
\end{equation}where $d_n = a_{n,1}^* +a_{n,2}^*$. Hence, $\nhalf (\thetahatprk - \btheta_0) \stackrel{d}{\rightarrow} \mathcal{N}_{(p+1)}[\mathbf{0},\bSigma\{\mu(\cdot\;;\bP_r)\}]$.
\end{theorem}
\begin{remark}\label{rem_thm4}
\emph{Note that the  term $a_{n,2}^*$ \emph{always} requires $q > r/2$ in order to converge to $0$, thus showing the contrasting behavior of the case $\K = 1$ compared to $\K \geq 2$ where no such higher order kernel restriction is required. Nevertheless, when $\alpha_n = O(\nnhalf)$ or $\alpha_n = 0$, the optimal bandwidth order:  $h_{opt} = O(n^{-1/(2q+r)})$ can indeed be \emph{still} used as long as $q > r/2$ is satisfied. Despite these facts and all the theoretical guarantee in Theorem \ref{thm4}, empirical evidence however seems to suggest that $\bthetahat_{(\bP_r,1)}$ can be substantially \emph{biased} in finite samples, in part due to over-fitting.
}
\end{remark}
\begin{remark}\label{rem_tech_benefits}
Technical benefits of refitting and CV: \emph{
Suppose that $\bP_r = I_p$, so that the SNP imputation with $\Tsc :=$ KS is indeed sufficient. Further, assume that all of Theorems \ref{thm1}-\ref{thm4} hold, so that the estimators $\bthetahat_{np}$, $\bthetahat_{(\bP_r,1)}$, and $\thetahatprk$ ($\K \geq 2$) are comparable and all asymptotically optimal. However, their constructions are quite different which can significantly affect their finite sample performances. $\bthetahat_{np}$ is based on KS only, and requires stringent under-smoothing and a kernel of order $q > p$ (Remark \ref{rem_thm1_bad}); $\bthetahat_{(\bP_r,1)}$ is based on KS and refitting (\emph{although} the KS itself is certain to be sufficient), and requires no under-smoothing but needs a (weaker) kernel order condition $(q > p/2)$ (Remark \ref{rem_thm4}); while $\thetahatprk$ ($\K \geq 2$) additionally involves CV, and requires no under-smoothing or higher order kernel conditions (Remark \ref{rem_thm2_2}). This highlights the critical role played by refitting and CV, apart from their primary roles in the SNP imputation, in removing any under-smoothing and/or higher order kernel restrictions when $\Tsc :=$ KS, and this continues to hold for any other $(r,\bP_r)$ as well. In particular, it shows, rather surprisingly, that refitting should be performed in order to avoid under-smoothing \emph{even if} the smoothing is known to be sufficient.}
\end{remark}

\begin{remark}\label{other_smth_mthd}
\emph{As mentioned in Section \ref{snp_est}, $\Tsc :=$ KS along with possible dimension reduction is just \emph{one} reasonable choice of $\Tsc$ for implementing the SNP imputation, all technical requirements for which have been thoroughly established in Section \ref{kern_smth_impl}. In general, other smoothing methods, as long as the requirements are satisfied, can also be equally used as choices of $\Tsc$. One such choice could be kernel machine (KM) regression (with possibly no use of dimension reduction, as KM uses penalization to effectively regularize the target even with $\bP_r = I_p$). We leave its implementation details to the reader as they are readily available in a multitude of references, and also skip any theoretical treatment, considering the primary goal and scope of this paper. However, detailed numerical results are presented in Section \ref{num_analysis} for this choice of $\Tsc$ as well to illustrate the wider applicability of our proposed methods.
}

\end{remark}
\section{Choices of $\bP_r$: Dimension Reduction Techniques}\label{dim_red}

We next discuss choosing and estimating the matrix $\bP_r$ $(r < p)$ to be used for dimension reduction, if required, in the SNP imputation, and which can play an important role in the sufficiency of the imputation. Simple choices of $\bP_r$ include $r$ leading principal component directions of $\bX$ or any $r$ canonical directions of $\bX$. Note that under the SS setting, $\bP_r$ is effectively known if it only involves the distribution of $\bX$, as is true for these choices.
We now focus primarily on the case where $\bP_r$ also depends on the distribution of $Y$ and hence, is unknown. Such a choice of $\bP_r$ is often desirable to ensure that the imputation is as `sufficient' as possible for predicting $Y$. Several reasonable choices of such $\bP_r$ and their estimation are possible based on 
sufficient dimension reduction (s.d.r.) methods like Sliced Inverse Regression (SIR) \citep{K-C_Li_1991}, Principal Hessian Directions (PHD) \citep{K-C_Li_1992,Cook_1998}, Sliced Average Variance Estimation (SAVE) \citep{Cook_1991,Cook_1999} etc. 

In particular, we focus here on SIR where the choice of $\bP_r$ is given by: $\bP_r^0$ $= \bSigma^{-\half} \bPbar_r$, with $\bPbar_r$ being the $r$ leading eigenvectors 
of $\M = \Var\{\E(\X \given Y)\}$, where $\X = \bSigma^{-\half}(\bX - \bmu)$, with $\bmu = \E(\bX)$, denotes the standardized version of $\bX$. It is well known \citep{K-C_Li_1991} that these directions lead to an optimal (in some appropriate sense) $r$-dimensional linear transformation of $\bX$ that can be predicted by $Y$. Apart from these general optimality, they also have deeper implications in the context of s.d.r. We refer the reader to \citet{K-C_Li_1991} and other relevant references in the s.d.r. literature for further details.

For estimating $\bP_r^0$, we consider the SIR algorithm of \citet{K-C_Li_1991} and further propose a SS modification to it. With $\K$ and $\{\Lsc_k^-,\Isc_k^-,\Phat_{r,k}\}_{k=1}^\K$ as before, let $(\bmuhat_k,\bSigmahat_k)$ denote the estimates of $(\bmu,\bSigma)$ based on $\Lsc_k^-$ and define $\X^{(k)} = \bSigmahat_k^{-\half}(\bX - \bmuhat_k)$. Then, the original SIR algorithm 
estimates $\bP_r^0$ based on $\Lsc_k^-$ as follows:
(i) Divide the range of $\{Y_i\}_{i\in\Isc_k^-}$ into $H$ slices $\{I_1,..,I_H\}$, where $H$ may depend on $n_\K^-$. For $1\leq h \leq H$, let $\widehat{p}_{h,k}$ denote the proportion of $\{Y_i\}_{i\in\Isc_k^-}$ in slice $I_h$; (ii) For each $I_h$, let $\Mhat_{h,k}$ denote the sample average of the set: $\{\X_i^{(k)}\in\Lsc_k^-: Y_i \in I _h\}$; (iii) Estimate $\M$ as: $\Mhat_{k}$ $= \sum_{h=1}^{H}\widehat{p}_{h,k}\Mhat_{h,k}\Mhat_{h,k}'$ and $\bP_r^0$ as: $\Phat^0_{r,k} = \bSigmahat_k^{-\half}\Phat_{r,k}$, where $\Phat_{r,k}$ denotes the $r$ leading eigenvectors 
of $\Mhat_{k}$.
However, the SIR algorithm often tends to give unstable estimates of $\bP_r^0$, especially for the directions corresponding to the smaller eigenvalues of $\M$. To improve the efficiency in estimating $\bP_r^0$, we now propose a semi-supervised SIR (SS-SIR) algorithm as follows.

\subsection*{SS-SIR Algorithm}\label{SS_SIR} \noindent Given $\{\Lsc_k^-,\Isc_k^-,\Phat_{r,k}\}_{k=1}^\K$, let $(\bmuhat_k^*,\bSigmahat_k^*)$ denote the estimates of $(\bmu,\bSigma)$ based on $\Lsc_k^- \cup \Usc$ and define $\X^{(k*)} = \bSigmahat_k^{*-\half}(\bX - \bmuhat_k)$. Then the SS-SIR proceeds as follows. Step (i) stays the same as in SIR. In step (ii), for each $k$, and each $j \in \{n+1, ..., n+N\}$, we impute $Y_{j}$ as $Y_{j,k}^* = Y_{\ihat_{j,k}}$, where $\ihat_{j,k} = \mbox{argmin}_{i \in \Isc_k^{-}}{\|\X_i^{(k*)} - \X_j^{(k*)}\|^2}$. For each $I_h$, let $\Mhat_{h,k}^*$ be the sample average of the set: $\{\X_i^{(k*)}\in\Lsc_k^-: Y_i \in I _h\} \cup \{\X_j^{(k*)}\in\mathcal{U}: Y_{j,k}^* \in I_h\}$. Then in step (iii), we estimate $\M$ as: $\Mhat_{k}^* = \sum_{h=1}^{H}\widehat{p}_{h,k}\Mhat_{h,k}^*\Mhat_{h,k}^{*'}$ and then, 
$\bP_r^0$ as: $\Phat_{r,k}^{0*} = \bSigmahat^{* -\half}_k \Phat_{r,k}^{*}$, where $\Phat_{r,k}^{*}$ denotes the $r$ leading eigenvectors 
of $\Mhat_{k}^*$.

The SS-SIR algorithm aims to improve the estimation of $\bP_r^0$ by making use of $\Usc$ in step (ii) through a nearest neighbour approximation for the unobserved $Y$ in $\mathcal{U}$ using $\mathcal{L}_k^-$. With $n_\K^-$ large enough and $m(\cdot)$ smooth enough, the imputed and the true underlying $Y$ should belong to the same slice with a high probability. Thus, the set of $\X$'s 
belonging to a particular slice is now `enriched' and consequently, improved estimation of $\M$ and $\bP_r^{0}$ is expected. The proposed method based on a nearest neighbor approximation is also highly scalable and while other smoothing based approximations may be used, they can be computationally intensive. The SS-SIR algorithm is fairly robust to the choice of $H$, and $H = O(\nhalf\log n)$ seems to give fairly satisfactory performance. The slices may be chosen to have equal width or equal number of observations. For SIR, $\nhalf $-consistency of the estimates are well established \citep{K-C_Li_1991,Duan_1991,Li-Xing_Zhu_1995} for various formulations under fairly general settings (without any model based assumptions). The theoretical properties of SS-SIR, although not derived here, are expected to follow similarly. Our simulation results (not shown here) further suggest that SS-SIR significantly outperforms SIR, leading to substantially improved estimation of $\btheta_0$ from the proposed methods.

\section{Numerical Studies}\label{num_analysis}
\subsection{Simulation Studies}\label{sim}

We conducted extensive simulation studies to examine the finite sample performance of our proposed point and interval estimation procedures as well as to compare with existing methods. Throughout we let $n = 500$, $N = 10000$, and considered $p= 2, 10$ and $20$. For our CV based methods, we let $\K = 5$. The true values of the target parameter $\btheta_0$  were estimated via monte carlo with a large sample size of $50,000$. For each configuration, the results were summarized based on 500 replications. Results for $p=2$ are summarized in the \hyperref[supp_mat]{Supplementary Material}, and the discussions below focus primarily on $p=10$ and $20$.

We generated $\bX \sim \mathcal{N}_{p}[\mathbf{0}, I_{p}]$ and restricted $\bX$ to $[-5, 5]^p$ to ensure its boundedness. Given $\bX=\bx$, we generated $Y  \sim \mathcal{N}_{1}[m(\bx), 1]$, where we considered four different choices of $m(\bx)$ :
\begin{enumerate}
\item[] (i) \emph{Linear}: $m(\bx) = \bx'\bb_p$;
\item[] (ii) \emph{Non-linear one component (NL1C):} $m(\bx)= (\bx'\bb_p) + (\bx'\bb_p)^2$;
\item[] (iii) \emph{Non-linear two component (NL2C):} $m(\bx)= (\bx'\bb_p)(1 + \bx'\boldsymbol{\delta}_p)$; and
\item[] (iv) \emph{Non-linear three component (NL3C):} $m(\bx)= (\bx'\bb_p)(1 + \bx'\boldsymbol{\delta}_p) + (\bx'\boldsymbol{\omega}_p)^2$;
\end{enumerate}
where, for each setting, we considered $\bb_p=\bb_p^{(1)} \equiv (\bone_{p/2}', \bzero_{p/2}')'$ 
and $\bb_p=\bb_p^{(2)} \equiv \bone_p$, 
and set $\boldsymbol{\delta}_p = (\bzero_{p/2}', \bone_{p/2}')'$ and $\boldsymbol{\omega}_p = (1, 0, 1, 0,\hdots, 1, 0)'_{p\times 1}$, where for any $a$, $\bone_a = (1,\hdots, 1)'_{a \times 1}$ and $\bzero_a = (0,\hdots, 0)'_{a\times 1}$. 
Through appropriate choices of $\bb_p$, $\boldsymbol{\delta}_p$ and $\boldsymbol{\omega}_p$, as applicable, these models can incorporate commonly encountered linear, quadratic and interaction effects.

For each setting, we used two choices of the smoothing method:
(a) $\Tsc := \mbox{KS}_{2,\bP_2}$ denoting KS with 2-dimensional smoothing over $\bP_r'\bX \equiv \bP_2'\bX$, where $\bP_2$ was estimated via SIR with $H = 100$ slices of equal width, following which $\{\mhat_k(\bx;\Phat_{r,k})\}_{k=1}^\K$ were obtained via KS using a  Gaussian kernel; 
(b) $\Tsc := \mbox{KM}$ where we let $\bP_r = I_p$ and then estimated $\{\mhat_k(\bx;I_p)\}_{k=1}^\K$ using kernel machine (KM) regression based on a radial basis function (RBF) kernel. Throughout, $h$ for KS, and all tuning parameters for KM were selected via least squares CV. 
For (a), with $\bX \sim \Nsc_p[\bzero, I_p]$, results from \cite{K-C_Li_1991} imply that the SNP imputation with $r =2$ is sufficient for models (i)-(iii), and insufficient for model (iv).
For comparison, we also implemented two other SS estimators: the density ratio based ``DRESS" estimator of \citet{Kawakita_2013} and the estimator of \citet{Sokolovska_2008} called ``MSSL" by \citet{Kawakita_2013}. The density ratio estimation for the DRESS estimator was implemented using either (i) linear bases $\{1, (\bX_{[j]})_{j=1}^p\}$ ($\mbox{DRESS}_1$); or (ii) cubic bases $\{1, (\bX_{[j]}^d)_{j=1, d = 1}^{p, \; 3}\}$ ($\mbox{DRESS}_3$). 

First, we compare the various estimators with respect to their efficiencies based on empirical mean squared error. In Table \ref{tab_rel-eff}, we present the efficiencies of the proposed SNP and EASE estimators as well as other SS estimators relative to the OLS. As expected, under model mis-specification, our estimators are substantially more efficient than the OLS with the relative efficiency (RE) as high as near 5 fold when $p=10$ and 3 fold when $p =20$, for the non-linear models. The efficiency gain is generally lower for $p=20$ than for $p=10$, likely a consequence of overfitting of the non-parametric estimators involved in the SNP imputation for larger $p$. 
Comparing EASE to SNP, the EASE generally perform better for both linear and non-linear settings, as expected. Comparing the two smoothers, it appears that $\Tsc := \mbox{KM}$ generally attains higher efficiency compared to that of $\Tsc := \mbox{KS}_{2,\bP_2}$. This is in part due to the high variability in the SIR direction estimation which impacts the performance of the resulting SS estimator in finite samples. Interestingly, none of the existing SS estimators perform well with REs ranging only from about 0.9 to 1.1 across all settings.

\renewcommand\thetable{\arabic{table}}
\renewcommand\thefigure{\arabic{figure}}
%
\begin{table}[H]
\centering
\caption{Efficiencies of $\thetahatprk$ (SNP) and $\thetahatprk^E$ (EASE) using $\Tsc := \mbox{KS}_{2,\bP_2}$ or $\Tsc := \mbox{KM}$, as well as $\mbox{DRESS}_1$, $\mbox{DRESS}_3$ and MSSL,  relative to $\bthetahat$ (OLS) with respect to the empirical mean squared error (MSE) under models (i), (ii), (iii) and (iv) each with:  (I) $\bb_p = \bb_p^{(1)}$ or, (II) $\bb_p = \bb_p^{(2)}$. } \label{tab_rel-eff}
\centerline{(a) $p=10$}
\vspace{0.05in}
\resizebox{\textwidth}{!}{
\begin{tabular}{c|c|c||rr||rr||ccc}\hline\hline
& \multicolumn{1}{c|}{} &\multicolumn{1}{c||}{OLS} &\multicolumn{1}{c}{SNP}&\multicolumn{1}{c||}{EASE} &\multicolumn{1}{c}{SNP}&\multicolumn{1}{c||}{EASE} &\multicolumn{3}{c}{Other SS Estimators}\\
\raisebox{1ex}[0cm][0cm]{Setting} & \raisebox{1ex}[0cm][0cm]{Models} &(Ref.)    &\multicolumn{2}{c||}{$(\Tsc:= \mbox{KS}_{2, \bP_2})$}  &\multicolumn{2}{c||}{$(\Tsc := \mbox{KM})$}          &$\mbox{DRESS}_1$ &$\mbox{DRESS}_3$ &MSSL   \\
\cline{1-10}
&  Linear  &1    &0.895  &0.983  &0.772 &0.985   &0.982  &0.927 &0.982 \\
\raisebox{-1ex}[0cm][0cm]{(I)}  & NL1C    &1    &4.481  &4.424  &4.501 &5.543   &1.136  &1.110 &1.135 \\
& NL2C    &1    &2.683  &2.700  &4.268 &5.055   &1.120  &1.016 &1.119 \\
& NL3C    &1    &2.772  &2.795  &4.481 &5.560   &1.102  &1.025 &1.103 \\ \hline
&  Linear  &1    &0.841  &0.989  &0.657 &0.993   &0.981  &0.924 &0.981 \\
 \raisebox{-1ex}[0cm][0cm]{(II)}   & NL1C    &1    &4.511  &4.585  &4.416 &5.471   &1.132  &1.030 &1.130 \\
&  NL2C    &1    &3.596  &3.634  &4.405 &5.497   &1.127  &1.042 &1.128 \\
&  NL3C    &1    &3.280  &3.301  &4.636 &5.566   &1.110  &1.079 &1.109 \\ \hline
 \end{tabular}
}

\vspace{0.1in}
\centerline{(b) $p=20$}
\vspace{0.05in}
\resizebox{\textwidth}{!}{
\begin{tabular}{c|c|c||rr||rr||ccc}\hline\hline
&\multicolumn{1}{c|}{} &\multicolumn{1}{c||}{OLS} &\multicolumn{1}{c}{SNP}&\multicolumn{1}{c||}{EASE} &\multicolumn{1}{c}{SNP}&\multicolumn{1}{c||}{EASE} &\multicolumn{3}{c}{Other SS Estimators}\\
\raisebox{1ex}[0cm][0cm]{Setting} & \raisebox{1ex}[0cm][0cm]{Models} &(Ref.)    &\multicolumn{2}{c||}{$(\Tsc:= \mbox{KS}_{2, \bP_2})$}  &\multicolumn{2}{c||}{$(\Tsc := \mbox{KM})$}          &$\mbox{DRESS}_1$ &$\mbox{DRESS}_3$ &MSSL   \\
\cline{1-10}
&  Linear    &1    &0.673  &0.986  &0.740 &0.981   &0.956  &0.866 &0.956 \\
\raisebox{-1ex}[0cm][0cm]{(I)} & NL1C      &1    &2.256  &2.288  &2.680 &3.630   &1.035  &0.920 &1.035 \\
&  NL2C      &1    &1.414  &1.388  &2.661 &3.544   &1.032  &0.922 &1.033 \\
& NL3C      &1    &1.539  &1.531  &2.605 &3.510   &1.049  &0.931 &1.051 \\ \hline
& Linear    &1    &0.519  &0.991  &0.609 &0.989   &0.958  &0.872 &0.958 \\
\raisebox{-1ex}[0cm][0cm]{(II)} &  NL1C      &1    &2.290  &2.346  &2.669 &3.660   &1.032  &0.908 &1.031 \\
& NL2C      &1    &1.899  &1.917  &2.766 &3.963   &1.036  &0.917 &1.036 \\
& NL3C      &1    &1.937  &1.949  &2.682 &3.702   &1.046  &0.958 &1.046 \\ \hline
 \end{tabular}
}
\end{table}

We next examine the performance of the proposed inference procedures. In Table \ref{tab_p10}(a) and (b), we present the bias, empirical standard error (ESE), the average of the estimated standard error (ASE) and the coverage probability (CovP) of the 95\%  CIs for each component of $\btheta_0$ when $p=10$ under the linear and NL2C models. In general, the EASE with both the KS and the KM smoothers have negligible biases although the KM based estimator appears to have slightly lower biases. The ASEs are close to the ESEs and the CovPs are close to the nominal level, suggesting that the  variance estimators work well in practice with $\K = 5$. 
%
%
\begin{table}[H]
\centering
\centerline{(a) OLS and EASE for the linear model.}
\vspace{.05in}
\resizebox{\textwidth}{!}{
\begin{tabular}{c|rr||rrrr||rrrr}\hline\hline
&\multicolumn{2}{c||}{OLS $(\bthetahat)$} &\multicolumn{4}{c||}{EASE $(\thetahatprk^E; \Tsc := \mbox{KS}_{2,\bP_2})$} &\multicolumn{4}{c}{EASE $(\thetahatprk^E; \Tsc := \mbox{KM})$} \\
\raisebox{2ex}[0cm][0cm]{Parameter} &Bias&ESE    &Bias&ESE&ASE&CovP           &Bias&ESE&ASE&CovP   \\
\cline{1-11}
 $\alpha_0   =0$    &-0.001 & 0.043    &-0.001 & 0.043 & 0.044 & 0.95    & 0.000  &0.043 &0.044 &0.96 \\    
 $\beta_{01} =1$    & 0.002 & 0.044    &-0.003 & 0.045 & 0.044 & 0.94    & 0.004  &0.047 &0.044 &0.93 \\    
 $\beta_{02} =1$    &-0.001 & 0.044    &-0.005 & 0.044 & 0.044 & 0.94    & 0.000  &0.045 &0.044 &0.95 \\    
 $\beta_{03} =1$    &-0.001 & 0.046    &-0.005 & 0.046 & 0.044 & 0.95    &-0.004  &0.045 &0.044 &0.94 \\    
 $\beta_{04} =1$    &-0.002 & 0.045    &-0.006 & 0.045 & 0.044 & 0.94    & 0.001  &0.047 &0.044 &0.94 \\    
 $\beta_{05} =1$    &-0.004 & 0.048    &-0.008 & 0.049 & 0.044 & 0.92    &-0.001  &0.046 &0.044 &0.95 \\    
 $\beta_{06} =0$    &-0.000 & 0.045    &-0.001 & 0.045 & 0.044 & 0.94    & 0.001  &0.045 &0.044 &0.95 \\    
 $\beta_{07} =0$    & 0.003 & 0.046    & 0.003 & 0.046 & 0.044 & 0.93    & 0.001  &0.043 &0.044 &0.96 \\    
 $\beta_{08} =0$    &-0.001 & 0.045    &-0.001 & 0.045 & 0.044 & 0.95    &-0.000  &0.048 &0.044 &0.94 \\    
 $\beta_{09} =0$    &-0.002 & 0.047    &-0.002 & 0.048 & 0.044 & 0.94    & 0.000  &0.045 &0.045 &0.95 \\    
 $\beta_{010}=0$    & 0.003 & 0.045    & 0.003 & 0.045 & 0.044 & 0.94    &-0.002  &0.045 &0.045 &0.94 \\    
 \hline
\end{tabular}
}

\vspace{.1in}
\centering
\centerline{(b) OLS and EASE for the NL2C model.}
\vspace{.05in}
\resizebox{\textwidth}{!}{
\begin{tabular}{c|rr||rrrr||rrrr}\hline\hline
&\multicolumn{2}{c||}{OLS $(\bthetahat)$} &\multicolumn{4}{c||}{EASE $(\thetahatprk^E; \Tsc := \mbox{KS}_{2,\bP_2})$} &\multicolumn{4}{c}{EASE $(\thetahatprk^E; \Tsc := \mbox{KM})$} \\
\raisebox{2ex}[0cm][0cm]{Parameter} &Bias&ESE     &Bias&ESE&ASE&CovP            &Bias&ESE&ASE&CovP   \\
\cline{1-11}
 $\alpha_0   =0$    &-0.015  & 0.239   &-0.016 & 0.146 & 0.136 & 0.93     & 0.013 &0.105 &0.096 &0.93 \\ 
 $\beta_{01} =1$    & 0.000  & 0.260   & 0.015 & 0.159 & 0.160 & 0.96     &-0.004 &0.124 &0.112 &0.93 \\ 
 $\beta_{02} =1$    &-0.004  & 0.269   & 0.017 & 0.173 & 0.158 & 0.93     & 0.010 &0.127 &0.113 &0.93 \\ 
 $\beta_{03} =1$    &-0.015  & 0.249   & 0.018 & 0.156 & 0.158 & 0.95     &-0.000 &0.118 &0.113 &0.95 \\ 
 $\beta_{04} =1$    &-0.001  & 0.267   & 0.016 & 0.164 & 0.159 & 0.94     & 0.007 &0.124 &0.113 &0.93 \\ 
 $\beta_{05} =1$    & 0.013  & 0.260   & 0.019 & 0.164 & 0.158 & 0.94     & 0.002 &0.120 &0.113 &0.94 \\ 
 $\beta_{06} =0$    &-0.010  & 0.281   & 0.008 & 0.164 & 0.155 & 0.94     & 0.005 &0.119 &0.112 &0.94 \\ 
 $\beta_{07} =0$    & 0.006  & 0.277   & 0.002 & 0.166 & 0.155 & 0.93     & 0.011 &0.116 &0.111 &0.95 \\ 
 $\beta_{08} =0$    &-0.008  & 0.277   &-0.004 & 0.167 & 0.156 & 0.94     &-0.001 &0.120 &0.112 &0.95 \\ 
 $\beta_{09} =0$    & 0.002  & 0.279   & 0.003 & 0.160 & 0.157 & 0.95     & 0.007 &0.118 &0.113 &0.95 \\ 
 $\beta_{010}=0$    &-0.008  & 0.272   & 0.002 & 0.160 & 0.155 & 0.95     & 0.004 &0.130 &0.111 &0.91 \\ 
  \hline
\end{tabular}
}

\vspace{0.1in}
\centering
\centerline{(c) All other SS estimators for the models in (a) and (b) above.} 
\vspace{.05in}
\resizebox{\textwidth}{!}{
\begin{tabular}{rr|rr|rr||rr|rr|rr}\hline\hline
\multicolumn{6}{c||}{Linear Model}  &\multicolumn{6}{c}{NL2C Model} \\
\multicolumn{2}{c|}{$\mbox{DRESS}_1$} &\multicolumn{2}{c|}{$\mbox{DRESS}_3$} &\multicolumn{2}{c||}{$\mbox{MSSL}$} &\multicolumn{2}{c|}{$\mbox{DRESS}_1$} &\multicolumn{2}{c|}{$\mbox{DRESS}_3$} &\multicolumn{2}{c}{$\mbox{MSSL}$} \\
\multicolumn{1}{c}{Bias} &\multicolumn{1}{c|}{ESE} &\multicolumn{1}{c}{Bias} &\multicolumn{1}{c|}{ESE} &\multicolumn{1}{c}{Bias} &\multicolumn{1}{c||}{ESE} &\multicolumn{1}{c}{Bias} &\multicolumn{1}{c|}{ESE}  &\multicolumn{1}{c}{Bias} &\multicolumn{1}{c|}{ESE} &\multicolumn{1}{c}{Bias} &\multicolumn{1}{c}{ESE} \\
\cline{1-12}
  -0.001 &0.043    &-0.001 &0.044    &-0.001 &0.043          &-0.004 &0.223   &-0.003 &0.226     &-0.004 &0.223  \\
  -0.002 &0.044    &-0.001 &0.046    &-0.002 &0.044          &-0.014 &0.266   &-0.009 &0.279     &-0.014 &0.266  \\
   0.000 &0.045    & 0.001 &0.046    & 0.000 &0.045          & 0.005 &0.257   & 0.006 &0.266     & 0.006 &0.257  \\
   0.006 &0.045    & 0.006 &0.047    & 0.006 &0.045          &-0.013 &0.256   &-0.019 &0.281     &-0.011 &0.256  \\
   0.003 &0.045    & 0.003 &0.046    & 0.003 &0.045          &-0.005 &0.262   &-0.007 &0.274     &-0.005 &0.262  \\
  -0.004 &0.047    &-0.004 &0.049    &-0.004 &0.047          & 0.002 &0.250   &-0.007 &0.266     & 0.002 &0.252  \\
  -0.001 &0.045    &-0.001 &0.046    &-0.001 &0.045          &-0.017 &0.239   &-0.009 &0.247     &-0.017 &0.239  \\
  -0.000 &0.048    &-0.001 &0.050    &-0.000 &0.048          &-0.022 &0.260   &-0.019 &0.270     &-0.022 &0.260  \\
  -0.004 &0.043    &-0.003 &0.044    &-0.004 &0.043          &-0.011 &0.241   &-0.013 &0.261     &-0.010 &0.241  \\
  -0.001 &0.048    &-0.001 &0.049    &-0.001 &0.048          &-0.020 &0.256   &-0.019 &0.259     &-0.020 &0.256  \\
  -0.003 &0.047    &-0.003 &0.049    &-0.003 &0.047          &-0.020 &0.252   &-0.022 &0.269     &-0.020 &0.252  \\ \hline
\end{tabular}
}
\caption{Coordinate-wise bias, ESE, ASE and CovP of EASE, obtained using $\Tsc := \mbox{KS}_{2,\bP_2}$ or $\Tsc := \mbox{KM}$, for estimating $\btheta_0 $ under the linear and NL2C models with $p = 10$  and $\bb_p = \bb_p^{(1)}$. Shown also are the corresponding bias and ESE of the OLS, as well as the $\mbox{DRESS}_1$, $\mbox{DRESS}_3$ and $\mbox{MSSL}$ estimators. } \label{tab_p10}
\end{table}As shown in Table \ref{tab_p10}(c), the other SS estimators tend to have slightly larger biases and substantially larger standard errors (SEs) compared to our estimators under the NL2C model.

\subsection{Application to EMR Data}\label{data_ex}

We applied our proposed SS procedures to an EMR study of rheumatoid arthritis (RA), a systemic auto-immune disease (AD), conducted at the Partners HealthCare \citep{Liao_2010}.  The study cohort consists of $3854$ RA patients with blood samples stored. The outcome of interest is the (logarithm of) anti-CCP (antibodies to cyclic citrullinated polypeptide), a biomarker that is often used to determine subtypes of RA. Due to cost constraints, anti-CCP was measured only for a random subset of $n = 355$ patients, thereby leading to a SS set-up. To investigate the validity of the MCAR assumption, we report in Table \ref{tab_covcomp_data_ex} in the \hyperref[supp_mat]{Supplementary Material} summary measures of the distributions in the labeled and unlabeled data for each of the predictors, as well as p-values from various tests for assessing equality of those distributions. The results suggest that the MCAR assumption is appropriate in this study.


We relate the log anti-CCP level to a set of $p = 24$ clinical variables $\bX$ related to ADs, including age, gender, race; total counts of codified and/or narrative mentions extracted from physicians' notes via natural language processing (NLP) for various RA related conditions including RA, Lupus, Polymyalgiarheumatica (PmR), Spondyloarthritis (SpA), as well as various  RA  medications; indicators of seropositivity and radiological evidence of erosion;  mentions of rheumatoid factor (RF), as well as anti-CCP positivity from prior medical history. Since the tests for RF and anti-CCP were not always ordered, missing indicators for these variables were also included. All count variables were transformed as: $x \to \log(1+x)$ to increase stability of the model fitting. All predictors were normalized to have unit variance.


We obtained the OLS, the EASE using both $\Tsc := \mbox{KS}_{2,\bP_2}$ and $\Tsc := \mbox{KM}$ in the smoothing step, as well as the $\mbox{DRESS}_1$ estimator for comparison. For EASE, we again used $\K=5$ and for the $\mbox{KS}_{2,\bP_2}$ smoother, $\bP_2$ was obtained using SIR with $H=80$ slices of equal width. In Table \ref{tab_data_ex_new}, we present the coordinate-wise estimates of the regression parameters along with their estimated SEs and the corresponding p-values based on these estimates. Overall, the point estimators from all methods are quite close to each other. Our proposed EASE, with both KS and KM smoothers, is substantially more efficient than the OLS across all coordinates with efficiency ranging from about 1.4 to 2.4. The $\mbox{DRESS}_1$ estimator improved estimation for a few coordinates but the efficiency remains comparable to the OLS for most coordinates. This again suggests the advantage of our proposed estimators compared to both OLS and other SS estimators.
%
%
\begin{table}[H]
\centering
\resizebox{\textwidth}{!}{
\begin{tabular}{l|rrr||rrr|r}
\hline\hline
\multicolumn{1}{l|}{} &\multicolumn{3}{c||}{OLS $(\bthetahat)$} &\multicolumn{3}{c|}{EASE $(\mbox{KS}_{2,\bP_2})$} &\multicolumn{1}{c}{} \\
\multicolumn{1}{l|}{\raisebox{1ex}[0cm][0cm]{Predictors}} &\multicolumn{1}{c}{Est} &\multicolumn{1}{c}{SE} &\multicolumn{1}{c||}{Pval} &\multicolumn{1}{c}{Est} &\multicolumn{1}{c}{SE} &\multicolumn{1}{c|}{Pval} &\multicolumn{1}{c}{\raisebox{1ex}[0cm][0cm]{RE}} \\
\cline{1-8}
Age                        & .105 & .076 & .168   & .106 & .064 & .099 & 1.40  \\
Gender                     &-.032 & .059 & .589   &-.028 & .050 & .570 & 1.41  \\
Race                       &-.041 & .065 & .534   &-.042 & .055 & .452 & 1.40  \\
Lupus                      & .038 & .066 & .563   & .048 & .052 & .359 & 1.59  \\
PmR                        &-.075 & .044 & .088   &-.076 & .031 & .013 & 2.07  \\
RA                         & .015 & .089 & .862   & .012 & .076 & .879 & 1.37  \\
SpA          &-.137 & .102 & .177   &-.133 & .072 & .063 & 2.02  \\
Other ADs                  &-.022 & .078 & .775   &-.024 & .058 & .679 & 1.79  \\
Erosion                    & .076 & .070 & .278   & .078 & .059 & .184 & 1.44  \\
Seropositivity             & .056 & .062 & .370   & .054 & .053 & .310 & 1.37  \\
$\mbox{Anti-CCP}_{prior}$  & .572 & .136 & .000   & .557 & .110 & .000 & 1.54  \\
$\mbox{Anti-CCP}_{miss}$   & .527 & .123 & .000   & .520 & .097 & .000 & 1.61  \\
RF                         & .128 & .081 & .113   & .125 & .066 & .059 		& 1.49  \\
$\mbox{RF}_{miss}$         & .085 & .085 & .316   & .084 & .070 & .233 & 1.46  \\
Azathioprine               &-.080 & .071 & .263   &-.074 & .056 & .185 & 1.62  \\
Enbrel                     & .138 & .070 & .048   & .133 & .058 & .021 & 1.48  \\
Gold salts                 & .138 & .050 & .006   & .136 & .043 & .002 & 1.37  \\
Humira                     &-.051 & .068 & .453   &-.049 & .057 & .391 & 1.43  \\
Infliximab                 & .003 & .069 & .968   & .008 & .057 & .887 & 1.50  \\
Leflunomide                &-.027 & .069 & .697   &-.023 & .058 & .693 & 1.40  \\
Methotrexate               &-.021 & .073 & .775   &-.024 & .061 & .699 & 1.42  \\
Plaquenil         &-.043 & .069 & .540   &-.038 & .057 & .503 & 1.47  \\
Sulfasalazine              &-.114 & .074 & .125   &-.116 & .063 & .064 & 1.39  \\
Other meds.          &-.042 & .074 & .570   &-.052 & .060 & .385 & 1.52 \\ \hline
\end{tabular}\vspace{.1in}

\begin{tabular}{||rrr|r||rrr|r}\hline\hline
\multicolumn{3}{||c|}{$\mbox{DRESS}_1$} &\multicolumn{1}{c||}{} &\multicolumn{3}{c|}{EASE $(\mbox{KM})$} &\multicolumn{1}{c}{}\\
\multicolumn{1}{||c}{Est} &\multicolumn{1}{c}{SE} &\multicolumn{1}{c|}{Pval} &\multicolumn{1}{c||}{\raisebox{1ex}[0cm][0cm]{RE}} &\multicolumn{1}{c}{Est} &\multicolumn{1}{c}{SE} &\multicolumn{1}{c|}{Pval} &\multicolumn{1}{c}{\raisebox{1ex}[0cm][0cm]{RE}}\\
\cline{1-8}
.094 &.073 &.199 &1.09   & .104 &.064 &.103 &1.42 \\
-.027 &.058 &.638 &1.04   &-.031 &.049 &.524 &1.44 \\
-.044 &.067 &.511 &.95   &-.040 &.055 &.462 &1.41 \\
.021 &.063 &.731 &1.11   & .037 &.051 &.464 &1.70 \\
-.074 &.031 &.016 &2.10   &-.075 &.030 &.014 &2.04 \\
.008 &.080 &.923 &1.23   & .016 &.075 &.832 &1.30 \\
-.128 &.075 &.089 &1.82   &-.136 &.066 &.038 &2.37 \\
-.018 &.067 &.792 &1.35   &-.022 &.056 &.692 &1.93 \\
.085 &.069 &.221 &1.03   & .076 &.058 &.189 &1.47 \\
.041 &.061 &.496 &1.05   & .055 &.052 &.296 &1.41 \\
.567 &.123 &.000 &1.23   & .568 &.107 &.000 &1.60 \\
.508 &.115 &.000 &1.15   & .523 &.096 &.000 &1.64 \\
.149 &.079 &.059 &1.05   & .127 &.066 &.054 &1.49 \\
.137 &.080 &.088 &1.12   & .084 &.070 &.231 &1.48 \\
-.075 &.062 &.225 &1.33   &-.079 &.053 &.132 &1.83 \\
.136 &.073 &.064 &.91   & .137 &.057 &.017 &1.49 \\
.147 &.050 &.003 &1.01   & .137 &.042 &.001 &1.40 \\
-.057 &.067 &.389 &1.03   &-.051 &.056 &.360 &1.49 \\
 .000 &.067 &.994 &1.07   & .003 &.055 &.959 &1.57 \\
-.031 &.071 &.660 &.93   &-.026 &.057 &.644 &1.45 \\
-.025 &.073 &.728 &1.01   &-.022 &.060 &.720 &1.46 \\
-.044 &.070 &.532 &.98   &-.042 &.057 &.460 &1.48 \\
-.105 &.072 &.145 &1.06   &-.113 &.061 &.065 &1.45 \\
-.052 &.071 &.466 &1.10   &-.042 &.059 &.473 &1.59 \\ \hline
\end{tabular}
}
\caption{Estimates (Est) of the regression coefficients based on OLS, EASE obtained using either $\Tsc := \mbox{KS}_{2,\bP_2}$ or $\Tsc :=$ KM, as well as $\mbox{DRESS}_1$, along with their estimated standard errors (SE) and the corresponding p-values (Pval.)  for testing the null effect of each predictor. Shown also are the relative efficiencies (RE) of all the estimators compared to the OLS.}\label{tab_data_ex_new}
\end{table}

We also estimated the prediction errors (PEs) for each of the fitted linear models based on the aforementioned estimation methods via CV. To remove potential randomness in the CV partitions, we averaged over 10 replications of leave-5-out CV estimates. The PE was about 1.28 for EASE with both smoothers, 1.29 for OLS and 1.30 for DRESS$_1$. For prediction purposes, we may also directly employ non-parametric estimates of the conditional mean rather than the fitted linear models. The PE in fact is slightly larger when we use $\mhat(\bx;\Phat_r)$ or $\widehat{\mu}(\bx; \Pschat_{r,\K})$. The PE was 1.34 for KS and 1.33 for KM based on $\mhat(\bx;\Phat_r)$, and 1.30 for KS and 1.28 for KM based on  $\widehat{\mu}(\bx; \Pschat_{r,\K})$. This confirms that while the linear model may be mis-specified, it may often be preferable to non-parametric models in practice as it may achieve simplicity without substantial loss in prediction performance.

\section{Discussion}\label{discus}

We have developed in this paper an efficient and adaptive estimation strategy for the SS linear regression problem.
The adaptive property possessed by the proposed EASE  is crucial for advocating `safe' use of the unlabeled data and is often unaddressed in the existing literature. In general, the magnitude of the efficiency gain with EASE depends on the inherent degree of non-linearity in $\E(Y \medgiven\bX)$ and the extent of sufficiency of the underlying SNP imputation. In particular, if the imputation is sufficient or the working linear model is correct, $\thetahatprk^E$ is further optimal among a wide class of estimators. We obtained theoretical results along with IF 
expansions for $\thetahatprk$ and $\thetahatprk^E$ substantiating all our claims and also validated them based on numerical studies. The double CV method further facilitates accurate inference, overcoming potential over-fitting issues in finite samples due to smoothing. An R code for implementing EASE is available upon request.

The proposed SNP imputation, the key component of EASE, apart from being flexible and scalable, enjoys several useful properties. The refitting step and CV play a crucial role in reducing the bias of $\thetahatprk$, and for $\Tsc :=$ KS in particular, eradicate any under-smoothing or higher order kernel requirements: two undesirable, yet often inevitable, conditions required for $\nhalf$-consistency of two-step estimators based on a first step of smoothing. Theorem \ref{thm4}, apart from showing the distinct behaviour of $\bthetahat_{(\bP_r,1)}$ compared to $\thetahatprk$ for $\K \geq 2$, also highlights the key role of CV in completely removing kernel order restrictions, apart from addressing over-fitting issues. The error rates in the results of Theorems \ref{thm3}-\ref{thm4} are quite sharp and account for any estimation error from $\Phat_r$. The regularity conditions required are also fairly mild and standard in the literature. The continuity assumption on $\bX$ in Sections \ref{fnp_est} and \ref{kern_smth_impl} is mostly for the convenience of proofs, and the results continue to hold for more general $\bX$. Lastly, while we have focussed here on linear regression for simplicity, our methods can indeed be easily adapted to other regression problems such as logistic regression for binary outcomes.

When the goal is solely that of \emph{prediction}, one obviously does not have to employ linear regression models, and models that incorporate non-linear effects can be helpful. For such settings, the estimators $\mhat(\bx; \Phat_r)$ or $\widehat{\mu}(\bx; \Pschat_{r,\K})$, obtained as by-products of our SNP imputation, can themselves serve as potentially useful non-linear predictors. These SNP estimators may substantially outperform naive non-parametric estimators such as a $p$-dimensional KS estimator, as demonstrated in Table \ref{tab_oos-pe-np} of the \hyperref[supp_mat]{Supplementary Material} for the models considered in our simulation studies. In practice, when the covariates are substantially correlated and the dimension of $p$ is not small as in the EMR example, it is unclear whether non-linear models necessarily provide better prediction performance than the linear models. Under such settings, the linear model also has a clear advantage due to its simplicity. Furthermore, while prediction is a vitally important goal of predictive modeling, association analysis under interpretable models is key to clinical studies for discovery research and efficient estimation of the corresponding model parameters remains an important task.

We end with a comment on the \emph{choice of  $\K \geq 2$} in $\thetahatprk$. While (\ref{snp_k2_exp}) holds for any $\K \geq 2$, the error term in (\ref{snp_k2_exp}) depends on $\K$ through $\cnkminus$ and more precisely, through $\widetilde{c}_{\nkminus}$ $= \Khalf\cnkminus$. Since $\K$ is fixed, $\cnkminus$ and $\widetilde{c}_{\nkminus}$ are asymptotically equivalent. But for a given $n$, $\cnkminus$ is expected to decrease with $\K$, while $\widetilde{c}_{\nkminus}$ is likely to increase. It is however desirable that both are small since $\cnkminus$ inherently controls the efficiency of the SNP imputation, while $\widetilde{c}_{\nkminus}$ directly controls the bias of $\thetahatprk$. Hence, a reasonable choice of $\K \geq 2$ may be based on minimizing: ($\cnkminus^2$ $+ \lambda\widetilde{c}_{\nkminus}^2$) for some $\lambda \geq 0$. Since the (first order) asymptotic variance of $\thetahatprk$ is independent of $\K$, this is equivalent to a penalized minimization of the asymptotic MSE of $\thetahatprk$ with $\lambda$ denoting the weightage of the (lower order) bias relative to the (first order) variance. In general, the optimal $\K$ should be inversely related to $\lambda$. Conversely, choice of any $\K$ may be viewed to have an associated regularization effect (through $\lambda$) resulting in a `variance-bias trade-off' with smaller $\K$ leading to lower bias at the cost of some efficiency, and higher $\K$ leading to improved efficiency in lieu of some bias. In practice, we find that $\K$ $= 5$ works well, and $\K=10$ tends to give slightly smaller MSE at the cost of increased bias.

\begin{appendix}  
\numberwithin{equation}{section}

\section{}\label{app}

\subsection{Preliminaries}\label{app-lemmas}
The following Lemmas \ref{lemma0}-\ref{lemma2} would be useful in the proofs of the main theorems. The proofs of these lemmas, as well as Theorems \ref{thm1}, \ref{thm3} and \ref{thm4}, can be found in the \hyperref[supp_mat]{Supplementary Material}.

\begin{lemma}\label{lemma0}
Let $\bZ \in \mathbb{R}^{l}$ be any random vector and $\bg(\bZ) \in \mathbb{R}^{d}$ be any measurable function of $\bZ$, where $l$ and $d$ are fixed. Let $\S_n = \{\bZ_i\}_{i = 1}^{n}$ $\ind \S_m = \{\bZ_j\}_{j = 1}^{m}$ be two random samples of $n$ and $m$ i.i.d. observations of $\bZ$ respectively. Let $\bghat_{n}(\cdot)$ be any estimator of $\bg(\cdot)$ based on $\S_{n}$ such that the random sequence: $\That_n \equiv \mbox{sup}_{\bz \in \boldsymbol{\chi}}  \|\bghat_{n}(\bz)\|$ is $ O_p(1)$, where $\boldsymbol{\chi} \subseteq \mathbb{R}^{l}$ denotes the support of $\bZ$. Let $\bGhat_{n,m}$ denote the (double) random sequence: $m^{-1}\sum_{\bZ_j \in \S_m} \bghat_n(\bZ_j)$, and let $\bGbar_n$ denote the random sequence: $\E_{\S_m} (\bGhat_{n,m}) = \E_{\bZ}\{\bghat_n(\bZ)\}$, where $\E_{\bZ}(\cdot)$ denotes expectation w.r.t. $\bZ \in \S_m \ind \S_n$, and all expectations involved are assumed to be finite almost surely (a.s.) $[\S_n]$ $\forall \; n$.

Then: (a) $\bG_{n,m} - \bGbar_n = O_p(m^{-\frac{1}{2}})$, and (b) as long as $g(.)$ has finite $2^{nd}$ moments, $m^{-1}\sum_{\bZ_j \in \S_m} \bg(\bZ_j) - \E_{\bZ} \left\{\bg(\bZ)\right\} = O_p(m^{-\frac{1}{2}})$.
\end{lemma}
\paragraph*{Controlling Empirical Processes Indexed by KS Estimators} The next two lemmas would be useful in the proof of Theorem \ref{thm4}. They may also be of more general use in other applications that involve controlling empirical processes indexed by kernel estimators - both linear and ratio-type estimators, where the smoothing is further allowed to be performed over a possibly lower dimensional and estimated transformation of the original covariate $\bX$. These allowances make the technical analyses of such processes considerably more involved and nuanced. The results of these lemmas and the techniques used in their proofs may therefore be of independent general interest. 

Suppose 
Assumption \ref{basic_assmpn} (a) holds, 
and consider the KS framework introduced in Section \ref{kern_smth_impl}. 
Let $\vphiPrtil^{(\varrho)}(\bw) = (nh^r)^{-1}\sum_{i=1}^n K_h(\bw,\bP_r'\bX_i) Y_i^{\varrho}$, for $\varrho=0,1$. Let $\fPrtil (\cdot) = \vphiPrtil^{(0)} (\cdot)$, $\lPrtil (\cdot) = \vphiPrtil^{(1)} (\cdot)$ and $\mPrtil (\cdot) = \lPrtil (\cdot) /\fPrtil (\cdot)$. Next, let $\vphiPr^{(0)}(\cdot) = \fPr(\cdot)$ and $\vphiPr^{(1)}(\cdot) = \lPr(\cdot)$, where $\lPr(\cdot) = \mPr(\cdot)\fPr(\cdot)$. For each $\varrho \in \{0,1\}$, let $\varphi^{(\varrho)}(\bx;\bP_r) = \vphiPr^{(\varrho)}(\bP_r'\bx)$ and $\vphitil^{(\varrho)}(\bx;\bP_r) = \vphiPrtil^{(\varrho)}(\bP_r'\bx)$. 
Further, let $\ftil(\cdot) = \vphitil^{(0)}(\cdot)$, $\ltil(\cdot) = \vphitil^{(1)}(\cdot)$ and $\mtil(\cdot) = \ltil(\cdot)/\ftil(\cdot)$.

Lastly, let $\P_n$ denote the empirical probability measure on $\mathbb{R}^p$ based on $\{\bX_i\}_{i=1}^n$, and for any measurable (and possibly vector-valued) function $\boldsymbol{\gamma}(\cdot)$ of $\bX$, where $\boldsymbol{\gamma}(\cdot)$ can be random itself, let $\G_n^*(\boldsymbol{\gamma}) = \nhalf \int\boldsymbol{\gamma}(\bx)(\P_n-\P_{\bX})(d\bx)$, the (centered) $\nhalf$-scaled empirical process indexed by $\boldsymbol{\gamma}(\cdot)$. Lemmas \ref{lemma1}-\ref{lemma2} together, among other more general implications, establish explicit rates of convergence of the quantity $\G_n^*\{ \widehat{g}(\cdot) - g(\cdot)\}$, for any linear or ratio-type kernel estimator $\widehat{g}(\cdot)$ of the type discussed above and its corresponding target $g(\cdot)$.
%
\begin{lemma}\label{lemma1}
Consider the set-up introduced above. For any fixed integer $d \geq 1$, let $\blambda(\cdot)$ be any $\mathbb{R}^d$-valued measurable function of $\bX$ that is bounded a.s. [$\P_{\bX}$]. Define: $b_n^{(1)} = \nnhalf h^{-r} + h^q$ and $a_{n,2} = (\log n)^{\half}(nh^r)^{-\half} + h^q$. Assume $b_n^{(1)} = o(1)$ for (\ref{lemma1_eqn1}) and $\nhalf a_{n,2}^2 = o(1)$ for (\ref{lemma1_eqn2}) below. Then, under Assumption \ref{assmpn_unifconv_snp} (i)-(v), and $\forall \; \varrho \in \{0,1\}$,
\begin{eqnarray}
&& \G_n^*[\blambda(\cdot) \{\vphitil^{(\varrho)}(\cdot\;;\bP_r)-\varphi^{(\varrho)}(\cdot\;;\bP_r)\}] = O_p(b_n^{(1)}) = o_p(1), \;\; \mbox{and} \label{lemma1_eqn1} \\
&& \G_n^*[\blambda(\cdot) \{\mtil(\cdot\;;\bP_r)-m(\cdot\;;\bP_r)\}] = O_p(\nhalf a_{n,2}^2) = o_p(1). \label{lemma1_eqn2}
\end{eqnarray}
\end{lemma}
\noindent Let $\vphihat^{(\varrho)}(\bx;\Phat_r) = (nh^r)^{-1}\sum_{i=1}^n K_h(\Phat_r'\bx,\Phat_r'\bX_i) Y_i^{\varrho}$ $\forall \; \varrho \in \{ 0, 1\}$, where $\Phat_r$ is as in Section \ref{smooth_step_snp} and all other notations are the same as in the set-up of Lemma \ref{lemma1}. Let $\fhat(\bx;\Phat_r) = \vphihat^{(0)}(\bx;\Phat_r)$ and $\lhat(\bx;\Phat_r) = \vphihat^{(1)}(\bx;\Phat_r)$. Then: 
\begin{lemma}\label{lemma2}
Consider the set-up of Lemma \ref{lemma1}. Let $\vphihat^{(\varrho)}(\bx;\Phat_r)$ be as above, and let $\blambda(\cdot)$ be as in Lemma \ref{lemma1}.
Suppose $(\Phat_r -\bP_r) = O_p(\alpha_n)$ for some $\alpha_n = o(1)$. Assume $b_n^{(2)} = o(1)$, where $b_n^{(2)} =$ $\alpha_n + \nnhalf \alpha_n h^{-(r+1)} + \nhalf \alpha_n^2(h^{-2} + n^{-1}h^{-(r+2)})$. Then, under Assumption \ref{assmpn_unifconv_snp},
\begin{equation}\label{lemma2_eqn1}
\G_n^*[\blambda(\cdot) \{\vphihat^{(\varrho)}(\cdot\;;\Phat_r) - \vphitil^{(\varrho)}(\cdot\;;\bP_r)\}] = O_p(b_n^{(2)}) = o_p(1) \;\; \forall \; \varrho \in\{0, 1\}.
\end{equation}
\end{lemma}

\subsection{Proof of Theorem \ref{thm2}}\label{appB}

Let $\bGamma_n = \frac{1}{n}\sum_{i=1}^n\bXv_i\bXv_i'$, and
$$
\bT^{(1)}_{n} = \frac{1}{n}  \sum_{i=1}^n \bXv_i\left\{Y_i - \mu(\bX_i;\bP_r)\right\},  \bT_{n,\K}^{(2)} = \frac{1}{n} \sum_{k=1}^{\K} \sum_{i \in \Isc_k} \bXv_i\Deltahat_k(\bX_i;\bP_r,\Phat_{r,k}).
$$
Then, using (\ref{eq-etahat})-(\ref{deltakhat_def}), it is straightforward to see that:
\begin{align}
&\E[\bXv\{Y- \mu(\bX;\bP_r)\}] \equiv \E[\bXv\{Y- m(\bX;\bP_r) - \bXv'\etapr\}] = \mathbf{0}, \;\; \text{and}\label{b_eqn1}\\
&\bGamma_n\left(\etahatprk - \etapr\right) = \bT^{(1)}_{n} - \bT_{n,\K}^{(2)}.\label{b_eqn2}
\end{align}
Under (\ref{b_eqn1}), Assumptions \ref{basic_assmpn} (a) and \hyperref[thm2]{(i)}, it follows from Lemma \ref{lemma0} (b) that $\bT^{(1)}_{n} = O_p(\nnhalf)$. Next, due to assumption \hyperref[thm2]{(ii)} and boundedness of $\bX$,
$$\|\bT_{n,\K}^{(2)}\| \leq \ninv \sum_{k=1}^\K \sum_{i\in\Isc_k} \text{sup}_{\bx\in\mathcal{X}}\{\|\bxv\|\;|\Deltahat_k(\bx;\bP_r,\Phat_{r,k})|\} = O_p(\cnkminus).$$
Finally, under Assumption \ref{basic_assmpn} (a), we have: $\bGamma_n = \bGamma + O_p(\nnhalf)$ using Lemma \ref{lemma0} (b). Further, since $\bGamma_n$ is invertible a.s., $\bGamma_n^{-1} = \bGammainv + O_p(\nnhalf)$. Using all these facts, we then have: $(\etahatprk-\etapr) =  \bGamma_n^{-1}(\bT^{(1)}_{n}-\bT_{n,\K}^{(2)}) = \bGammainv (\bT^{(1)}_{n}-\bT_{n,\K}^{(2)}) + O_p\{\nnhalf (\nnhalf+\cnkminus)\}$. Thus,
\begin{align}
&(\etahatprk - \etapr) = \bGammainv(\bT^{(1)}_{n} - \bT_{n,\K}^{(2)}) + O_p(\ninv + \nnhalf \cnkminus) .\label{b_eqn3}
\end{align}
Next, let us define:
\begin{eqnarray*}
&& \bGamma_N = N^{-1}\sum_{j=n+1}^{n+N}\bXv_j\bXv_j', \;\; \bR_{N}^{(1)} = \Ninv \sum_{j=n+1}^{n+N}\bXv_j\{\mu(\bX_j;\bP_r)-\bXv_j'\btheta_0\},\\
&& \mbox{and} \;\; \bRhat_{N,n}^{(\K)} = \Ninv \sum_{j=n+1}^{n+N}\bXv_j\{\widehat{\mu}(\bX_j;\Pschat_{r,\K})-\mu(\bX_j;\bP_r)\}.
\end{eqnarray*}
Then, using (\ref{thetahat_snp_defn}), we have:
$$\bGamma_N(\thetahatprk-\btheta_0)= \Ninv\sum_{j=n+1}^{n+N}\bXv_j[\widehat{\mu}(\bX_j;\Pschat_{r,\K})-\bXv_j'\btheta_0] = \bR_{N}^{(1)} + \bRhat_{N,n}^{(\K)}.$$
Next, using (\ref{eq-etahat})-(\ref{deltakhat_def}), we have: $\bRhat_{N,n}^{(\K)} =  \bGamma_N(\etahatprk-\etapr) + \bShat_{N,n}^{(\K)}$, where
$$\bShat_{N,n}^{(\K)} = {\textstyle \Kinv \sum_{k=1}^\K\{\Ninv \sum_{j=n+1}^{n+N}} \bXv_j\Deltahat_k(\bX_j;\bP_r,\Phat_{r,k})\}.$$ Hence, we have: $\bGamma_N(\thetahatprk-\btheta_0)= \bGamma_N(\etahatprk-\etapr) + \bR_{N}^{(1)} + \bShat_{N,n}^{(\K)}$.\vspace{0.05in}

\noindent Now, under assumptions \hyperref[thm2]{(i)-(ii)} and Assumption \ref{basic_assmpn} (a), we have:
$$\mbox{(I)} \quad {\textstyle \sum_{k=1}^{\K}\text{sup}_{\bx\in\mathcal{X}} } \; \|\bxv \Deltahat_k(\bx;\bP_r,\Phat_{r,k})\| = O_p(1),$$
so that using Lemma \ref{lemma0} (a), $\bShat_{N,n}^{(\K)} = \Kinv\sum_{k=1}^\K \bShat_{n,k}^* + O_p(\Nnhalf)$, where $\bShat_{n,k}^* = \E_{\bX}\{\bXv\Deltahat_k(\bX;\bP_r,\Phat_{r,k})\} \;\; \forall \; 1\leq k \leq \K$;
$$\mbox{(II)} \quad \bR_N^{(1)} = \E[\bXv\{\mu(\bX;\bP_r)-\bXv'\btheta_0\}] +  O_p(\Nnhalf) = O_p(\Nnhalf)$$
from Lemma \ref{lemma0} (b) and $\E[\bXv\{\mu(\bX;\bP_r)-\bXv'\btheta_0\}] = \mathbf{0}$ due to (\ref{b_eqn1}) and \ref{tp_defn}; and lastly, (III) $\bGamma_N^{-1} = \bGammainv + O_p(\Nnhalf)$. It then follows from (I)-(III) that
\begin{align}
\label{b_eqn4} & \thetahatprk-\btheta_0 = (\etahatprk-\etapr) + \Kinv \bGammainv \sum_{k=1}^\K \bShat_{n,k}^* + O_p(\Nnhalf).
\end{align}
Using (\ref{b_eqn3}) and (\ref{Gkhat_def}) in (\ref{b_eqn4}), we then have:
$$\thetahatprk-\btheta_0 = \frac{1}{n}
\sum_{i=1}^n\bpsi(\bZ_i; \bP_r)- \; \bGammainv \frac{1}{\K} 
\sum_{k=1}^{\K}\Big\{\frac{1}{\nk}
\sum_{i \in \Isc_k} \bGhat_k(\bX_i)\Big\} + O_p(b_{n,\K}),$$
where $b_{n,\K} = \ninv + \nnhalf \cnkminus + \Nnhalf$. It follows, as claimed in (\ref{snp_fund_exp}), that
\begin{equation}\label{b_eqn5}
  \nhalf (\thetahatprk - \btheta_{0}) = \nnhalf \sum_{i=1}^n\bpsi(\bZ_i; \bP_r) - \bGammainv \G_{n,\K} + O_p(c_{n,\K}^{*}) \qed
\end{equation}

We next show that $\G_{n,\K} = O_p(\cnkminus)$ for any fixed $\K \geq 2$. To this end, let $\T_k^{(n)}$ $= (\nk)^{-\half}\sum_{i \in \Isc_k}\bGhat_k(\bX_i)$, $\Dhat_k = \text{sup}_{\bx\in\Xsc}\; |\Deltahat_k(\bx;\bP_r,\Phat_{r,k})|$ and $C = \text{sup}_{\bx\in\Xsc} \|\bxv\| < \infty$.
For any subset $\mathcal{A} \subseteq \Lsc$, let $\P_{\mathcal{A}}$ denote the joint distribution of the observations in $\mathcal{A}$, and let $\E_{\mathcal{A}}(\cdot)$ denote expectation w.r.t. $\P_{\mathcal{A}}$.
By definition, $\G_{n,\K} = \Knhalf \sum_{k=1}^\K \T_k^{(n)} = O_p(\cnkminus)$ if and only if given any $\epsilon > 0$, $\exists$ $M_\epsilon > 0$ such that $\P\left(\|\G_{n,\K}\| > M_\epsilon \cnkminus \right)$ $\leq \epsilon$ $\forall$ $n$. Note that for any $M > 0$,
\begin{eqnarray}
\nonumber &&\P\left(\|\G_{n,\K}\| > M\cnkminus\right) 
\leq \; \P\left(\Knhalf \sum_{k=1}^\K \|\T_k^{(n)}\| > M\cnkminus \right) \; \\
\nonumber  &&\leq\;  \sum_{k=1}^\K \P\left(\Knhalf \|\T_k^{(n)}\| > \frac{M\cnkminus}{\K}\right)
\leq \; \sum_{k=1}^\K \sum_{l=1}^{p+1} \P\left\{|\T_{k [l]}^{(n)}| > \frac{M\cnkminus}{\Khalf (p+1)^{\half}} \right\} \\
&& \;\;\;=\;\sum_{k=1}^\K \sum_{l=1}^{p+1} \E_{\Lsc_k^-}\left[\P_{\Lsc_k}\left\{|\T_{k [l]}^{(n)}| > \frac{M\cnkminus}{\Khalf (p+1)^{\half}} \given[\bigg] \Lsc_k^- \right\}\right], \label{b_eqn6}
\end{eqnarray} where the steps follow from repeated use of Bonferroni's inequality and other standard arguments. Now, conditional on $\Lsc_k^-$ $(\ind \Lsc_k, \; \mbox{with} \; \K \geq 2)$, $\nk^{\half}\T_k^{(n)}$ is a centered sum of the i.i.d. random vectors $\{\bXv_i\Deltahat_k(\bX_i;\bP_r,\Phat_{r,k})\}_{i \in \Isc_k}$ which, due to assumption \hyperref[thm2]{(ii)} and the compactness of $\Xsc$, are bounded by: $C\Dhat_k < \infty$ a.s. [$\P_{\Lsc_k^-}$] $\forall \; k, n$. Hence, applying Hoeffding's inequality to $\T_{k[l]}^{(n)} \; \forall \; l$, we have: 
\begin{align}
&\P_{\Lsc_k}\left\{|\T_{k [l]}^{(n)}| > \frac{M\cnkminus}{\Khalf (p+1)^{\half}} \given[\bigg] \Lsc_k^- \right\} \;\leq \; 2\;\mbox{exp} \left\{-\;\frac{M^2\cnkminus^2}{2 (p+1) \K C^2 \Dhat_k^2}\right\}\label{b_eqn7}\\
\nonumber &\mbox{a.s.} \; [\P_{\Lsc_k^-}] \; \forall \; n; \;\mbox{for each} \; k \in \{1,...,\K\} \; \mbox{and} \; \forall\; 1 \leq l \leq (p+1).
\end{align}
Now, since $\Dhat_k = O_p(\cnkminus)$, $(\cnkminus/\Dhat_k) \geq  0$ is stochastically bounded away from $0$. Thus, $\forall \; k$, and for any given $\epsilon > 0$, $\exists \; \delta(k,\epsilon) > 0$ (independent of $n$) such that: $\P_{\Lsc_k^-}\{(\cnkminus/\Dhat_k) \leq \delta(k,\epsilon)\} \leq \epsilon^* \; \forall \; n$, 
where $\epsilon^* = \epsilon/\{4\K(p+1)\} > 0$. Let $\deltatil(\K,\epsilon) = \mbox{min}\{\delta(k,\epsilon): k = 1,...,\K\} > 0$ (as $\K$ is fixed). Let $\A(k,\epsilon)$ denote the event: $\{(\cnkminus/\Dhat_k) \leq \deltatil(\K,\epsilon)\}$, and let $\A^c(k,\epsilon)$ be its complement. Then, $\P_{\Lsc_k^-}\left\{\A(k,\epsilon)\right\} \leq \epsilon^*$, while on $\A^c(k,\epsilon), (\cnkminus/\Dhat_k) >  \deltatil(\K,\epsilon)$. Thus, the bound in (\ref{b_eqn7}) is dominated by: $2\;\mbox{exp}[-M^2\deltatil^2(\K,\epsilon)/\{2 (p+1) \K C^2\}]$ on $\A^c(k,\epsilon)$, and trivially by $2$ on $\A(k,\epsilon)$ $\forall \; k$. Plugging the bound of (\ref{b_eqn7}) into (\ref{b_eqn6}) and using all these facts, we then have:
\begin{eqnarray}
  \nonumber &&\P\left(\|\G_{n,\K}\| > M\cnkminus\right) \; \leq \; \sum_{k=1}^\K \sum_{l=1}^{p+1}
  \E_{\Lsc_k^-} \left[2\;\mbox{exp} \left\{-\;\frac{M^2\cnkminus^2}{2 (p+1) \K C^2 \Dhat_k^2}\right\} \right] \\
  \nonumber &&= \; \sum_{k=1}^\K \sum_{l=1}^{p+1}
  \E_{\Lsc_k^-} \left[2\;\mbox{exp} \left\{-\;\frac{M^2\cnkminus^2}{2 (p+1) \K C^2 \Dhat_k^2}\right\}\left\{1_{\A^c(k,\epsilon)} + 1_{\A(k,\epsilon)}\right\}\right]\\
 \nonumber &&\leq \; \sum_{k=1}^\K \sum_{l=1}^{p+1} \left[
  2\;\mbox{exp}\left\{-\; \frac{M^2\deltatil^2(\K,\epsilon)}{2 (p+1) \K C^2}\right\}\P_{\Lsc_k^-}\left\{\A^c(k,\epsilon)\right\} +
  2\;\P_{\Lsc_k^-}\left\{\A(k,\epsilon)\right\}\right]\\
  \nonumber &&\leq \; 2\K(p+1)\left[\mbox{exp}\left\{-\; \frac{M^2\deltatil^2(\K,\epsilon)}{2 (p+1) \K C^2}\right\} +
  \epsilon^*\right] \\
  && \quad \;\; \leq \; \frac{\epsilon}{2} + \frac{\epsilon}{2} = \epsilon \;\; (\mbox{with some suitable choice} \; M_\epsilon \; \mbox{for} \; M),\label{b_eqn8}
\end{eqnarray} where the last step follows from noting the definition of $\epsilon^*$ and choosing $M_\epsilon$ to be any $M$ large enough such that $4\;\mbox{exp}[-\; M^2\deltatil^2(\K,\epsilon)/\{2 (p+1) \K C^2\}] \leq \epsilon/\{\K(p+1)\}$. Thus, (\ref{b_eqn8}) shows $\G_{n,\K} = O_p(\cnkminus)$ for any fixed $\K \geq 2$. This further establishes (\ref{snp_k2_exp}) and all its associated implications. The proof of Theorem \ref{thm2} is now complete. \qed
\section*{Acknowledgements}\label{acknowledge}
The authors would like to thank Dr. James Robins and Dr. Eric Tchetgen Tchetgen for many helpful discussions throughout the progress of this paper, as well as the editor Dr. Edward George, the anonymous associate editor and the two referees for their useful comments and suggestions that helped significantly in improving and revising the original version of this article. 





\begin{supplement}\label{supp_mat}
\stitle{Supplement to ``Efficient and Adaptive Linear Regression in Semi-Supervised Settings"}%
\slink[doi]{COMPLETED BY THE TYPESETTER}
\sdatatype{.pdf}
\sdescription{The supplement includes: (i) Supplementary results for the simulation studies and the real data analysis; (ii) Discussions on generalization of the proposed SS estimators to MAR settings; (iii) Proof of Lemma \ref{lemma0}; (iv) Proof of Theorem \ref{thm1}; (v) Proof of Theorem \ref{thm3}; and (vi) Proofs of Lemmas \ref{lemma1}-\ref{lemma2} and Theorem \ref{thm4}.}
\end{supplement}

\end{appendix} 


\clearpage
\newpage

\begin{center}
{\bf \uppercase{Supplement to ``Efficient and Adaptive Linear Regression in Semi-Supervised Settings''}}
\end{center}
\vspace{0.02in}

\begin{center}
{\uppercase{By Abhishek Chakrabortty and Tianxi Cai}} \par\smallskip
{\em University of Pennsylvania and Harvard University}
\end{center}
\vspace{0.01in}


\renewcommand{\thesection}{\Roman{section}}
\setcounter{section}{0}

\pagenumbering{roman}
\setcounter{page}{1}


\renewcommand{\theequation}{\arabic{equation}}
\setcounter{equation}{0}    


\renewcommand{\thefigure}{\Roman{figure}}
\setcounter{figure}{0}


\renewcommand{\thetable}{\Roman{table}}
\setcounter{table}{0}

\noindent This supplementary document contains additional numerical results as well as technical materials, including proofs etc., that could not be accommodated in the main article. 

\section*{I. Numerical Studies: Supplementary Results}\label{supp_num_results}   
\subsection*{\emph{I.1.} Simulation Results for p = 2}\label{sim_p2}             
\phantomsection
\addcontentsline{toc}{subsection}{Simulation Results for p = 2}
For $p = 2$, we investigated three choices of $m(\bx)$ as follows: 
\begin{itemize}
\item[] (Linear): $m(\bx) = x_1 + x_2$;
\item[] ($\mbox{NL-I}_{\lambda^{(k)}}$):  $m(\bx)= x_1 + x_2 + \lambda^{(k)} x_{1}x_{2}$ for $\lambda^{(1)} = 0.5$ and $\lambda^{(2)} = 1$; and
\item[] $(\mbox{NL-Q}_{\gamma^{(k)}})$: $m(\bx) = x_1 + x_2 + \gamma^{(k)} (x_1^2 + x_2^2)$ for $ \gamma^{(1)} = 0.3$ and $\gamma^{(2)} = 1$.
\end{itemize}
Since the dimension is low, we implemented EASE using the KS and KM smoothers with $\bP_2 = I_2$ for both, i.e. without any dimension reduction. For comparison, the other SS estimators were also obtained. In Table \ref{tab_p2}, we summarize the efficiencies of all the estimators relative to OLS, based on the empirical mean squared error (MSE), where for any estimator $\widetilde{\btheta}$, the empirical MSE is summarized as $\|\widetilde{\btheta}- \btheta_0\|^2$ averaged over the 500 replications.
\begin{table}[H]
\centering
\resizebox{\textwidth}{!}{
\begin{tabular}{l|c||rr||rr||ccc}\hline\hline
\multicolumn{1}{c|}{}  &\multicolumn{1}{c||}{OLS} &\multicolumn{1}{c}{SNP} &\multicolumn{1}{c||}{EASE} &\multicolumn{1}{c}{SNP} &\multicolumn{1}{c||}{EASE} &\multicolumn{3}{c}{Other SS Estimators} \\
\multicolumn{1}{l|}{\raisebox{1ex}[0cm][0cm]{Models}} &\multicolumn{1}{c||}{(Ref.)} &\multicolumn{2}{c||}{$(\Tsc := \mbox{KS})$} &\multicolumn{2}{c||}{$(\Tsc := \mbox{KM})$}
&\multicolumn{1}{c}{$\mbox{DRESS}_1$} &\multicolumn{1}{c}{$\mbox{DRESS}_3$} &\multicolumn{1}{c}{$\mbox{MSSL}$} \\
\cline{1-9}
$\mbox{Linear}$                    &1  &0.897 &0.995  &0.920 &0.988  &0.993 &0.963 &0.993 \\
$\mbox{NL-I}_{\lambda^{(1)}}$   &1  &1.229 &1.243  &1.338 &1.355  &1.072 &1.039 &1.072 \\
$\mbox{NL-I}_{\lambda^{(2)}}$   &1  &2.261 &2.261  &2.301 &2.267  &1.217 &1.181 &1.216 \\
$\mbox{NL-Q}_{\gamma^{(1)}}$   &1  &2.241 &2.215  &2.500 &2.550  &1.187 &2.063 &1.187 \\
$\mbox{NL-Q}_{\gamma^{(2)}}$   &1  &4.096 &4.144  &4.612 &4.641  &1.352 &3.217 &1.352 \\
\hline
\end{tabular}
}
\caption{Efficiencies of SNP and EASE, obtained using $\Tsc :=$ KS or KM, as well as $\mbox{DRESS}_1$, $\mbox{DRESS}_3$ and MSSL, relative to OLS with respect to the empirical MSE under the various models considered with $p=2$.}\label{tab_p2}
\end{table}
For this setting, all estimators have comparable efficiency under the linear model, as expected. Under the non-linear models, the EASE estimators are substantially more efficient than the OLS and also more efficient than the other SS estimators. For the non-linear models with quadratic effects, the $\mbox{DRESS}_3$ is also substantially more efficient than the OLS while our EASE estimator performs even better. For the non-linear models with interaction effects, the efficiency gain was very modest when employing existing SS estimation procedures while it was quite substantial for EASE.

\subsection*{\emph{I.2.} Supplementary Results for the Data Example}\label{supp_data_ex}
\phantomsection
\addcontentsline{toc}{subsection}{Supplementary Results for the Data Example}
We present in Table \ref{tab_covcomp_data_ex} some summary measures of the distributions in the labeled and unlabeled data for each of the predictors in the data example, and also report p-values for diagnostic tests aimed at detecting any possible differences in the labeled and unlabeled data distributions for each of the predictors.
\begin{table}[H]
\centering
\resizebox{\textwidth}{!}{
\begin{tabular}{l|cc|cc||ccc}\hline\hline
\multicolumn{1}{c|}{} &\multicolumn{2}{c|}{Labeled Data} &\multicolumn{2}{c||}{Unlabeled Data} &\multicolumn{3}{c}{P-values from Diagnostic Tests}\\
\multicolumn{1}{c|}{\raisebox{1ex}[0cm][0cm]{Predictors}} &\multicolumn{1}{r}{Mean}&\multicolumn{1}{c|}{Sd} &\multicolumn{1}{r}{Mean}&\multicolumn{1}{c||}{Sd}  &\multicolumn{1}{c}{T-test} &\multicolumn{1}{c}{Wilcoxon Test} &\multicolumn{1}{c}{PS Model}\\
\cline{1-8}
Age                           & 4.090 & 0.241   & 4.070 & 0.272   & 0.151 & 0.373 & 0.475 \\
Gender                        & 0.786 & 0.411   & 0.799 & 0.401   & 0.556 & 0.547 & 0.574 \\
Race                          & 0.696 & 0.461   & 0.673 & 0.469   & 0.371 & 0.378 & 0.456 \\
Lupus                         & 0.230 & 0.520   & 0.251 & 0.600   & 0.461 & 0.877 & 0.689 \\
PmR                           & 0.057 & 0.336   & 0.078 & 0.382   & 0.269 & 0.326 & 0.255 \\
RA                            & 4.171 & 1.071   & 4.084 & 1.079   & 0.144 & 0.055 & 0.255 \\
SpA                           & 0.073 & 0.343   & 0.066 & 0.313   & 0.716 & 0.932 & 0.780 \\
Other ADs                     & 0.251 & 0.642   & 0.271 & 0.690   & 0.582 & 0.780 & 0.540 \\
Erosion                       & 0.577 & 0.495   & 0.567 & 0.494   & 0.709 & 0.701 & 0.743 \\
Seropositivity                & 0.369 & 0.483   & 0.395 & 0.489   & 0.331 & 0.335 & 0.231 \\
$\mbox{Anti-CCP}_{prior}$     & 0.386 & 0.629   & 0.405 & 0.718   & 0.592 & 0.471 & 0.564 \\
$\mbox{Anti-CCP}_{miss}$      & 0.645 & 0.479   & 0.610 & 0.488   & 0.192 & 0.198 & 0.089 \\
RF                            & 0.949 & 0.772   & 0.897 & 0.846   & 0.227 & 0.184 & 0.199 \\
$\mbox{RF}_{miss}$            & 0.307 & 0.462   & 0.324 & 0.468   & 0.516 & 0.520 & 0.410 \\
Azathioprine                  & 0.121 & 0.397   & 0.137 & 0.419   & 0.463 & 0.449 & 0.734 \\
Enbrel                        & 0.738 & 0.858   & 0.722 & 0.822   & 0.740 & 0.921 & 0.574 \\
Gold salts                    & 0.346 & 0.568   & 0.336 & 0.555   & 0.729 & 0.791 & 0.962 \\
Humira                        & 0.856 & 0.833   & 0.917 & 0.835   & 0.193 & 0.190 & 0.177 \\
Infliximab                    & 0.386 & 0.673   & 0.400 & 0.672   & 0.711 & 0.607 & 0.789 \\
Leflunomide                   & 0.549 & 0.740   & 0.555 & 0.743   & 0.895 & 0.912 & 0.973 \\
Methotrexate                  & 1.417 & 0.638   & 1.389 & 0.669   & 0.435 & 0.584 & 0.549 \\
Plaquenil                     & 0.248 & 0.464   & 0.273 & 0.496   & 0.331 & 0.463 & 0.398 \\
Sulfasalazine                 & 0.535 & 0.752   & 0.554 & 0.734   & 0.661 & 0.458 & 0.724 \\
Other meds.             & 0.163 & 0.380   & 0.189 & 0.403   & 0.223 & 0.192 & 0.322 \\
(Intercept)                   & --    & --      & --    & --      & --    & --    & 0.000 \\ \hline
\end{tabular}
}
\caption{Comparison of the means and standard deviations (sd) from the labeled and unlabeled data for each predictor in the data example. Shown also are the p-values obtained from various diagnostic tests, testing for possible differences in the distributions of each of the predictors in the labeled and unlabeled data, including a two-sample T-test (with possibly unequal variances in the two populations), a Wilcoxon rank sum test, and a test obtained by fitting a parametric logistic regression model for the propensity score (PS) of missingness, with all the predictors included as covariates, and then testing for the null effect of each of the predictors in the fitted model.}\label{tab_covcomp_data_ex}
\end{table}

\subsection*{\emph{I.3.} Simulation Results on the Prediction Error}
\phantomsection
\addcontentsline{toc}{subsection}{Simulation Results on the Prediction Error}
In Table \ref{tab_oos-pe-np}, we present the out-of-sample mean squared prediction error for various SNP imputation estimators under all the models considered in the simulation studies for $p = 2$, $10$ and $20$. The results suggest that the SNP imputation estimators for both KS and KM based smoothers perform substantially better than the naive non-parametric estimator based on a $p$-dimensional kernel smoothing.
\begin{table}[H]
\centering
\centerline{(a) $p=2$}
\vspace{0.1in}
\resizebox{0.43\textwidth}{!}{
\begin{tabular}{l|rr||rr}\hline\hline
&\multicolumn{2}{c||}{$\Tsc := \mbox{KS}$} &\multicolumn{2}{c}{$\Tsc := \mbox{KM}$}\\
\multicolumn{1}{l|}{\raisebox{1ex}[0cm][0cm]{Models}} &$\mhat_{\Tsc}$ &$\muhat_{\Tsc}$ &$\mhat_{\Tsc}$ &$\muhat_{\Tsc}$ \\ \hline
$\mbox{Linear}$                    &0.36 &0.35  &0.35 &0.35 \\
$\mbox{NL-I}_{\lambda^{(1)}}$   &0.34 &0.33  &0.32 &0.32 \\
$\mbox{NL-I}_{\lambda^{(2)}}$    &0.29 &0.28  &0.27 &0.27 \\
$\mbox{NL-Q}_{\gamma^{(1)}}$   &0.29 &0.29  &0.27 &0.27 \\
$\mbox{NL-Q}_{\gamma^{(2)}}$    &0.20 &0.20  &0.17 &0.18 \\ \hline
\end{tabular}
}\vspace{.2in}

\centerline{(b) $p=10$ and $20$}
\vspace{0.1in}
\resizebox{\textwidth}{!}{
\begin{tabular}{cc|rr|rr|r||rr|rr|r}\hline\hline
&\multicolumn{1}{c|}{} &\multicolumn{5}{c||}{$p=10$} &\multicolumn{5}{c}{$p=20$}\\ \cline{3-12}
&\multicolumn{1}{c|}{} &\multicolumn{2}{c|}{$\Tsc := \mbox{KS}_{2, \bP_2}$} &\multicolumn{2}{c|}{$\Tsc := \mbox{KM}$} & &\multicolumn{2}{c|}{$\Tsc := \mbox{KS}_{2, \bP_2}$} &\multicolumn{2}{c|}{$\Tsc := \mbox{KM}$} &\\
&\raisebox{1ex}[0cm][0cm]{Models}     &$\mhat_{\Tsc}$ &$\muhat_{\Tsc}$ &$\mhat_{\Tsc}$ &$\muhat_{\Tsc}$
& \raisebox{1ex}{$\mbox{KS}_{p}$}  &$\mhat_{\Tsc}$ &$\muhat_{\Tsc}$&$\mhat_{\Tsc}$ &$\muhat_{\Tsc}$ &  \raisebox{1ex}{$\mbox{KS}_{p}$}  \\ \hline
& Linear    &0.186  &0.174   &0.214  &0.204   &0.260    &0.143  &0.099   &0.130  &0.119   &0.698 \\
(I)&  NL1C      &0.143  &0.131   &0.147  &0.154   &0.290    &0.337  &0.332   &0.280  &0.317   &0.921 \\
&  NL2C      &0.241  &0.228   &0.150  &0.156   &0.554    &0.557  &0.566   &0.271  &0.299   &1.000 \\
&  NL3C      &0.275  &0.254   &0.138  &0.146   &0.458    &0.543  &0.552   &0.282  &0.327   &0.974 \\ \hline
& Linear    &0.106  &0.096   &0.133  &0.126   &0.313    &0.095  &0.054   &0.083  &0.073   &0.755 \\
(II) & NL1C      &0.147  &0.131   &0.135  &0.143   &0.658    &0.341  &0.336   &0.283  &0.331   &1.095 \\
& NL2C      &0.179  &0.163   &0.135  &0.142   &0.487    &0.423  &0.422   &0.272  &0.302   &0.980 \\
& NL3C      &0.218  &0.199   &0.137  &0.144   &0.527    &0.424  &0.424   &0.282  &0.318   &0.996 \\ \hline
 \end{tabular}
}
\caption{Mean squared prediction errors (PEs), relative to $\Var(Y)$, 
of the SNP imputation estimators $\mhat_{\Tsc} = \mhat(\bx; \bPhat_r)$ and $\muhat_{\Tsc} = \muhat(\bx; \Pschat_{r,\K})$, with $\Tsc := \mbox{KS}_{2,\bP_2}$ or $\Tsc := \mbox{KM}$, under the various models discussed in the simulation studies. Shown also are the corresponding PEs for the fully non-parametric KS estimator, $\mbox{KS}_p$, for comparison, in the case of $ p = 10$ and $20$.} \label{tab_oos-pe-np}
\end{table}

\section*{II. Generalization to the Missing at Random (MAR) Case}\label{sec_MAR}

Our SS estimation methods proposed so far assume that the underlying $Y$ for subjects in $\Usc$ are MCAR, a standard (and often implicit) assumption made in SSL. In this section, we provide some discussions on possible generalizations of our SS methods to the MAR case. Such generalizations might be desirable for settings where the availability of $Y$ is not determined by design. To this end, let $\Nbar=N+n$ denote the sample size of the entire data $\Sbb = \Lsc \cup \Usc$. Then, $\Sbb = \{\bZ_i \equiv (R_i, R_iY_i, \bX_i): i = 1, \hdots, \Nbar\}$ consists of $\Nbar$ i.i.d. realizations of $\bZ = (R, RY, \bX)$, where $R \in \{0,1\}$ denotes the indicator of $Y$ being observed. As opposed to the stronger MCAR setting with the assumption $R \ind (Y, \bX)$ and the probability law of $\Sbb$ being determined by the law of $(Y,\bX)$, we now have: under the MAR setting, $R \ind Y \given \bX$ and the probability law of $\Sbb$ is determined by $\P_{\bZ}$, the law of $\bZ$. For notational ease, we also let $\P_{\Nbar}$ denote the empirical
measure for $\Sbb$, and for any function $\be(\cdot)$ of $\bZ$, possibly random and vector-valued, we let $\PNbar(\be) = \Nbarinv\sumiNbar \be(\bZ_i)$, and $\PZ(\be) = \E_{\bZ}\{\be(\bZ)\} = \int \be(\bz) d\P_{\bZ}(\bz)$.

Under a SS set-up as above, we have: $n = \sumiNbar R_i$ is a random quantity and $n/\Nbar \to 0$ in probability. It is important to note that $\pi_{\Nbar} \equiv \P(R =1)$ \emph{must} depend on $\Nbar$. Let $\pi_{\Nbar}(\bX) = \P(R = 1 \mid \bX)$ be the ``propensity score", assumed to be strictly greater than 0 almost surely (a.s.) for any given $\Nbar$ and let $b_\Nbar = [\E\{\piN^{-2}(\bX)\}]^{-\half}$. Then, under the above set-up, we assume that $\E\{\pi_{\Nbar}(\bX) \} = \pi_{\Nbar} \to 0$, $b_{\Nbar} \to 0$ and $\Nbar b_\Nbar \to \infty$ as $\Nbar \to \infty$. This decaying sampling probability is the main factor that distinguishes SSL from standard missing data problems and contributes to the complexity of devising and analyzing 
SS estimators which, even for the MCAR setting, were seen to only have a convergence rate of $(\Nbar b_{\Nbar})^{-\half}$, rather than $\Nbarnhalf$, with $b_{\Nbar} = (n/\Nbar)$. For simplicity, we shall first assume that $\pi_{\Nbar}(\bX)$ is known and next detail how we may extend our proposed procedures to obtain SS estimators of $\btheta_0$, the solution to: $\bphi(\btheta) \equiv \E\{\bXv(Y - \bXv'\btheta)\} = \bzero$, under the MAR setting.

To derive efficient SS estimators 
of $\btheta_0$ based on $\Sbb$ under MAR, we first note
that our proposed SNP estimator $\thetahatprk$ in Section \ref{snp_est} in fact remains valid even under the MAR setting, whenever the imputation is sufficient i.e. $\mu(\bX;\bP_r)$ equals the true conditional mean $m(\bX) \equiv \E(Y \medgiven \bX)$. 
For the general case allowing for insufficient imputation, we need to modify our SNP imputation to account for the MAR setting. To this end, we note that
\begin{align}
\bphi(\btheta) & \equiv \E\{\bXv(Y - \bXv'\btheta)\} = \E\left\{\frac{R}{\pi_\Nbar(\bX)}\bXv(Y - \bXv'\btheta)\right\}, \label{MAR_eqn1}\\
& = \E[\bXv\{m(\bX) - \bXv'\btheta\}], \;\; \label{MAR_eqn2} \\
& = \E[\bXv\{m(\bX) - \bXv'\btheta\}] + \E\left[\frac{R}{\pi_\Nbar(\bX)}\bXv\{Y - m(\bX)\}\right]. \label{MAR_eqn3}
\end{align}
More generally, for any $\mu(\cdot) \in \Lsc_2(\P_{\bX})$ satisfying:
\begin{equation}
\E\left[\frac{\RN}{\pi_\Nbar(\bX)} \bXv\{Y - \mu(\bX)\} \right] \; \equiv \; \E [\bXv \{m(\bX)-\mu(\bX)\}] \; = \; \bzero, \label{eq-mu-MAR}
\end{equation}
it is easy to see that the following representation of $\bphi(\cdot)$ holds under MAR: 
\begin{equation}
\bphi(\btheta) = \E[\bXv\{\mu(\bX) - \bXv'\btheta\}] + \E\left[\frac{R}{\pi_\Nbar(\bX)} \bXv\{Y - \mu(\bX)\}\right]. \label{MAR_eqn4}
\end{equation}

Let $\muhat(\cdot)$ be any estimator of $\mu(\cdot)$ based on $\Sbb$.  Motivated by (\ref{MAR_eqn4}), we may then devise a SS estimator of $\btheta_0$, $\bthetahatmar \equiv \bthetahat_{\mbox{\tiny{MAR}},\;\mu(\cdot)}$, as the solution  to:
\begin{eqnarray}
&& \;\widehat{\bphi}_\Nbar(\btheta) \equiv \Nbarinv\sumiNbar \left[\bXv_i\{\muhat(\bX_i) - \bXv_i'\btheta\} +  \frac{R_i}{\piN(\bX_i)} \bXv_i \{Y_i - \muhat(\bX_i)\} \right]. \label{MAR_eqn5}
\end{eqnarray}
Then, letting $\bGamma_\Nbar = \Nbarinv \sumiNbar \bXv_i \bXv_i'$, it is straightforward to show that:
\begin{equation}
\bGamma_\Nbar \left( \bthetahatmar - \btheta_0 \right) =  \PNbar(\bT) + \PNbar(\bS) - \PNbar(\behat), \; \mbox{where} \label{MAR_eqn6}
\end{equation}
where $\bT(\bZ) = \bXv\{\mu(\bX) - \bXv'\btheta_0\}$, $\bS(\bZ) = \{R/\piN(\bX)\} \bXv \{Y - \mu(\bX)\}$, and
\begin{align*}
\widehat{\be}(\bZ) \equiv \left\{\frac{\RN}{\piN(\bX)} - 1 \right\}\bXv \{\muhat(\bX) -\mu(\bX)\}.
 \end{align*}

Convergence rates of the terms in (\ref{MAR_eqn6}) need careful analysis as the asymptotics here is \emph{non-standard}, with the dominating rate being slower than $N^{-\half}$. To this end, note that $\PNbar(\bT)$ is a simple centered i.i.d. average of variables with bounded variance. Hence, $\T_\Nbar = O(N^{-\half})$ indeed. On the other hand, $\PNbar(\bS)$ has a slower convergence rate since the variance of $\bS$, $\bV_\Nbar$, diverges due to the $\piN(\cdot) \downarrow 0$ appearing in the denominator. Under mild moment conditions, it can be shown that $b_\Nbar \bV_\Nbar$ converges to a positive definite matrix $\bV$ with $\|\bV\| < \infty$. Hence, using concentration inequalities, and assuming $\Nbar b_\Nbar\rightarrow \infty$, it can be shown that the convergence rate of $\PNbar(\bS)$ is $O\{(N b_\Nbar)^{-\half}\}$. Further, using CLT for triangular arrays, it can be shown under suitable conditions that $(N b_\Nbar)^{\half}\PNbar(\bS) \stackrel{d}{\rightarrow} \Nsc_{(p+1)}[\bzero,\bV]$. Lastly, to control the term $\PNbar(\behat)$, note that $\P_{\bZ}(\behat) = \bzero$. Therefore, $\G_\Nbar$ is a \emph{centered} empirical process indexed by $\muhat(\cdot) - \mu(\cdot)$. Hence, as long as $\E_{\bX}[\{\muhat(\bX) - \mu(\bX)\}^2] \stackrel{P}{\rightarrow} 0$, and $\muhat(\cdot) - \mu(\cdot)$ lies in a $\P-$Donsker class with probability $\rightarrow 1$, it can be shown using results from empirical process theory (see \citet{VdVaart_2000} for instance) that $(N b_\Nbar)^{\half}\PNbar(\behat)= o_p(1)$. Finally, note that $\bGamma_\Nbar \succ 0$ a.s., and $\bGammainv_\Nbar = \bGamma^{-1} + O_p(N^{-\half})$. Hence, under suitable regularity conditions, we have:
\begin{eqnarray*}
  (\Nbar b_\Nbar)^{\half} (\bthetahatmar - \btheta_0) &=& (\Nbar b_\Nbar)^{\half} \bGammainv \frac{1}{\Nbar} \sumiNbar \bS(\bZ_i) + o_p(1)\\
  &\stackrel{d}{\rightarrow}&  \Nsc_{(p+1)}[\bzero, \bGammainv \bV \bGammainv], \;\; \mbox{with} \; \bV \; \mbox{as defined above}.
\end{eqnarray*}

Having now provided an abstract sketch of the construction of the estimators and their properties, we next briefly discuss the \emph{choice} of the `imputation' function $\mu(\cdot)$, and its estimator $\muhat(\cdot)$ inherent in the construction of $\bthetahatmar \equiv \bthetahat_{\mbox{\tiny{MAR}},\;\mu(\cdot)}$. With $(r,\bP_r, \Phat_r)$ as defined in Section \ref{snp_est} and $\{K(\cdot), h, K_h(\cdot,\cdot)\}$ as in Section \ref{kern_smth_impl}, we may modify the SNP estimator in Section \ref{snp_est} under the MAR setting as follows: consider $\mu(\bX) \equiv \mu(\bX;\bP_r)  = m(\bX;\bP_r) + \bXv'\etapr$, where $m(\bX;\bP_r) = \E(Y \medgiven \bP_r'\bX)$, and $\etapr$ satisfies
$$
\E\left[\frac{\RN}{\piN(\bX)}\bXv\{Y - m(\bX;\bP_r) - \bXv'\etapr\}\right] = \bzero.
$$
This will ensure that (\ref{eq-mu-MAR}) holds. Then, we may estimate $\mu(\bX;\bP_r)$ as $\muhat(\bX;\Phat_r)$ $= \mhat(\bX;\Phat_r) + \bXv'\widehat{\boldsymbol{\eta}}_{\bP_r}$, where
\begin{align*}
&\mhat(\bx;\Phat_r) \;  = \; \frac{ \sumiNbar \frac{R_i}{\piN(\bX_i)} Y_i K_h(\Phat_r'\bX_i,\Phat_r'\bx)}{\sumiNbar \frac{R_i}{\piN(\bX_i)} K_h(\Phat_r'\bX_i,\Phat_r'\bx)}, \;\; \mbox{and}&\\
& \widehat{\boldsymbol{\eta}}_{\bP_r} \; \mbox{satisfies:} \;\; \Nbarinv\sumiNbar \frac{R_i}{\piN(\bX_i)}\bXv_i\{Y_i - \mhat(\bX_i;\Phat_r) - \bXv_i'\widehat{\boldsymbol{\eta}}_{\bP_r}\} = \bzero.&
\end{align*}
Thus, to accommodate the MAR setting, one essentially needs to implement appropriately \emph{weighted} versions of both the smoothing and the refitting steps in our original SNP imputation. Of course, while we have chosen the smoothing method $\Tsc$ to be the weighted KS here for illustration, other reasonable choices of $\Tsc$ such as an appropriately weighted KM may also be used. Under MCAR, with $\pi_{\Nbar}(\bX) \equiv \pi_{\Nbar} \equiv n/\Nbar$ and $b_{\Nbar} = (n/\Nbar)$, the estimator $\bthetahatmar$ indeed becomes (asymptotically) equivalent to the SNP estimators obtained earlier in Section \ref{snp_est}. Further, with various choices of $\mu(\cdot)$, the SNP imputation strategy again equips us with a \emph{family} of SS estimators of $\btheta_0$ under the MAR setting, with $\mu(\cdot) = m(\cdot)$ leading to the optimal estimator.

The above estimator $\bthetahatmar$ is derived with a known $\piN(\cdot)$, for simplicity. In practice, $\piN(\cdot)$ is typically unknown and a consistent estimator $\widehat{\pi}_\Nbar(\cdot)$ may be constructed. Then, one may modify $\bthetahatmar$ by replacing $\piN(\cdot)$ with $\widehat{\pi}_\Nbar(\cdot)$ in all the steps. The resulting estimator will have an expansion similar to (\ref{MAR_eqn6}) but with extra error terms accounting for the variability in $\widehat{\pi}_\Nbar(\cdot)$, which need to be properly controlled. The theoretical analysis will be more involved since establishing the convergence rates and asymptotic expansion for $\widehat{\pi}_\Nbar(\cdot) - \pi_\Nbar(\cdot)$ is also non-standard due to $b_\Nbar \to 0$. Lastly, the score equation (\ref{MAR_eqn4}) used to construct $\bthetahatmar$ has the additional benefit of \emph{`double robustness'}, in the sense that \emph{even if} $\widehat{\pi}_\Nbar(\cdot)$ is inconsistent for $\pi_\Nbar(\cdot)$, as long as $\muhat(\cdot)$ estimates the true $m(\cdot)$, $\bthetahat_{\mbox{\tiny{MAR}},\;\mu(\cdot)}$ is consistent for $\btheta_0$. On the other hand, as long as $\widehat{\pi}_\Nbar(\cdot)$ targets the true $\pi_\Nbar(\cdot)$, then for \emph{any} choice of $\mu(\cdot)$,  $\bthetahat_{\mbox{\tiny{MAR}},\;\mu(\cdot)}$ is consistent for $\btheta_0$. For the MCAR case, $\widehat{\pi}_\Nbar(\cdot) \equiv \widehat{\pi}_\Nbar = n/\Nbar$ is always consistent for $\pi_\Nbar(\cdot)$ and in fact this is exactly what allowed us to achieve a \emph{family} of SNP estimators, all consistent for $\btheta_0$, for various choices of the SNP imputation function $\mu(\cdot)$. For a specific choice of $\mu(\cdot)$, this also included the supervised estimator $\bthetahat$. Note however that under MAR, $\bthetahat$ simply denotes the naive `complete case' estimator and in general, is \emph{not} consistent for $\btheta_0$ unless $\pi_{\Nbar}(\cdot)$ is trivially a constant or $m(\cdot)$ is exactly linear. 


\section*{III. Proof of Lemma \ref{lemma0}}\label{app0} 

Firstly, since $d$ is fixed, it suffices to prove the result for any arbitrary scalar coordinate $\bGhat_{n,m}^{(j)} \equiv \Gschat_{n,m}$ (say) and $\bGbar_{n}^{(j)} \equiv \Gscbar_n$ (say) of $\bGhat_{n,m}$ and $\bGbar_n$ respectively, for any $j \in \{1,\hdots,d\}$. For any data $\S$ and $\S^*$, we let $\P_{\S}$ and $\P_{\S,\S^*}$ denote the joint probability distributions of the observations in $\S$ and $(\S,\S^*)$ respectively, $\E_{\S}(\cdot)$ denote the expectation w.r.t $\P_{\S}$, and $\P_{\S \smallgiven \S^*}$ denote the conditional probability distribution of the observations in $\S$ given $\S^*$.

To show that $\Gschat_{n,m} - \Gscbar_n = O_p(m^{-\frac{1}{2}})$, we first note that since $\S_n \ind \S_m$,
\begin{align*}
\P_{\S_n,\S_m}\left(|\Gschat_{n,m} - \Gscbar_n| > m^{-\frac{1}{2}} t\right) 
 \;=\; \E_{\S_n}\left\{\P_{\S_m}\left(|\Gschat_{n,m} - \Gscbar_n| > m^{-\frac{1}{2}} t \given[\big] \S_n\right)\right\},
\end{align*}
for any $t > 0$. Now, conditional on $\S_n$, $\bGhat_{n,m} - \bGbar_n$ is a centered average of $\{\bghat_n(\bZ_j)\}_{j=1}^m$ which are i.i.d. and bounded by $\That_n < \infty$ a.s. $[\P_{\S_n}]$ $\forall \; n$. Hence, applying Hoeffding's inequality, we have for any $n$ and $m$,
\begin{equation}\label{lemma0_eqn2}
\P_{\S_m}\left(|\Gschat_{n,m} - \Gscbar_n| > m^{-\frac{1}{2}} t \given[\big] \S_n\right) \;\leq \; 2\;\mbox{exp} \left(-\;\frac{2m^2t^2}{4m^2\That_n^2}\right)
\hspace{3mm} \mbox{a.s.} \; [\P_{\S_n}].
\end{equation}
Now, since $\That_n \geq 0$ is $O_p(1)$, we have: for any given $\epsilon > 0$, $\exists$ $\delta(\epsilon) > 0$ such that: $\P_{\S_n}\{\That_n > \delta(\epsilon)\} \leq \epsilon/4$ $\forall$ $n$. Let $\A(\epsilon)$ denote the event: $\{\That_n > \delta(\epsilon)\}$ and let $\A^c(\epsilon)$ denote its complement. Then, using (\ref{lemma0_eqn2}), we have: $\forall \; n$ and $m$,
\begin{eqnarray*}
  \nonumber && \P_{\S_n,\S_m}\left(|\Gschat_{n,m} - \Gscbar_n| > m^{-\frac{1}{2}} t\right) \; \leq \; \E_{\S_n}\left\{2\;\mbox{exp} \left(-\;\frac{2m^2t^2}{4m^2\That_n^2}\right)\right\}\\
  \nonumber && = \E_{\S_n}\left\{2\;\mbox{exp} \left(-\;\frac{t^2}{2\That_n^2}\right)\right\} = \E_{\S_n}\left[2\;\mbox{exp} \left(-\;\frac{t^2}{2\That_n^2}\right)\left\{1_{\A^c(\epsilon)} + 1_{\A(\epsilon)}\right\}\right] \\
  \nonumber && \leq \left[2\;\mbox{exp}\left\{-\; \frac{t^2}{2\delta^2(\epsilon)}\right\}\P_{\S_n}\left\{\A^c(\epsilon)\right\} +
  2\;\P_{\S_n}\left\{\A(\epsilon)\right\}\right] \\
  \nonumber && \leq 2\;\mbox{exp}\left\{-\; \frac{t^2}{2\delta^2(\epsilon)}\right\} +
  \frac{\epsilon}{2} \;\; \leq \frac{\epsilon}{2} + \frac{\epsilon}{2} = \epsilon \; \mbox{(for some suitable choice of} \; t), 
\end{eqnarray*} where the last step follows by choosing $t \equiv t_\epsilon$ to be any large enough $t$ such that $\mbox{exp}\{- t^2/2\delta^2(\epsilon)\}$ $\leq \epsilon/4$. Such a choice of $t_\epsilon$ clearly exists. This establishes the first claim (a) in Lemma \ref{lemma0}. The second claim (b) in Lemma \ref{lemma0} is a trivial consequence of the Central Limit Theorem (CLT). \qed

\section*{IV. Proof of Theorem \ref{thm1}}\label{pf_thm1} 

To show Theorem \ref{thm1}, we first note that under Assumption \ref{assmpn_np} (i)-(v), and letting $a_n=(\log n)^{\half}(nh^p)^{-\half}+h^q$, the following holds:
\begin{equation}\label{a_eqn1}
\text{sup}_{\bx\in\mathcal{X}}|\widehat{m}(\bx)-m(\bx)| = O_p(a_n) = \text{sup}_{\bx\in\mathcal{X}}|\widehat{f}(\bx)-f(\bx)|.
\end{equation}
(\ref{a_eqn1}) is a fairly standard result and we only provide a sketch of its proof as follows. Under Assumption \ref{assmpn_np} (ii)-(iii), using Theorem 2 of \citet{Hansen_2008},
${\text{sup}}_{\bx\in\mathcal{X}} |\widehat{l}(\bx)-\E_{\Lsc}\{\widehat{l}(\bx)\}| = O_p(a_n^*) = {\text{sup}}_{\bx\in\mathcal{X}} |\widehat{f}(\bx)-\E_{\Lsc}\{\widehat{f}(\bx)\}|$, where $a_n^*= (\log n)^{\half}(nh^p)^{-\half}$. Next, using standard arguments based on Taylor series expansions of $l(\cdot)$ and $m(\cdot)$ under their assumed smoothness, and noting that $K(\cdot)$ is a $q^{th}$ order kernel having finite $q^{th}$ moments, we obtain:
$$\text{sup}_{\bx\in\mathcal{X}}|\E_{\Lsc}\{\widehat{l}(\bx)\}-l(\bx)| =  O(h^q) = \text{sup}_{\bx\in\mathcal{X}}|\E_{\Lsc}\{\widehat{f}(\bx)\}-f(\bx)|.$$
Combining these two results, and the definitions of $m(.)$ and $\mhat(.)$ along with Assumption \ref{assmpn_np} (iv), we have (\ref{a_eqn1}). Next, note that using (\ref{thetahat_np_defn}), we have:
\begin{align*}
\bGamma_N(\bthetahat_{np}-\btheta_0)  &= \E_{\Usc}[\Ninv\sum_{j=n+1}^{n+N}\bXv_j\{\mhat(\bX_j)-\bXv_j'\btheta_0\}] + O_p(\Nnhalf) \\
 &= \E_{\bX}[\bXv\{\mhat(\bX)-m(\bX)\}] + O_p(\Nnhalf),
\end{align*}
where the first step is due to Lemma \ref{lemma0} (a) with $\text{sup}_{\bx\in\mathcal{X}}\|\bxv\{\mhat(\bx)-\bxv'\btheta_0\}\|$ $\leq \text{sup}_{\bx\in\mathcal{X}}[\|\bxv\|\{|\mhat(\bx)-m(\bx)| + |m(\bx)-\bxv'\btheta_0|\}] = O_p(1)$ due to (\ref{a_eqn1}) and the boundedness of $\bX$ and $m(\cdot)$, while the last step uses: $\E_{\bX}[\bXv\{m(\bX)-\bXv'\btheta_0\}] = \mathbf{0}$ which follows from the definitions of $\btheta_0$ and $m(\cdot)$. It then follows further, using $\bGamma_N^{-1} = \bGammainv + O_p(\Nnhalf)$, that
$$
\nhalf(\bthetahat_{np}-\btheta_0) \; = \; \nhalf\;\bGammainv\E_{\bX}[\bXv\{\mhat(\bX)-m(\bX)\}] + O_p\left(\frac{n}{N}\right)^{\half}.
$$
Letting $\phi_n(\bX) = (nh^p)^{-1}\sum_{i=1}^{n}K\{(\bX-\bX_i)/h\}\{Y_i - m(\bX)\}$, and expanding the first term in the above equation, we now obtain:
\begin{equation}
\nhalf\left(\bthetahat_{np}-\btheta_0\right) \;= \; \bGammainv\left(\bT^{(1)}_{n,1} + \bT^{(2)}_{n,1}\right) + O_p\left(\frac{n}{N}\right)^{\half},\label{a_eqn2}
\end{equation}
where $\bT^{(1)}_{n,1}= \nhalf\;\E_{\bX}\{ \bXv\phi_{n}(\bX) / f(\bX) \}$ and
\begin{align}
\bT^{(2)}_{n,1} &= \nhalf\;\E_{\bX}\left[\bXv\phi_n(\bX) \{\widehat{f}(\bX)^{-1}- f(\bX)^{-1}\} \right] \nonumber \\
&=  \nhalf\;\E_{\bX}[\bXv\left\{\mhat(\bX)-m(\bX)\right\}\{f(\bX)-\widehat{f}(\bX)\}/f(\bX)] \nonumber \\
&\leq  \nhalf\;\text{sup}_{\bx\in\mathcal{X}}
\left\{\|\bxv\| \left|\mhat(\bx)-m(\bx)\right| \left|\widehat{f}(\bx)/f(\bx)-1\right|\right\} \;= O_p\left(\nhalf a_n^2\right), \label{a_eqn3}
\end{align}
where the last step in (\ref{a_eqn3}) follows from (\ref{a_eqn1}), Assumption \ref{assmpn_np} (iv) and the boundedness of $\bX$. For $\bT^{(1)}_{n,1}$, we have:
\begin{eqnarray}
\bT^{(1)}_{n,1} &=& \nhalf \int_{\mathcal{X}}\bxv\phi_n(\bx)d\bx = \nnhalf \sum_{i=1}^n \int_{\mathcal{X}}\bxv h^{-p}K_h(\bx-\bX_i) \left\{Y_i-m(\bx)\right\}d\bx \nonumber\\
  &=& \nhalf\sum_{i=1}^n \ninv\int_{\mathcal{A}_{i,n}}\overrightarrow {(\bX_i+h\bpsi_i)} \; K\left(\bpsi_i\right) \left\{Y_i-m(\bX_i+h\bpsi_i)\right\}d\bpsi_i,\label{a_eqn4}
\end{eqnarray}
where $\bpsi_i = (\bx-\bX_i)/h$ and $\mathcal{A}_{i,n} =\{\bpsi_i \in \mathbb{R}^p: (\bX_i+h\bpsi_i) \in \mathcal{X}\}$. Now, since $K(\cdot)$ is zero outside the bounded set $\mathcal{K}$, the $i^{th}$ integral in (\ref{a_eqn4}) only runs over $\left(\mathcal{A}_{i,n}\cap\mathcal{K}\right)$. Further, since $h = o(1)$, using Assumption \ref{assmpn_np} (vi), $\mathcal{A}_{i,n} \supseteq \mathcal{K}$ a.s. $[\P_{\Lsc}]$ or, $(\mathcal{A}_{i,n} \cap \mathcal{K}) = \mathcal{K}$ a.s. $[\P_{\Lsc}]$ $\forall$ $1 \leq i \leq n$ with $n$ large enough. Thus, for large enough $n$, (\ref{a_eqn4}) can be written as:
\begin{align}
\bT^{(1)}_{n,1} &=  \nnhalf\sum_{i=1}^n \int_{\mathcal{K}}\overrightarrow {(\bX_i+h\bpsi_i)} \; K\left(\bpsi_i\right) \left\{Y_i-m(\bX_i+h\bpsi_i)\right\}d\bpsi_i  \;\; \mbox{a.s.}\; [\P_{\Lsc}]   \nonumber \\
  &=\nhalf\sum_{i=1}^n \ninv\left[\bXv_i\left\{Y_i - m(\bX_i)\right\} + O_p(h^q)\right] \label{a_eqn5}\\
  &=\nnhalf\sum_{i=1}^{n}\bXv_i\left\{Y_i-m(\bX_i)\right\} + O_p\left(\nhalf h^q\right), \label{a_eqn6}
\end{align}
where (\ref{a_eqn5}), and hence (\ref{a_eqn6}), follows from standard arguments based on Taylor series expansions of $m(\bX_i+h\bpsi_i)$ around $m(\bX_i)$ under the assumed smoothness of $m(\cdot)$, and using the fact that $K(\cdot)$ is a $q^{th}$ order kernel. Combining (\ref{a_eqn2}), (\ref{a_eqn3}) and (\ref{a_eqn6}), and noting that under our assumptions, $(\nhalf a_n^2 + \nhalf h^q)$ $= O\{\nhalf h^q + (\log n)(\nhalf h^p)^{-1}\}$, the result of Theorem \ref{thm1} now follows. \qed

\clearpage
\newpage

\section*{V. Proof of Theorem \ref{thm3}}\label{appC} 

Let $a_{n,2}=(\log n)^{\half} (nh^r)^{-\half}+h^q$. Then, we first note that
\begin{equation}\label{c_eqn1}
\text{sup}_{\bw\in\chiPr}|\vphiPrtil^{(\varrho)}(\bw)-\vphiPr^{(\varrho)}(\bw)| = O_p(a_{n,2}), \quad \forall \; \varrho \in \{0, 1\}.
\end{equation}
To see this, note that under Assumption \ref{assmpn_unifconv_snp} (ii)-(iii), Theorem 2 of \citet{Hansen_2008} applies, and we have for $d_n = (\log n)^{\half}(nh^r)^{-\half}$,
$$\text{sup}_{\bw\in\chiPr}\;|\vphiPrtil^{(\varrho)}(\bw)-\E_{\Lsc}\{\vphiPrtil^{(\varrho)}(\bw)\}| = O_p(d_n) \quad \forall \; \varrho \in \{0, 1\}.$$
Next, using standard arguments based on a $q^{th}$ order Taylor series expansion of $\vphiPr^{(\varrho)}(\cdot)$ and noting that $K(\cdot)$ is a $q^{th}$ order kernel, we obtain:
$$\text{sup}_{\bw\in\chiPr} |\E_{\Lsc}\{\vphiPrtil^{(\varrho)}(\bw)\}-\vphiPr^{(\varrho)}(\bw)| = O(h^q) \quad \forall \; \varrho \in \{0, 1\}.$$
Combining these two results gives (\ref{c_eqn1}). Further,
\begin{eqnarray}
\nonumber && \text{sup}_{\bx\in\Xsc}\;\left|\mtil(\bx;\bP_r)- m(\bx;\bP_r)\right| \;=\; \text{sup}_{\bw\in\chiPr}\;\left|\mPrtil(\bw)- \mPr(\bw)\right| \\
&& \leq \; \text{sup}_{\bw\in\chiPr}\;\left|\frac{\lPrtil(\bw)- \lPr(\bw)}{\fPrtil(\bw)}\right| +
\nonumber \text{sup}_{\bw\in\chiPr}\;\left\{\left|\frac{\left|\lPr(\bw)\right|}{\fPr(\bw)} -\frac{\left|\lPr(\bw)\right|}{\fPrtil(\bw)} \right|\right\}\\
&& \; = \; O_p(a_{n,2}),\label{c_eqn2}
\end{eqnarray}%
\noindent where the last step follows from repeated use of (\ref{c_eqn1}) and Assumption \ref{assmpn_unifconv_snp} (iii)-(iv). Next, we aim to bound $\text{sup}_{\bx \in \Xsc}|\vphihat^{(\varrho)}(\bx;\Phat_r) - \vphitil^{(\varrho)}(\bx;\bP_r)|$ to account for the potential estimation error of $\Phat_r$. Using a first order Taylor series expansion of $K(.)$ under Assumption \ref{assmpn_unifconv_snp} (vi), we have: $\forall \; \varrho \in \{0,1\}$,
\begin{align}
\nonumber \vphihat^{(\varrho)}(\bx;\Phat_r) - \vphitil^{(\varrho)}(\bx;\bP_r) \; = \;
\frac{1}{nh^r} \sum_{i=1}^n  \bnabla K' (\bwix) (\Phat_r-\bP_r)' \left(\frac{\bx-\bX_i}{h}\right)   Y_i^{\varrho} \\
 = \; \mbox{trace}\left\{(\Phat_r-\bP_r)' \bMhat^{(1)}_{n,\varrho,\bx}\right\} +
\mbox{trace}\left\{(\Phat_r-\bP_r)' \bMhat^{(2)}_{n,\varrho,\bx}\right\}, \quad \quad \;\;\label{c_eqn3}
\end{align}where
\begin{align*}
\nonumber \bMhat^{(1)}_{n,\varrho,\bx}  &= \frac{1}{nh^{r+1}}\sum_{i=1}^n (\bx-\bX_i)
\left\{\bnabla K'\left(\frac{\bP_r'\bx-\bP_r'\bX_i}{h}\right)\right\} Y_i^{\varrho} \;\; \mbox{and}\\
\nonumber \bMhat^{(2)}_{n,\varrho,\bx}  &=  \frac{1}{nh^{r+1}}\sum_{i=1}^n (\bx-\bX_i)
\left\{\bnabla K'(\bwix)-\bnabla K'\left(\frac{\bP_r'\bx-\bP_r'\bX_i}{h}\right) \right\} Y_i^{\varrho},
\end{align*}
with $\bwix \in \mathbb{R}^r$ being `intermediate' points satisfying: $\|\bwix - \bP_r'(\bx-\bX_i)h^{-1}\|$ $\leq \|\Phat_r'(\bx-\bX_i)h^{-1} - \bP_r'(\bx-\bX_i)h^{-1}\| \leq O_p(\alpha_nh^{-1})$. The last bound, based on $(\Phat_r - \bP_r) = O_p(\alpha_n)$ and the compactness of $\Xsc$, is uniform in $(i,\bx)$. For any matrix $\mathbf{A} = [a_{\mathfrak{i}\mathfrak{j}}]$, let $\|\mathbf{A}\|_{\max}$ denote the max-norm of $\mathbf{A}$, and $|\mathbf{A}|$ denote the matrix $[|a_{\mathfrak{i}\mathfrak{j}}|]$. Now, Assumption \ref{assmpn_unifconv_snp} (viii) implies: $\|\bnabla K(\bw_1) - \bnabla K(\bw_2)\| \leq B\|\bw_1-\bw_2\| \; \forall \; \bw_1,\bw_2 \in \mathbb{R}^r$, for some constant $B < \infty$. Then using the above arguments, we note that $\forall \; \varrho \in \{0,1\}$, $\| \sup_{\bx \in \Xsc} |\bMhat^{(2)}_{n,\varrho,\bx}| \|_{\max}$ is bounded by:
\begin{align*}
& \phantom{\leq}  \text{sup}_{\bx \in \Xsc} \left\{\frac{B}{nh^{r+1}} \sum_{i=1}^n \|\bx-\bX_i\| \left\|\bwix- \frac{\bP_r'\bx-\bP_r'\bX_i}{h} \right\| |Y_i^{\varrho}| \right\} \\
& \leq \; \text{sup}_{\bx \in \Xsc} \left\{\frac{B}{nh^{r+1}}\sum_{i=1}^n \|\bx-\bX_i\|
\left\| \frac{(\Phat_r-\bP_r)'(\bx-\bX_i)}{h} \right\| |Y_i^{\varrho}| \right\} \\
& \leq  \underset{\bx \in \Xsc,\bX \in \Xsc}{\text{sup}}\left\{\|\bx-\bX\|
\|(\Phat_r-\bP_r)'(\bx-\bX)\|\right\} \frac{B}{nh^{r+2}}\sum_{i=1}^n |Y_i^{\varrho}| \; \leq  O_p\left(\frac{\alpha_n}{h^{r+2}}\right) .
\end{align*}
The first two steps above use the triangle inequality, the Lipschitz continuity of $\bnabla K(\cdot)$ and the definition of $\bwix$, while the next two use the compactness of $\Xsc$, the uniform bound obtained in the last paragraph, the Law of Large Numbers (LLN), and that $(\Phat_r - \bP_r) = O_p(\alpha_n)$. Thus, we have:
\begin{equation}
\text{sup}_{\bx \in \Xsc}\left|\mbox{trace}\left\{(\Phat_r-\bP_r)'\bMhat^{(2)}_{n,\varrho,\bx}\right\}\right| = O_p\left(\frac{\alpha_n^2}{h^{r+2}}\right) \quad \forall \; \varrho \in \{0,1\}.\label{c_eqn4}
\end{equation}

\hypertarget{Mhat1_def}{}Now for bounding $\bMhat^{(1)}_{n,\varrho,\bx}$, let us first write it as: $\bMhat^{(1)}_{n,\varrho,\bx} = \bMhat^{(1,1)}_{n,\varrho,\bx} - \bMhat^{(1,2)}_{n,\varrho,\bx}$, where $\bMhat^{(1,1)}_{n,\varrho,\bx}$ $= (nh^{r+1})^{-1} \sum_{i=1}^n \bx\bnabla K' \{\bP_r'(\bx-\bX_i)/h\} Y_i^{\varrho}$ and $\bMhat^{(1,2)}_{n,\varrho,\bx}$ $= (nh^{r+1})^{-1}\sum_{i=1}^n \bX_i \bnabla K' \{\bP_r'(\bx-\bX_i)/h\} Y_i^{\varrho}$ $\forall \; \varrho \in \{0,1\}$. Then, under Assumption \ref{assmpn_unifconv_snp} (iii), (vi) and (vii), using Theorem 2 of \citet{Hansen_2008} along with the compactness of $\Xsc$, we have: for each $s \in \{1,2\}$ and $\varrho \in \{0,1\}$,
\begin{eqnarray}
&& \left\|\text{sup}_{\bx \in \Xsc} \left| \bMhat^{(1,s)}_{n,\varrho,\bx} - \E_{\Lsc}\left(\bMhat^{(1,s)}_{n,\varrho,\bx}\right) \right| \right\|_{\max} \leq
O_p\left(\frac{\log n}{nh^{r+2}}\right)^{\half} . \label{c_eqn5}
\end{eqnarray}
Now, $\forall \; \varrho \in \{0,1\}$, let $\nu^{(\varrho)}(\bw) =$ $\E\{Y^{\varrho}\given\bXPr = \bw\}\fPr(\bw)$ and $\bxi^{(\varrho)}(\bw) =$ $\E\{\bX Y^{\varrho}\given \bXPr = \bw\}\fPr(\bw)$. Further, let $\{\bnabla\nu^{(\varrho)}(\bw)\}_{r \times 1}$ and $\{\bnabla\bxi^{(\varrho)}(\bw)\}_{p \times r}$ denote their respective first order derivatives. Then, $\forall \; \varrho \in \{0,1\}$, we have:
\begin{eqnarray}
\nonumber && \left\|\text{sup}_{\bx \in \Xsc}\left|\E_\Lsc\left(\bMhat^{(1,1)}_{n,\varrho,\bx}\right) \right| \right\|_{\max} \\
\nonumber &=& \left\|\text{sup}_{\bx \in \Xsc}\left|\frac{\bx}{h^{r+1}} \int \nu^{(\varrho)}(\bw) \; \bnabla K'\left(\frac{\bP_r'\bx - \bw}{h}\right)d\bw \right| \right\|_{\max}\\
&=& \left\|\text{sup}_{\bx \in \Xsc} \left|\bx \int \bnabla \nu^{(\varrho)'}\left(\bP_r'\bx+h\bpsi\right)\; K(\bpsi)d\bpsi\right| \right\|_{\max}
=  \; O(1), \label{c_eqn6}
\end{eqnarray}
\begin{eqnarray}
\nonumber && \left\|\text{sup}_{\bx \in \Xsc}\left|\E_\Lsc\left(\bMhat^{(1,2)}_{n,\varrho,\bx}\right) \right| \right\|_{\max} \\
\nonumber &=& \left\|\text{sup}_{\bx \in \Xsc}\left|h^{-(r+1)} \int \bxi^{(\varrho)}(\bw) \; \bnabla K'\left(\frac{\bP_r'\bx - \bw}{h}\right) d\bw \right| \right\|_{\max} \\
&=& \left\| \text{sup}_{\bx \in \Xsc}\left|\mathbf{D}(\bx) \int \bnabla\bxi^{(\varrho)} \left(\bP_r'\bx+h\bpsi\right) \; K(\bpsi)d\bpsi\right| \right\|_{\max} = \; O(1), \label{c_eqn7}
\end{eqnarray}
where, $\forall \; \bx \in \Xsc$, $\mathbf{D}(\bx)$ denotes the $p \times p$ diagonal matrix: $\mbox{diag}(\bx_{[1]},\hdots,\bx_{[p]})$. In both (\ref{c_eqn6}) and (\ref{c_eqn7}), the first step follows from definition, the second from standard arguments based on integration by parts (applied coordinate-wise) and change of variable, while the last one is due to compactness of $\Xsc$ and a medley of the conditions in Assumption \ref{assmpn_unifconv_snp} namely, boundedness and integrability of $K(\cdot)$ and $\bnabla K(\cdot)$, (iii) and (v) for (\ref{c_eqn6}) so that $\bnabla\nu^{(\varrho)}(\cdot)$ is bounded on $\chiPr$, and (ix) for (\ref{c_eqn7}). It now follows that for each $\varrho \in \{0,1\}$,
\begin{eqnarray}
\left\|\text{sup}_{\bx \in \Xsc}\left|\E_\Lsc\left(\bMhat^{(1)}_{n,\varrho,\bx}\right)\right| \right\|_{\max} = \; O(1). \label{c_eqn8}
\end{eqnarray}
Letting $d_n^* = (\log n)^{\half}(nh^{r+2})^{-\half}$, we now have from (\ref{c_eqn5}) and (\ref{c_eqn8}):
\begin{equation}\label{c_eqn9}
\text{sup}_{\bx \in \Xsc}\left|\mbox{trace}\left\{(\Phat_r'-\bP_r')\bMhat^{(1)}_{n,\varrho,\bx}\right\}\right|  = \;  O_p\left(\alpha_nd_n^*+\alpha_n\right) \;\; \forall \; \varrho \in \{0,1\}.
\end{equation}
Applying (\ref{c_eqn9}) and (\ref{c_eqn4}) to (\ref{c_eqn3}) using the triangle inequality, we have $\forall \; \varrho$,
\begin{equation}\label{c_eqn10}
\text{sup}_{\bx \in \Xsc}|\vphihat^{(\varrho)}(\bx;\Phat_r) - \vphitil^{(\varrho)}(\bx;\bP_r)| = O_p\left\{\frac{\alpha_n^2}{h^{r+2}}+\alpha_n \frac{(\log n)^{\half}}{(nh^{r+2})^{\half}} + \alpha_n\right\}.
\end{equation}
Finally, note that $\mhat(\bx;\Phat_r) = \lhat(\bx;\Phat_r)/\fhat(\bx;\Phat_r) = \vphihat^{(1)}(\bx;\Phat_r)/\vphihat^{(0)}(\bx;\Phat_r)$. Repeated use of (\ref{c_eqn10}), along with (\ref{c_eqn2}) and Assumption \ref{assmpn_unifconv_snp} (iii)-(iv),  leads to:
\begin{eqnarray}
\nonumber && \text{sup}_{\bx\in\Xsc}\;\left|\mhat(\bx;\Phat_r)- m(\bx;\bP_r)\right| \\
\nonumber && \leq \text{sup}_{\bx\in\Xsc}\;\left|\mhat(\bx;\Phat_r)- \mtil(\bx;\bP_r)\right| + \text{sup}_{\bx\in\Xsc}\;\left|\mtil(\bx;\bP_r)- m(\bx;\bP_r)\right| \\
\nonumber && \leq  \text{sup}_{\bx \in \Xsc}\left\{\left|\frac{\lhat(\bx;\Phat_r)- \ltil(\bx;\bP_r)}{\fhat(\bx;\Phat_r)}\right| +
\left|\frac{\ltil(\bx;\bP_r)}{\ftil(\bx;\bP_r)} - \frac{\ltil(\bx;\bP_r)}{\fhat(\bx;\Phat_r)} \right|\right\} + O_p(a_{n,2}) \\
&& \; \leq O_p\left\{\frac{\alpha_n^2}{h^{r+2}}+\alpha_n \frac{(\log n)^{\half}}{(nh^{r+2})^{\half}} + \alpha_n\right\} + O_p(a_{n,2}) \;\; =O_p(a_{n,1}+a_{n,2}). \label{c_eqn11}
\end{eqnarray}
The proof of Theorem \ref{thm3} is now complete. \qed

\section*{VI. Proofs of Lemmas \ref{lemma1}-\ref{lemma2} and Theorem \ref{thm4}}\label{appC_4} 

\subsection*{\emph{VI.1.} Proof of Lemma \ref{lemma1}}\label{appC_4_lemma1}
\phantomsection
\addcontentsline{toc}{subsection}{Proof of Lemma \ref{lemma1}}
First note that for each $\varrho \in \{0,1\}$,
$$\int \vphitil^{(\varrho)}(\bx;\bP_r)\P_n (d\bx) = n^{-2} \sum_{{i_1}=1}^n\sum_{{i_2}=1}^n \bH_{i_1,i_2}^{(n,\varrho)}$$ is a V-statistic, where $\bH_{i_1,i_2}^{(n,\varrho)} = h^{-r}\blambda(\bX_{i_1})Y_{i_2}^{\varrho}K\{\bP_r'(\bX_{i_1}-\bX_{i_2})/h\}$. Using the V-statistic projection result given in Lemma 8.4 of \citet{Newey_Book_1994}, it then follows that for each $\varrho \in \{0,1\}$,
\begin{eqnarray}\label{c_eqn15}
\nonumber && \G_n^*\left\{\blambda(\cdot)[\vphitil^{(\varrho)}(\cdot\;;\bP_r) - \E_{\Lsc}\{\vphitil^{(\varrho)}(\cdot\;;\bP_r)\}]\right\} \\
&=& \nnhalf O_p\left[\E(\|\bH_{i_1,i_1}^{(n,\varrho)}\|) + \{\E(\|\bH_{i_1,i_2}^{(n,\varrho)}\|^2)\}^{\half}\right] = O_p\left(\nnhalf h^{-r}\right),
\end{eqnarray}
The last step follows from $K(\cdot)$ and $\blambda(\cdot)$ being bounded and $Y^{\varrho}$ having finite $2^{nd}$ moments. Now, observe that  $\nhalf \G_n^*\left\{\blambda(\cdot)[\E_{\Lsc}\{\vphitil_{\star}^{(\varrho)}(\cdot\;;\bP_r)\} ]\right\}$  is a centered sum of i.i.d. random vectors bounded by: 
$$D_{n,\varrho} = 
\text{sup}_{\bx \in \Xsc} 
\left\{\|\blambda(\bx)\| \; |\E_{\Lsc}\{\vphitil_{\star}^{(\varrho)}(\bx;\bP_r)\} |\right\} = O(h^q)  \quad \forall \; \varrho \in \{0,1\},$$
where throughout, for any estimator $\tilde{\xi}(\cdot)$ with population limit $\xi(\cdot)$, we use the notation $\tilde{\xi}_{\star}(\cdot)$ to denote its centered version given by: $\tilde{\xi}_{\star}(\cdot) = \tilde{\xi}(\cdot) - \xi(\cdot)$.
Here, $D_{n,\varrho}=O(h^q)$ since $\blambda(\cdot)$ is bounded and $\text{sup}_{\bx \in \Xsc}|\E_{\Lsc}\{\vphitil_{\star}(\bx;\bP_r)\}| = \text{sup}_{\bw \in \chiPr} |\E_{\Lsc}\{\vphiPrtil^{(\varrho)}(\bw)\} - \vphiPr^{(\varrho)}(\bw)| = O(h^q)$, as argued while proving (\ref{c_eqn1}). Hence, $\exists$ a constant $\kappa_{\varrho} > 0$ such that $h^q/D_{n,\varrho} \geq \kappa_{\varrho} \; \forall \; n$. Then, using Hoeffding's Inequality, we have: $\forall \; n$, given any $\epsilon > 0$ and any $M = M(\epsilon)$ large enough,
\begin{gather}
\nonumber \sum_{l=1}^d\P\left[ \left|\G_n^*\left\{\blambda_{[l]}(\cdot)[\E_{\Lsc}\{\vphitil_{\star}^{(\varrho)}(\cdot\;;\bP_r)\} ]\right\}\right| > \frac{Mh^q}{d^{\half}} \right]
\leq  2d\;\mbox{exp} \left(-\;\frac{M^2h^{2q}}{2dD_{n,\varrho}^2}\right) \Rightarrow  \\
\nonumber \P\left[\left\|\G_n^*\left\{\blambda(\cdot)[\E_{\Lsc}\{\vphitil_{\star}^{(\varrho)}(\cdot\;;\bP_r)\} ]\right\}\right\| > Mh^q \right] \leq 2d\;\mbox{exp} \left(-\;\frac{M^2\kappa_{\varrho}^2}{2d}\right) \leq \epsilon \;\; \Rightarrow\\
\G_n^*\left\{\blambda(\cdot)[\E_{\Lsc}\{\vphitil_{\star}^{(\varrho)}(\cdot\;;\bP_r)\} ]\right\} = O_p(h^q) \quad \forall \; \varrho \in \{0,1\}. \label{c_eqn16}
\end{gather}
Combining (\ref{c_eqn15}) and (\ref{c_eqn16}) using the linearity of $\G_n^*(\cdot)$, we then have (\ref{lemma1_eqn1}). \qed
\par\smallskip
Next, to show (\ref{lemma1_eqn2}), let $f(\bx;\bP_r) = \varphi^{(0)}(\bx;\bP_r)$ and $l(\bx;\bP_r) = \varphi^{(1)}(\bx;\bP_r)$. Then, we write
\begin{equation*}
\G_n^*[\blambda(\cdot)\{\mtil_{\star}(\cdot\;;\bP_r)\}] \; = \; \G_n^*[\blambda(\cdot)\{\bTtil_{n,\bP_r}^{(1)}(\cdot) - \bTtil_{n,\bP_r}^{(2)}(\cdot) - \bTtil_{n,\bP_r}^{(3)}(\cdot) + \bTtil_{n,\bP_r}^{(4)}(\cdot)\}],
\end{equation*}
where
\begin{alignat}{2}
\nonumber & \bTtil_{n,\bP_r}^{(1)}(\bx) =  \frac{\ltil_{\star}(\bx;\bP_r) }{f(\bx;\bP_r)}, &\;\;& \bTtil_{n,\bP_r}^{(2)}(\bx) = \frac{\ftil_{\star}(\bx;\bP_r)l(\bx;\bP_r)}{f(\bx;\bP_r)^2},\\
& \bTtil_{n,\bP_r}^{(3)}(\bx) = \frac{\ltil_{\star}(\bx;\bP_r) \ftil_{\star}(\bx;\bP_r)} {\ftil(\bx;\bP_r)f(\bx;\bP_r)}, &\;\;\mbox{and}\;\;& \bTtil_{n,\bP_r}^{(4)}(\bx) = \frac{l(\bx;\bP_r)\ftil_{\star}(\bx;\bP_r)^2} {\ftil(\bx;\bP_r)f(\bx;\bP_r)^2}. \label{c_eqn17}
\end{alignat}
Since $\blambda^{(1)}_{\bP_r}(\bx) \equiv \blambda(\bx)f(\bx;\bP_r)^{-1}$ and $\blambda^{(2)}_{\bP_r}(\bx) \equiv \blambda(\bx)l(\bx;\bP_r)f(\bx;\bP_r)^{-2}$ are  bounded a.s. [$\P_{\bX}$] due to Assumption \ref{assmpn_unifconv_snp} (iii)-(iv) and the boundedness of $\blambda(\cdot)$, using these as choices of `$\blambda(\cdot)$' in (\ref{lemma1_eqn1}), we have:
\begin{align*}
& \G_n^*\{\blambda_{\bP_r}^{(1)}(\cdot)\ltil_{\star}(\cdot\;;\bP_r)\} = \G_n^*\{\blambda(\cdot)\bTtil_{n,\bP_r}^{(1)}(\cdot)\} = O_p(b_n^{(1)}), \\
& \G_n^*\{\blambda_{\bP_r}^{(2)}(\cdot)\ftil_{\star}(\cdot\;;\bP_r)\} = \G_n^*\{\blambda(\cdot)\bTtil_{n,\bP_r}^{(2)}(\cdot)\} = O_p(b_n^{(1)}).
\end{align*}
Further,  for each $s \in \{3,4\}$, $\sup_{\bx \in \Xsc} \;\|\bTtil_{n,\bP_r}^{(s)}(\bx)\| \leq O_p(a_{n,2}^2)$ which follows from repeated use of (\ref{c_eqn1}) along with Assumption \ref{assmpn_unifconv_snp} (iii)-(iv). Consequently, with $\blambda(\cdot)$ bounded a.s. [$\P_{\bX}$], for each $s \in \{3,4\}$, $\G_n^*\{\blambda(\cdot)\bTtil_{n,\bP_r}^{(s)}(\cdot)\}$ is bounded by: $O_p(\nhalf a_{n,2}^2)$. Combining all these results using the linearity of $\G_n^*(\cdot)$, we finally obtain: $\G_n^*\{\blambda(\cdot)\mtil_{\star}(\cdot\;;\bP_r)\} = O_p(b_n^{(1)} + \nhalf a_{n,2}^2)$ $= O_p(\nhalf a_{n,2}^2)$, thus leading to (\ref{lemma1_eqn2}). The proof of the lemma is now complete. \qed

\subsection*{\emph{VI.2.} Proof of Lemma \ref{lemma2}}\label{appC_4_lemma2_and_final_proof}
\phantomsection
\addcontentsline{toc}{subsection}{Proof of Lemma \ref{lemma2}}

Throughout this proof, all additional notations introduced, if not explicitly defined, are understood to have been adopted from the proof of Theorem \ref{thm3} in Section \hyperref[appC]{V}. Now, using (\ref{c_eqn3}), $\vphihat^{(\varrho)}(\bx;\Phat_r) - \vphitil^{(\varrho)}(\bx;\bP_r) = \mbox{trace}\{(\Phat_r'-\bP_r')\bMhat^{(1)}_{n,\varrho,\bx}\} + \mbox{trace}\{(\Phat_r'-\bP_r') \bMhat^{(2)}_{n,\varrho,\bx}\}$, and $ \bMhat^{(1)}_{n,\varrho,\bx}= \bMhat^{(1,1)}_{n,\varrho,\bx} - \bMhat^{(1,2)}_{n,\varrho,\bx}$, as \hyperlink{Mhat1_def}{defined} in Section \hyperref[appC]{V}. Thus,
$$
\G_n^*[\blambda(\cdot) \{\vphihat^{(\varrho)}(\cdot\;;\Phat_r) - \vphitil^{(\varrho)}(\cdot\;;\bP_r)\}] = \G_n^*\left\{\bzetahat^{(1,1)}_{n,\varrho,\blambda}(\cdot) - \bzetahat^{(1,2)}_{n,\varrho,\blambda}(\cdot) +  \bzetahat^{(2)}_{n,\varrho,\blambda}(\cdot)\right\},
$$
where $ \forall \; (\omega) \in \{(1,1), (1,2), (2)\}$, $\varrho \in \{0,1\}$, and $\bx \in \Xsc$,
\begin{equation}
\bzetahat^{(\omega)}_{n,\varrho,\blambda}(\bx) = \blambda(\bx)\;\mbox{trace}\left\{(\Phat_r'-\bP_r')\bMhat^{(\omega)}_{n,\varrho,\bx}\right\}. \label{c_eqn18}
\end{equation}
Then, $\forall \; s \in \{1,2\}$ and $l \in \{1,\hdots,d\}$,  each element of
$$\int \blambda_{[l]}(\bx) \bMhat^{(1,s)}_{n,\varrho,\bx}  \P_n (d\bx) = n^{-2}\sum_{{i_1}=1}^n\sum_{{i_2}=1}^n \Hbb_{l,\varrho}^{(n,s)}(i_1,i_2)$$
is a V-statistic, where
$$\Hbb_{l,\varrho}^{(n,s)}(i_1,i_2) = h^{-(r+1)} \blambda_{[l]}(\bX_{i_1}) Y_{i_2}^{\varrho}\bU^{(s)} (i_1,i_2) \bnabla K'\{\bP_r'(\bX_{i_1}-\bX_{i_2})/h\} $$
with $\bU^{(1)}(i_1,i_2) = \bX_{i_1}$ and $\bU^{(2)}(i_1,i_2) = \bX_{i_2}$. Hence, similar to the proof of (\ref{c_eqn15}), using Lemma 8.4 of \citet{Newey_Book_1994} with $\Xsc$ compact, $\bnabla K(\cdot)$ and $\blambda(\cdot)$ bounded, and $Y^{\varrho}$ having finite $2^{nd}$ moments, we have: for each $l \in \{1,\hdots,d\}$, $s \in \{1, 2\}$ and $\varrho \in \{0,1\}$,
$$ \left\| \G_n^*\left[\blambda_{[l]}(\cdot) \bMhat^{(1,s)}_{n,\varrho,(\cdot)} - \E_\Lsc\left\{\blambda_{[l]}(\cdot) \bMhat^{(1,s)}_{n,\varrho,(\cdot)} \right\}\right] \right\|_{\max}= O_p\left(\nnhalf h^{-(r+1)}\right).$$
It then follows from $(\Phat_r -\bP_r) = O_p(\alpha_n)$ that for each $s$ and $\varrho$,
\begin{eqnarray}
&& \G_n^*\left[\bzetahat^{(1,s)}_{n,\varrho,\blambda}(\cdot) - \E_\Lsc\left\{\bzetahat^{(1,s)}_{n,\varrho,\blambda}(\cdot)\right\}\right] = O_p\left(\alpha_n\nnhalf h^{-(r+1)} \right). \label{c_eqn19}
\end{eqnarray}
Next, for any given $l$, $s$ and $\varrho$, each element of $\nhalf \G_n^*[\E_\Lsc\{\blambda_{[l]}(\cdot) \bMhat^{(1,s)}_{n,\varrho,(\cdot)} \}]$ is a centered sum of i.i.d. random variables which are bounded by:
$$
\left\|\text{sup}_{\bx \in \Xsc}\left\{\|\blambda(\bx)\| \; |\E_{\Lsc} (\bMhat^{(1,s)}_{n,\varrho,\bx})| \right\} \right\|_{\max}= O(1),
$$
where the order follows from (\ref{c_eqn6}), (\ref{c_eqn7}) and the boundedness of $\blambda(\cdot)$. Hence, similar to the proof of (\ref{c_eqn16}), using Hoeffding's inequality and that $(\Phat_r -\bP_r) = O_p(\alpha_n)$, we have: $\forall \; l \in \{1,\hdots,d\}$, $s \in \{1,2\}$ and $\varrho \in \{0,1\}$,
{\small
\begin{equation}
\left\| \G_n^*\left[\E_\Lsc\left\{\blambda_{[l]}(\cdot) \bMhat^{(1,s)}_{n,\varrho,(\cdot)}  \right\}\right] \right\|_{\max}= O_p(1) \Rightarrow \G_n^*\left[\E_\Lsc\left\{\bzetahat^{(1,s)}_{n,\varrho,\blambda}(\cdot)\right\}\right] =O_p(\alpha_n). \label{c_eqn20}
\end{equation}
}\vspace{-.1in}

For any matrix $\mathbf{A}$, let us denote by $\mathbf{A}_{[a,b]}$ the $(a,b)^{th}$ element of $\mathbf{A}$. Now, to control  $\G_n^*\{\bzetahat^{(2)}_{n,\varrho,\blambda}(.)\}$ in (\ref{c_eqn18}), note that $\|\G_n^*\{\bzetahat^{(2)}_{n,\varrho,\blambda}(\cdot)\}\|$ is bounded by:
\begin{eqnarray}
&& \nhalf \; \text{sup}_{\bx \in \Xsc}\left\|\blambda(\bx)\right\| \sum_{a,b} \int \left|(\Phat_r'-\bP_r')_{[b,a]}
\left(\bMhat^{(2)}_{n,\varrho,\bx}\right)_{[a,b]}\right|  (\P_n + \P_{\bX})  (d\bx) \nonumber \\
&& \;\; \leq \nhalf rp \underset{\bx \in \Xsc, \bX \in \Xsc}{\text{sup}} \{\left\|\blambda(\bx)\right\|\|\bx -\bX\|\} \left\| \Phat_r -\bP_r\right\|_{\max} \Zhat_{n}^{\varrho *} \nonumber \\
&& \;\; \leq O_p\left(\nhalf \alpha_n\right)  \Zhat_{n}^{\varrho *} , \label{c_eqn21}
\end{eqnarray}
where the last step follows from $(\Phat_r -\bP_r) = O_p(\alpha_n)$ and the boundedness of $\Xsc$ and $\blambda(\cdot)$, and
$\Zhat_{n}^{\varrho *} = \int\Zhat_{n}^{(\varrho)}(\bx) \hspace{0.75mm} (\P_n + \P_{\bX}) (d\bx)$ with
$$
\Zhat^{(\varrho)}_n(\bx) 
=\ninv \sum_{i=1}^n \frac{|Y_i^{\varrho}|}{h^{r+1}}\left\|\bnabla K (\bwix) - \bnabla K \left\{\frac{\bP_r'(\bx - \bX_i)}{h}\right\}\right\|.
$$
Now, $\|\bwix - \bP_r'(\bx-\bX_i)h^{-1}\| \leq \|(\Phat_r - \bP_r)'(\bx-\bX_i)h^{-1}\| \leq O_p(\alpha_nh^{-1})$ uniformly in $(i,\bx)$, as noted while proving (\ref{c_eqn4}). Further, with $L^*$, as defined in Assumption \ref{assmpn_unifconv_snp} (vii), let $\A_n$ denote the event: $\{\|(\Phat_r - \bP_r)'(\bx-\bX_i)h^{-1}\| \leq L^* \;\; \forall \; \bx \in \Xsc, \; i = 1,..,n \}$. Then, with $(\Phat_r - \bP_r) = O_p(\alpha_n)$, $\Xsc$ compact and $\alpha_nh^{-1} = o(1)$ since $\nhalf \alpha_n^2h^{-2} = o(1)$ as assumed, it follows that $\P(\A_n) \rightarrow 1$. Using these along with Assumption \ref{assmpn_unifconv_snp} (vii) and the function $\phi(.)$ defined therein, we have: on $\A_n$ with $\P(\A_n) \rightarrow 1$,
\begin{align*}
 \Zhat^{(\varrho)}_n(\bx) & \leq  \sum_{i=1}^n \frac{|Y_i^{\varrho}|}{nh^{r+1}} \left\|\frac{(\Phat_r-\bP_r)'(\bx-\bX_i)}{h}\right\| \phi\left\{\frac{\bP_r'(\bx-\bX_i)}{h}\right\} \\
& \leq  
\sqrt{rp}
\underset{\bx \in \Xsc,
\bX \in \Xsc
}{\text{sup}} 
\left\|\bx -\bX\right\| \left\|\Phat_r -\bP_r\right\|_{\max} \sum_{i=1}^n \frac{|Y_i^{\varrho}|}{nh^{r+2}}\hspace{0.5mm} \phi\left\{\frac{\bP_r'(\bx-\bX_i)}{h}\right\}.
\end{align*}
Thus, $\Zhat_n^{\varrho *} \leq O_p\left(\alpha_n \Ztil_{n}^{\varrho *}\right)$, where $\Ztil_{n}^{\varrho *} = \int\Ztil_{n}^{(\varrho)}(\bx)  (\P_n + \P_{\bX}) (d\bx)$,
$$
\Ztil_{n}^{(\varrho)}(\bx) = \ninv \sum_{i=1}^n \Ztil_{n}^{(\varrho)}(\bx;\bZ_i),  \ \mbox{and} \;\;
\Ztil_{n}^{(\varrho)}(\bx;\bZ)  =  \frac{|Y^{\varrho}|}{h^{r+2}} \phi\left\{\frac{\bP_r'(\bx-\bX)}{h}\right\} .
$$
Let $\bZ^0 \equiv (Y^0,\bX^{0'})' \sim \P_{\bZ}$ be generated independent of $\Lsc$, and define:
\begin{alignat*}{2}
&\Util_{n,\varrho}^{(1)} = \ninv \sum_{i=1}^n \E_{\bX^0}\{\Ztil_{n}^{(\varrho)}(\bX^0;\bZ_i)\}, &\quad&
\Util_{n,\varrho}^{(2)} = \ninv \sum_{i=1}^n \E_{\bZ^0}\{\Ztil_{n}^{(\varrho)}(\bX_i;\bZ^0)\},   \\
&\Util_{n,\varrho}^{(1,1)}  = \E\{\Ztil_{n}^{(\varrho)}(\bX^0;\bZ^0)\}, \quad \mbox{and} \ &\quad&
\Vtil_{n,\varrho}^{(k)} = \E\{\Ztil_{n}^{(\varrho)}(\bX^0; \bZ)^k\} \; \mbox{for}\; k =1,2.
\end{alignat*}
Then, first note that: $\int\Ztil_{n}^{(\varrho)}(\bx) \P_{\bX} (d\bx) = \Util_{n,\varrho}^{(1)}$. Further, since
$$\int \Ztil_{n}^{(\varrho)}(\bx) \P_n(d\bx) = n^{-2} \sum_{i_1=1}^n\sum_{i_2=1}^n \Ztil_{n}^{(\varrho)}(\bX_{i_1};\bZ_{i_2})$$
is a V-statistic, we have:
$$
\int \Ztil_{n}^{(\varrho)}(\bx) \P_n(d\bx) = \Util_{n,\varrho}^{(1)} + \Util_{n,\varrho}^{(2)} - \Vtil_{n,\varrho}^{(1)} + O_p\{\ninv \Util_{n,\varrho}^{(1,1)}  +
\ninv (\widetilde{\mathbb{V}}_{n,\varrho}^{(2)})^{\half}\}
$$
using Lemma 8.4 of \citet{Newey_Book_1994}. Then, with all notations as above, we have:
\begin{equation}
 \ninv \Util_{n,\varrho}^{(1,1)} +  \ninv (\widetilde{\mathbb{V}}_{n,\varrho}^{(2)})^{\half} \leq O_p\left(\ninv h^{-(r+2)}\right), \qquad \mbox{and} \label{c_eqn23}
\end{equation}
\begin{eqnarray}
\nonumber  \Util_{n,\varrho}^{(1)} && \hspace{-.2in} = \frac{1}{nh^{r+2}} \sum_{i=1}^n |Y_i^{\varrho}| \int_{\chiPr}\phi\left(\frac{\bw - \bP_r'\bX_i}{h}\right)\fPr(\bw) d\bw \\
\nonumber && \hspace{-.2in} \leq 
\frac{B_{\bP_r}}{nh^{2}} \sum_{i=1}^n \left\{|Y_i^{\varrho}| \int_{A_{\bX_i}^n}\phi(\bpsi_i)d\bpsi_i\right\},  \\
&& \hspace{-.2in} \leq \frac{B_{\bP_r}}{h^{2}} \left\{\int_{\mathbb{R}^r} \phi(\bpsi)d\bpsi \right\} \left\{\ninv \sum_{i=1}^n |Y_i^{\varrho}|\right\} \leq O_p\left(h^{-2}\right), \label{c_eqn24}
\end{eqnarray}
where $\bpsi_i = h^{-1}(\bw-\bP_r'\bX_i)$ $\forall \; i$, $A_{\bx}^n = \{\bpsi: (\bP_r'\bx + h\bpsi) \in \chiPr\}$ $\forall \; \bx \in \Xsc$, and $B_{\bP_r} = \text{sup}_{\bw \in \chiPr} \fPr(\bw) < \infty$.
The error rate in (\ref{c_eqn23}) follows since $\phi(\cdot)$ is bounded and $Y^{\varrho}$ has finite $2^{nd}$ moments, while that of $\Util_{n,\varrho}^{(1)}$ follows from Assumption \ref{assmpn_unifconv_snp} (iii), integrability of $\phi(\cdot)$, and LLN applied to the sequence $\{Y_i^{\varrho}\}_{i=1}^n$ having finite $2^{nd}$ moments. Now, note that $\Util_{n,\varrho}^{(2)} - \Vtil_{n,\varrho}^{(1)}$ is a centered average of
$[\E_{\bZ^0}\{\Ztil_{n}^{(\varrho)}(\bX_i;\bZ^0)\}]_{i=1}^n$ which are i.i.d. and bounded by:
$$\underset{\bx \in \Xsc}{\text{sup}}\;\E_{\bZ}\{\Ztil_{n}^{(\varrho)}(\bx;\bZ)\} =
\underset{\bx \in \Xsc}{\text{sup}}\;\frac{1}{h^{r+2}}\int_{\chiPr} \phi\left(\frac{\bP_r'\bx - \bw}{h}\right)\overline{m}_{\bP_r}^{(\varrho)}(\bw)\fPr(\bw) d\bw,$$
where $\overline{m}_{\bP_r}^{(\varrho)}(\bw) = \E(|Y|^{\varrho}\given \bXPr = \bw)$ $\forall \; \varrho \in \{0,1\}$ and $\bw \in \chiPr$.
Using the integrability of $\phi(\cdot)$, we then have:
\begin{align*}
& \ \underset{\bx \in \Xsc}{\text{sup}} \; \E_{\bZ}\{\Ztil_{n}^{(\varrho)}(\bx;\bZ)\} \; \leq \; \underset{\bx \in \Xsc}{\text{sup}} \; \frac{C_{\bP_r}^{(\varrho)}}{h^{r+2}}\int_{\chiPr} \phi\left(\frac{\bP_r'\bx - \bw}{h}\right)d\bw  \\
\leq & \ \underset{\bx \in \Xsc}{\text{sup}}\;\frac{C_{\bP_r}^{(\varrho)}}{h^2}\int_{A^n_\bx}\phi( - \bpsi)\;d\bpsi
\leq \frac{C_{\bP_r}^{(\varrho)}}{h^2} \left\{\int_{\mathbb{R}^r} \phi(\bpsi)d\bpsi\right\} \; = \; O\left(h^{-2}\right),\\
\end{align*}
where $C_{\bP_r}^{(\varrho)} = \text{sup}_{\bw \in \chiPr} \overline{m}_{\bP_r}^{(\varrho)}(\bw)\fPr(\bw) < \infty$ due to Assumption \ref{assmpn_unifconv_snp} (iii), and $A^n_\bx = \{\bpsi: (\bP_r'\bx + h\bpsi) \in \chiPr\}$, as before.
It then follows, similar to the proof of (\ref{c_eqn16}), from a simple application of Hoeffding's inequality that
\begin{equation}
\Util_{n,\varrho}^{(2)} - \Vtil_{n,\varrho}^{(1)} = O_p\left(\nnhalf h^{-2}\right).\label{c_eqn25}
\end{equation}
Using (\ref{c_eqn23})-(\ref{c_eqn25}), we finally have: $\Ztil_{n}^{\varrho *}  =  O_p(h^{-2}+ \ninv h^{-(r+2)})$. Hence,
\begin{eqnarray}
&& \Zhat_n^{\varrho *} = \int\Zhat^{(\varrho)}_n(\bx) \hspace{0.75mm} (\P_n + \P_{\bX}) (d\bx) \leq O_p\left(\alpha_n \Ztil_{n}^{\varrho *}\right) = O_p\left(\frac{\alpha_n}{h^2}+ \frac{\alpha_n}{nh^{r+2}}\right), \label{c_eqn26} \\
&& \mbox{and}\quad \left\|\G_n^*\left\{\bzetahat^{(2)}_{n,\varrho,\blambda}(\cdot)\right\}\right\| \leq O_p\left(\frac{\nhalf\alpha_n^2}{h^2}+ \frac{\nhalf \alpha_n^2}{nh^{r+2}}\right) \;\; \forall \; \varrho \in \{0,1\},\label{c_eqn27}
\end{eqnarray}where the final bound in (\ref{c_eqn27}) follows from (\ref{c_eqn21}). The desired result in (\ref{lemma2_eqn1}) now follows by applying (\ref{c_eqn19}), (\ref{c_eqn20}) and (\ref{c_eqn27}) to (\ref{c_eqn18}) using the linearity of $\G_n^*(\cdot)$. The proof of the lemma is now complete. (Note that conditions (i), (iv) and (viii) in Assumption \ref{assmpn_unifconv_snp} were actually not used in this proof).  \qed

\subsection*{\emph{VI.3.} Proof of Theorem \ref{thm4}}\label{appC_4_final_proof}
\phantomsection
\addcontentsline{toc}{subsection}{Proof of Theorem \ref{thm4}}

Finally, to establish the result of Theorem \ref{thm4}, let $\blambda_0(\bx) = \bxv$ which is measurable and bounded on $\Xsc$. Further, with $\G_n^*(\cdot)$ as defined in Appendix \ref{app-lemmas}, note that $\G_{n,\K}$ for $\K = 1$ is given by:
\begin{equation}
\G_{n,\K} = \G_n^*\{\blambda_0(\cdot)\mtil_{\star}(\cdot\;;\bP_r)\} + \G_n^*[\blambda_0(\cdot)\{\mhat(\cdot\;;\Phat_r)-\mtil(\cdot\;;\bP_r)\}], \label{c_eqn28}
\end{equation}
due to linearity of $\G_n^*(\cdot)$. Now, using Lemma \ref{lemma1}, we have:
\begin{eqnarray}
&& \G_n^*\{\blambda_0(\cdot)\mtil_{\star}(\cdot\;;\bP_r)\} = O_p(\nhalf a_{n,2}^2) = O_p(a_{n,2}^*). \label{c_eqn29}
\end{eqnarray}
The second term $\G_n^*[\blambda_0(\cdot)\{\mhat(\cdot\;;\Phat_r)-\mtil(\cdot\;;\bP_r)\}]$ in (\ref{c_eqn28}) can be written as:
\begin{align}
  &\G_n^*[\blambda_0(\cdot)\{\bThat_{n,\bP_r}^{(1)}(\cdot) - \bThat_{n,\bP_r}^{(2)}(\cdot) - \bThat_{n,\bP_r}^{(3)}(\cdot) + \bThat_{n,\bP_r}^{(4)}(\cdot)\}] \nonumber \\
= \ & O_p\left(b_n^{(2)} + \nhalf a_{n,1}^2 + \nhalf  a_{n,1}a_{n,2}\right) = O_p\left(a_{n,1}^*\right),  \label{c_eqn30}
\end{align}
where with slight abuse of notation,
\begin{align*}
& \bThat_{n,\bP_r}^{(1)}(\bx) =  \frac{\ahat - \atil}{b}, \quad  \bThat_{n,\bP_r}^{(2)}(\bx) = \frac{a(\bhat - \btil)}{b^2}, \\
& \bThat_{n,\bP_r}^{(3)}(\bx) = \frac{(\ahat-\atil)(\btil-b)}{b\;\btil} + \frac{(\ahat-\atil)(\bhat-\btil)}{\btil\;\bhat}, \quad \mbox{and}\\
& \bThat_{n,\bP_r}^{(4)}(\bx) = \frac{\atil(\bhat-\btil)^2}{\bhat\;b^2} \; - \; \frac{(\atil-a)(\bhat-\btil)}{b^2} + \frac{a(\bhat-\btil)(\btil-b)(b+\btil)}{(b\;\btil)(b\;\bhat)},
\end{align*}
with $(a,b) = \{l(\bx;\bP_r),f(\bx;\bP_r)\}$, $(\atil,\btil) = \{\ltil(\bx;\bP_r),\ftil(\bx;\bP_r)\}$ and $(\ahat,\bhat) = \{\lhat(\bx;\Phat_r),\fhat(\bx;\Phat_r)\}\}$.
\par\smallskip
For (\ref{c_eqn30}), the starting expansion is due to a linearization similar to (\ref{c_eqn17}), while the final rate is due to the following: note that $\blambda^{(1)}_{0,\bP_r}(\cdot) \equiv b^{-1}\blambda_0(\cdot)$ and $\blambda^{(2)}_{0,\bP_r}(\cdot) \equiv ab^{-2}\blambda_0(\cdot)$ are both bounded a.s. [$\P_{\bX}$] due to Assumption \ref{assmpn_unifconv_snp} (iii)-(iv) and the boundedness of $\blambda_0(\cdot)$. Hence, using these as choices of `$\blambda(\cdot)$' in Lemma \ref{lemma2}, we have: $\G_n^*\{(\ahat-\atil)\blambda_{0,\bP_r}^{(1)}(\cdot)\} = \G_n^*[\blambda_0(\cdot)\{\bThat_{n,\bP_r}^{(1)}(\cdot)\}] = O_p(b_n^{(2)})$ and $\G_n^*\{(\bhat-\btil)\blambda_{0,\bP_r}^{(2)}(\cdot)\} = \G_n^*[\blambda_0(\cdot)\{\bThat_{n,\bP_r}^{(2)}(\cdot)\}] = O_p(b_n^{(2)})$ respectively. Further, note that for each $s \in \{3,4\}$, $\text{sup}_{\bx \in \Xsc}\|\bThat_{n,\bP_r}^{(s)}(\bx)\| \leq O_p(a_{n,1}^2+ a_{n,1}a_{n,2})$ which follows from repeated use of (\ref{c_eqn1}), (\ref{c_eqn10}) along with Assumption \ref{assmpn_unifconv_snp} (iii)-(iv). Consequently, with  $\blambda_0(\bx)$ bounded a.s. [$\P_{\bX}$], for each $s \in$ $\{3,4\}$, $\G_n^*[\blambda_0(\cdot)\{\bThat_{n,\bP_r}^{(s)}(\cdot)\}]$ is bounded by: $O_p(\nhalf a_{n,1}^2 + \nhalf a_{n,1}a_{n,2})$. Combining all these results using the linearity of $\G_n^*(\cdot)$ and noting that with $a_{n,2}^* = o(1)$, $(b_n^{(2)}+ \nhalf a_{n,1}^2 + \nhalf a_{n,1}a_{n,2}) = O(a^*_{n,1})$, (\ref{c_eqn30}) now follows and, along with (\ref{c_eqn29}) and (\ref{c_eqn28}), implies: $\G_{n,\K} = O_p(a_{n,1}^*+a_{n,2}^*)$ as claimed in Theorem \ref{thm4}. Lastly, using this in (\ref{snp_fund_exp}), the expansion in (\ref{snp_k1_exp}) and its associated implications follow. The proof of Theorem \ref{thm4} is now complete. \qed

\bibliographystyle{imsart-nameyear}
\bibliography{P1_ArXiv_R1_Biblio}


\end{document}